\documentclass[showpacs,prd]{revtex4}
\usepackage{mathrsfs}
\usepackage{longtable,lscape} \usepackage{multirow}
\usepackage{txfonts,bm}
\usepackage{amssymb}
\usepackage{indentfirst}
\usepackage{graphicx,booktabs}
\usepackage{color}
\usepackage{hyperref}
\def\m{\mathcal}
\def\n{\nonumber}
\def\f{\frac}

\def\p{\partial}

\def\s{\sigma}
\def\S{\Sigma}
\def\a{\alpha}
\def\b{\beta}
\def\o{\omega}
\def\O{\Omega}
\def\L{\Lambda}
\def\l{\lambda}
\def\g{\gamma}

\begin{document}

\title{Hadronic Molecular States Composed of Heavy Flavor Baryons}
\author{Ning Li}\email{leening@pku.edu.cn}
\affiliation{Department of Physics
and State Key Laboratory of Nuclear Physics and Technology\\
Peking University, Beijing 100871, China}
\author{Shi-Lin Zhu}
\email{zhusl@pku.edu.cn} \affiliation{Department of Physics
and State Key Laboratory of Nuclear Physics and Technology\\
and Center of High Energy Physics, Peking University, Beijing
100871, China }
%\date{\today}

\begin{abstract}

We investigate the possible molecules composed of two heavy flavor baryons such
as ``$A_QB_Q$"($Q=b,~c$) within the one-pion-exchange model (OPE). Our results
indicate that the long-range $\pi$ exchange force is strong enough to form
molecules such as $[\Sigma_Q\Xi_Q^{'}]^{I=1/2}_{S=1}$($Q=b,~c$),
$[\Sigma_Q\Lambda_Q]^{I=1}_{S=1}$($Q=b,~c$), $[\Sigma_b\Xi_b^{'}]^{I=3/2}_{S=1}$
and $[\Xi_b\Xi^{'}_b]^{I=0}_{S=1}$ where the S-D mixing plays an important
role. In contrast, the $\pi$ exchange does not form the spin-singlet $A_QB_Q$
bound states. If we consider the heavier scalar and vector meson exchanges as
well as the pion exchange, some loosely bound spin-singlet S-wave states appear
while results of the spin-triplet $A_QB_Q$ system do not change significantly,
which implies the pion exchange plays an dominant role in forming the
spin-triplet molecules. Moreover, we perform an extensive coupled channel
analysis of the $\Lambda_Q\Lambda_Q$ system within the OPE and
one-boson-exchange (OBE) framework and find that there exist loosely bound
states of $\Lambda_Q\Lambda_Q$($Q=b,~c$) with quantum numbers $I(J^P)=0(0^+)$,
$0(0^-)$ and $0(1^-)$. The binding solutions of $\Lambda_Q\Lambda_Q$ system
mainly come from the coupled-channel effect in the flavor space. Besides the
OPE force, the medium- and short-range attractive force also plays a
significant role in the formation of the loosely bound $\Lambda_c\Lambda_c$ and
$\Lambda_b\Lambda_b$ states. Once produced, they will be very stable because
such a system decays via weak interaction with a very long lifetime around
$10^{-13}\sim10^{-12}$s.
\end{abstract}


\maketitle

%%%%%%%%%%%%%%%%%%%%%%%%%%%%%%%%%%%%%%%%%%%%%%%%%%%
\section{INTRODUCTION} \label{INTRODUCTION}
%%%%%%%%%%%%%%%%%%%%%%%%%%%%%%%%%%%%%%%%%%%%%%%%%%%

In the past few years, many exotic charmonium-like states have been
reported by the Belle, BARBAR, CDF and D0 collaborations, such as $X(3872)$
\cite{Choi:2003ue}, $X(4160)$ \cite{Abe:2007sya},$Y(4260)$ \cite{Aubert:2005rm},
 and $Z^+(4430)$ \cite{:2007wga}. Recently,
the BELLE Collaboration observed two charged bottomonium-like
resonances $Z_b(10610)$ and $Z_b(10650)$ in the hidden-bottom
decay channels $\pi^{\pm}\Upsilon(nS)$ ($n=1,~2,~3$) and
$\pi^{\pm}h_b(mP)$ ($m=1,~2$) of $\Upsilon(5S)$
\cite{Collaboration:2011gja}. Since many of these states do not
fit into the conventional $q\bar{q}$ picture in the quark model
easily, how to interpret these ``exotic" states becomes a
challenging problem. The prominent feature of these states is that
they are near the threshold of two charmed or bottomed mesons. For
example, $X(3872)$ lies close to the threshold of $D^0\bar{D}^{*0}$
while $Z_b(10610)$ and $Z_b(10650)$ are near the threshold of
$B\bar{B}^*$ and $B^*\bar{B}^*$ respectively. Inspired by this
striking feature of these exotic states, many physicists attempted
to interpret them as hadronic molecules composed of heavy mesons.

A hadronic molecular state is a loosely bound state of hadrons.
Voloshin and Okun began to investigate the existence of the bound
states composed of charmed meson and antimeson
\cite{Voloshin:1976ap}. De Rujula {\it et al} proposed
$\psi(4040)$ might be a $D^*\bar{D^*}$ molecular state \cite{De
Rujula:1976qd}. T\"ornqvist explored the possible deuteron-like
meson-antimeson bound states with the pion exchange potential
\cite{Tornqvist:1993vu,Tornqvist:1993ng}. Liu {\it et al.}
 investigated the possible molecular states composed of heavy mesons within the
framework of the one-boson-exchange model (OBE) \cite{Liu:2008tn}.
Ding {\it et al.} also gave a dynamic study of meson-meson
molecular states with the one-boson-exchange model at the quark
level \cite{Ding:2009vj}. Sun {\it et al.} interpreted the newly
observed $Z_b(10610)$ and $Z_b(10650)$ as $B\bar{B}^*$ and
$B^*\bar{B}^*$ molecular states respectively \cite{Sun:2011uh}.

Actually, the idea of the loosely bound molecular states is not
new in nuclear physics. It's well-known that the deuteron is a
very loosely bound state composed of a proton and neutron. The
interaction between the proton and neutron comes from the
color-singlet meson exchange. Besides the long-range attraction
from the pion exchange, the S-D mixing, the medium-range
attraction from the correlated two-pion exchange (or in the form
of the sigma meson exchange), and the short-range interaction in
terms of the vector meson exchange combine to form the loosely
bound deuteron.

It's quite natural to extend the same formalism to the heavy
baryon sector. Since the heavy baryon contains a charm or bottom
quark, its large mass reduces the kinetic energy and helps the
formation of the bound states. Fr\"oemel {\it et al} investigated
the bound states composed of heavy hyperon and nucleon by rescaling the
nucleon-nucleon potential in \cite{Froemel:2004ea}. Juli\'a-Diaz
{\it et al} explored the bound states composed of double-charmed
hyperons \cite{JuliaDiaz:2004rf}.
In our previous work \cite{Lee:2011rk}, we performed a study of
the systems $\L_c\L_c(\bar{\L}_c)$, $\Xi_c\Xi_c(\bar{\Xi}_c)$,
$\S_c\S_c(\bar{\S}_c)$, $\Xi_c^{'}\Xi_c^{'}(\bar{\Xi}_c^{'})$ and
$\O_c\O_c(\bar{\O}_c)$.

In the present work we shall study the systems with two different
heavy flavor baryons. For simplicity, we denote the systems with
two same baryons as ``$A_QA_Q$" and the systems with two different
baryons as ``$A_QB_Q$". The difference between
the two systems is that
the ``$A_QB_Q$" system contains the contributions coming from the
$K^{\pm}$, $K^0$, $\bar{K}^0$, $K^{*\pm}$, $K^{*0}$ and
$\bar{K}^{*0}$ exchange while the ``$A_QA_Q$" system does not.

Among the possible loosely bound states composed of a pair of
heavy baryons, the $\L_Q\L_Q$ ($Q=b,~c$) system is particularly
interesting since it is the heavy analogue of the well-known H
dibaryon. Since it was proposed by Jaffe in Ref.
\cite{Jaffe:1976yi}, there have been lots of theoretical and
experimental efforts. Recent investigations include the Lattice
QCD calculation~\cite{Beane:2010hg}, calculations using the chiral
effective field theory~\cite{Haidenbauer:2011ah} and the quark
model~\cite{Chen:2011zzb}. In this work, we shall perform an
extensive coupled channel analysis of the $\L_Q\L_Q$ ($Q=b,~c$)
system and investigate the role of the OPE and sigma/omega/rho meson
exchange in the formation of the possible loosely bound state.

This work is organized as follows. After the introduction, we
present the formalism in Section~\ref{FORMALISM} which contains
the Lagrangians, the coupling constants and the effective
interaction potentials. The formalism for the coupled channel
analysis of the $\L_Q\L_Q$ system is given in
Section~\ref{lambdaq}. In Sections~\ref{NUMERICAL}
and~\ref{lambdanum}, we show the numerical results for
``$A_QB_Q$" and $\L_Q\L_Q$ systems, respectively.
The last section~\ref{SUMMARY} is a brief summary. Some useful
formulae and functions are given in the Appendix. As a
byproduct, we also collect the numerical results for the loosely
bound states composed of a pair of heavy baryon and anti-baryon in
the Appendix.

%%%%%%%%%%%%%%%%%%%%%%%%%%%%%%%%%%%%%%%%%%%%%%
\section{FORMALISM}\label{FORMALISM}
%%%%%%%%%%%%%%%%%%%%%%%%%%%%%%%%%%%%%%%%%%%%%%

%%%%%%%%%%%%%%%%%%%%%%%%%%%%%%%
\subsection{The lagrangian}

% the heavy baryon matrices
The heavy flavor baryon contains a charm or bottom quark and a
diquark (two light quarks). In the heavy quark limit
($m_Q\rightarrow\infty$), the charm or bottom quark can be viewed
as a static color source. The SU(3) flavor symmetry of the baryon
is determined by the diquark. The heavy flavor baryons can be
classified in terms of the symmetry of the diquark. The symmetric
one belongs to the 6-representation while the antisymmetric one
belongs to the $\bar{3}$-representation. On the other hand, the
spin of the diquark is either 0 or 1 which is antisymmetric or
symmetric under the exchange of its two light quark spins. The
baryon is a fermion system. Its total wave function should be
antisymmetric under the exchange of its two light quarks.
Therefore, the spin and the flavor of the diquark are correlated
with each other. Taking the color wave function into account, the
diquark in the 6-representation should be spin-triplet while the
one in the $\bar{3}$-representation should be spin-singlet. The
spin of the baryon in the 6-representation is either $\f{1}{2}$ or
$\f{3}{2}$ while the spin of the baryon in the
$\bar{3}$-representation is only $\f{1}{2}$.

In the following, we follow the notations in Ref.
\cite{Yan:1992gz} and list the heavy flavor baryon matrices and
the exchanged meson matrices. The heavy flavor baryons are
\begin{eqnarray}
B_6 = \left(
\begin{array}{ccc}
      \S_Q^{+1}          & \f{1}{\sqrt{2}}\S_Q^0    & \f{1}{\sqrt{2}}\Xi_Q^{'+\f{1}{2}} \\
\f{1}{\sqrt{2}}\S_Q ^0   & \S_Q^{-1}                & \f{1}{\sqrt{2}}\Xi_Q^{'-\f{1}{2}} \\
\f{1}{\sqrt{2}}\Xi_Q^{'+\f{1}{2}}& \f{1}{\sqrt{2}}\Xi_Q^{'-\f{1}{2}}&\O_Q\\
\end{array}
\right),\quad
B_6^* = \left(
\begin{array}{ccc}
      \S_Q^{*+1}          & \f{1}{\sqrt{2}}\S_Q^{*0}    & \f{1}{\sqrt{2}}\Xi_Q^{*'+\f{1}{2}} \\
\f{1}{\sqrt{2}}\S_Q^{*0}   & \S_Q^{*-1}                & \f{1}{\sqrt{2}}\Xi_Q^{*'-\f{1}{2}} \\
\f{1}{\sqrt{2}}\Xi_Q^{*'+\f{1}{2}}& \f{1}{\sqrt{2}}\Xi_Q^{*'-\f{1}{2}}&\O_Q^*\\
\end{array}
\right),\quad
B_{\bar{3}} = \left(
\begin{array}{ccc}
     0                 & \L_Q                &\Xi_Q^{+\f{1}{2}} \\
   -\L_Q               &      0              &\Xi_Q^{-\f{1}{2}} \\
-\Xi_Q^{+\f{1}{2}}&-\Xi_Q^{-\f{1}{2}}        & 0     \\
\end{array}
\right). \label{heavy:baryons}
\end{eqnarray}
The spin $\f{3}{2}$ baryon is marked with $*$. The superscript
is the third component of its isospin. $Q=b$ or $c$ denotes
the corresponding heavy quark.
% the exchanged meson matrices
The exchanged bosons are
\begin{eqnarray}
\mathcal{M} = \left(
\begin{array}{ccc}
\f{\pi^0}{\sqrt{2}}+\f{\eta}{\sqrt{6}}&     \pi^+               &  K^+  \\
\pi^-                   &-\f{\pi^0}{\sqrt{2}}+\f{\eta}{\sqrt{6}}&  K^0   \\
K^-                  &\bar{K}^0                    &-\f{2}{\sqrt{6}}\eta  \\
\end{array}
\right),\quad
\mathcal{V}^{\mu} = \left(
\begin{array}{ccc}
\f{\rho^0}{\sqrt{2}}+\f{\o}{\sqrt{2}}&     \rho^{+}               &  K^{*+}  \\
\rho^-                    &-\f{\rho^0}{\sqrt{2}}+\f{\o}{\sqrt{2}} &  K^{*0}   \\
K^{*-}                  &\bar{K}^{*0}                    & \phi  \\
\end{array}
\right)^{\mu}. \label{light:mesons}
\end{eqnarray}
The exchanged bosons include the pseudoscalar and vector mesons
given in Eq. (\ref{light:mesons}) and the scalar meson $\sigma$.
The lagrangians built under the SU(3)-flavor symmetry read as
\begin{eqnarray*}
\m{L}&=&\m{L}_{phh}+\m{L}_{vhh}+\m{L}_{\s hh}, \n
\end{eqnarray*}
where
\begin{eqnarray}
\m{L}_{phh}&=&g_{pB_6B_6}\mbox{Tr}\left[\bar{B}_6i\g_5 \mathcal{M} B_6\right]
+g_{pB_{\bar{3}}B_{\bar{3}}}\mbox{Tr}\left[\bar{B}_{\bar{3}}i\g_5 \mathcal{M} B_{\bar{3}}\right]
+\left\{g_{pB_6B_{\bar{3}}}\mbox{Tr}\left[\bar{B}_6i\g_5 \mathcal{M} B_{\bar{3}}\right]+h.c.\right\},
\label{lagrangian:p}\\
\m{L}_{vhh}&=&g_{vB_6B_6}\mbox{Tr}\left[\bar{B}_6\g_{\mu}\mathcal{V}^{\mu}B_6\right]
+\f{f_{vB_6B_6}}{2m_6}\mbox{Tr}\left[\bar{B}_6\s_{\mu\nu}\p^{\mu}\mathcal{V}^{\nu}B_6\right]
+g_{vB_{\bar{3}}B_{\bar{3}}}\mbox{Tr}\left[\bar{B}_{\bar{3}}\g_{\mu}\mathcal{V}^{\mu}B_{\bar{3}}\right]
+\f{f_{vB_{\bar{3}}B_{\bar{3}}}}{2m_{\bar{3}}}\mbox{Tr}\left[\bar{B}_{\bar{3}}
\s_{\mu\nu}\p^{\mu}\mathcal{V}^{\nu}B_{\bar{3}}\right] \n \\
&+&\left\{g_{vB_6B_{\bar{3}}}\mbox{Tr}\left[\bar{B}_6\g_{\mu}\mathcal{V}^{\mu}B_{\bar{3}}\right]
+\f{f_{vB_6B_{\bar{3}}}}{2\sqrt{m_6m_{\bar{3}}}}\mbox{Tr}\left[\bar{B}_6\s_{\mu\nu}
\p^{\mu}\mathcal{V}^{\nu}B_{\bar{3}}\right]+h.c.\right\}, \label{lagrangian:v}\\
\m{L}_{\s hh}&=&g_{\s B_6B_6}\mbox{Tr}\left[\bar{B}_6\s B_6\right]+g_{\s B_{\bar{3}}B_{\bar{3}}}
\mbox{Tr}\left[\bar{B}_{\bar{3}}\s B_{\bar{3}}\right].   \label{lagrangian:s}
\end{eqnarray}
$\mathcal{M}$ and $\mathcal{V}^{\mu}$ are the exchanged pseudoscalar and
vector meson matrices, respectively, which are given in Eq.
(\ref{light:mesons}). $B_6$ and $B_{\bar{3}}$ are the heavy baryon
matrices shown in Eq. (\ref{heavy:baryons}). $m_6$ and
$m_{\bar{3}}$ are the masses of the heavy baryons belonging to the
6-representation and $\bar{3}$-representation, respectively.
``$g_{pB_6B_6},~g_{vB_6B_6}$, $\ldots$" are the coupling
constants.

%%%%%%%%%%%%%%%%%%%%%%%%%%%%%%%%%%%%%%%%%%%%%%%%%%%%%%
\subsection{Coupling Constants} \label{coup:const}

We derive the coupling constants in Eqs.
(\ref{lagrangian:p}-\ref{lagrangian:s}) from those between the nucleon and the
light meson within the Quark Model (QM). For the vector meson exchange, we
adopt the lagrangian without the anomalous magnetic term at the quark level as
in  Ref. \cite{Riska:2000gd}. One can refer to Ref. \cite{Lee:2011rk} for the
specific expressions of the couplings at the quark level in terms of those
between the nucleon and the exchanged mesons. We list the coupling constants we
need below and collect their numerical results in Table~\ref{re:coupling}.
\begin{eqnarray}
{\mbox{pseudoscalar~~exchange}},\quad
g_{pB_6B_6}=\f{4\sqrt{2}}{5}g_{\pi NN}\f{m_i+m_f}{2m_N},\quad
g_{pB_6B_{\bar{3}}}=-\f{2\sqrt{3}}{5}g_{\pi NN}\f{m_i+m_f}{2m_N},\quad
g_{pB_{\bar{3}}B_{\bar{3}}}=0.
\label{coupling:p}
\end{eqnarray}
\begin{eqnarray}
{\mbox{scalar~~exchange}},\quad
g_{\s B_6B_6}=\f{2}{3}g_{\s NN},\quad
g_{\s B_{\bar{3}}B_{\bar{3}}}=\f{1}{3}g_{\s NN}.
\label{coupling:s}
\end{eqnarray}
\begin{eqnarray}
{\mbox{vector~~exchange}},\quad
g_{vB_6B_6}&=&2\sqrt{2}g_{\rho NN},\quad
f_{vB_6B_6}=\f{4\sqrt{2}}{5}
(g_{\rho NN}+f_{\rho NN})\f{\sqrt{m_im_f}}{m_N}-2\sqrt{2}g_{\rho NN}, \n \\
g_{vB_{\bar{3}}B_{\bar{3}}}&=&\sqrt{2}g_{\rho NN},\quad
f_{v B_{\bar{3}}B_{\bar{3}}}=-\sqrt{2}g_{\rho NN}, \n \\
g_{vB_6B_{\bar{3}}}&=&0,\quad
f_{vB_6B_{\bar{3}}}=
-\f{2\sqrt{3}}{5}(g_{\rho NN}+f_{\rho NN})\f{\sqrt{m_im_f}}{m_N}.
\label{coupling:v}
\end{eqnarray}
In the above expressions, $g_{\pi NN}$, $g_{\s NN}$, $g_{\rho NN}$ and $f_{\rho
NN}$ are the coupling constants between the nucleon and the exchanged mesons.
Their numerical values are known quite well. $m_N$ is the mass of the nucleon.
$m_i$ and $m_f$ are the masses of the ingoing and outgoing baryons,
respectively. From Eqs. (\ref{coupling:p},\ref{coupling:v}), the value of the
same coupling constant is slightly different for different systems if one takes
into account the mass difference of the baryons of the same representation. For
example, $g_{pB_6B_6}=\f{4\sqrt{2}}{5}g_{\pi NN}\f{m_{\S_c}+m_{\S_c}}{2m_N}$
for the interaction vertex $\S_c\S_c\pi$ while
$g_{pB_6B_6}=\f{4\sqrt{2}}{5}g_{\pi NN}\f{m_{\Xi_c^{'}}+m_{\Xi_c^{'}}}{2m_N}$
for the $\Xi_c^{'}\Xi_c^{'}\pi$ vertex. The masses of the baryons and exchanged
mesons are summarized in Table~\ref{hadron:masses}. The values of the coupling
constants are \cite{Riska:2000gd,Machleidt:2000ge,Cao:2010km}: $g_{\pi
NN}^2/4\pi=13.6$, $g_{\eta NN}=0.4$, $g_{\s NN}^2/4\pi=5.69$, $g_{\rho
NN}^2/4\pi=0.84$, $f_{\rho NN}/g_{\rho NN}=\kappa_{\rho}=6.1$, $g_{\o
NN}^2/4\pi=20.0$, $f_{\o NN}/g_{\o NN}=\kappa_{\o}=0$. In our case, we assume
the SU(3) symmetry. Therefore we need only three couplings for the
pseudoscalar, scalar and vector meson exchange respectively. We adopt the three
independent ones as $g_{\pi NN}$, $g_{\s NN}$ and $g_{\rho NN}$, since they do
not vary much among different models. Generally speaking, the physical results
of the loosely bound deuteron system are very sensitive to the vector meson
coupling constants. The recently proposed renormalization approach, which uses
a regularized boundary condition, can decrease the dependence on the coupling
constants ~\cite{Cordon:2009pj}.

\begin{table}[htp]
\renewcommand{\arraystretch}{1.5}
\caption{The numerical results of the coupling constants. ``$\times$" means
such a vertex does not exist.}
\begin{tabular*}{17cm}{@{\extracolsep{\fill}}ccccccccc}
\toprule[1pt] \addlinespace[3pt]
  Vertex~      &\multicolumn{4}{c}{Q=c}&\multicolumn{4}{c}{Q=b}\\
\toprule[1pt]
               &$g_{pB_6B_6}$ & $g_{\s B_6B_6}$ & $g_{v B_6B_6}$ & $f_{v B_6B_6}$
               &$g_{pB_6B_6}$ & $g_{\s B_6B_6}$ & $g_{v B_6B_6}$ & $f_{v B_6B_6}$ \\
\specialrule{0.5pt}{3pt}{3pt}
$\S_Q\S_Q(p,\s,v)$&     38.69 & 5.64            &  9.19          &  59.08
               & 91.64        &  5.64           &  9.19          &  152.51        \\
$\S_q\Xi^{'}_Q(p,\s,v)$&39.65 & $\times$        &  9.19          &  60.76
               & 92.66        & $\times$        &  9.19          &  154.30        \\
$\S_Q\O_Q(p,\s,v)$& 40.61     & $\times$        &  9.19          &  62.39
               &  93.68       & $\times$        &  9.19          &  156.08        \\
$\Xi_Q^{'}\Xi^{'}_Q(p,\s,v)$&40.62    & 5.64            &  9.19          &  62.48
               &  93.68       & 5.64            &  9.19          &  156.18        \\
$\Xi_Q^{'}\O_Q(p,\s,v)$& 41.58& $\times$        &  9.19          &  64.15
               &  94.71       & $\times$        &  9.19          &  157.91        \\
$\O_Q\O_Q(p,\s,v)$& 42.53     & 5.64            &  9.19          &  65.86
               &  95.73       & 5.64            &  9.19          &  159.73        \\
\specialrule{0.5pt}{3pt}{3pt}
               &$g_{pB_6B_{\bar{3}}}$ & $g_{\s B_6B_{\bar{3}}}$ & $g_{v B_6B_{\bar{3}}}$
               & $f_{v B_6B_{\bar{3}}}$
               &$g_{pB_6B_{\bar{3}}}$ & $g_{\s B_6B_{\bar{3}}}$ & $g_{v B_6B_{\bar{3}}}$
               & $f_{v B_6B_{\bar{3}}}$ \\
\specialrule{0.5pt}{3pt}{3pt}
$\S_Q\L_Q(p,\s,v)$&$-22.89$   & $\times$        &   0            & $-40.36$
               & $-55.19$     & $\times$        &   0            & $-97.37$       \\
$\S_Q\Xi_Q(p,\s,v)$&$-23.77$  & $\times$        &   0            & $-41.95$
               & $-56.01$     & $\times$        &   0            & $-98.84$       \\
$\Xi_Q^{'}\L_Q(p,\s,v)$&$-23.48$&$\times$       &   0            & $-41.35$
               & $-55.82$     & $\times$        &   0            & $-98.46$       \\
$\Xi_Q^{'}\Xi_Q(p,\s,v$)&$-24.36$& $\times$     &   0            & $-42.98$
               & $-56.64$     & $\times$        &   0            & $-99.94$       \\
$\O_Q\Xi_Q(p,\s,v)$& $-24.95$ & $\times$        &   0            & $-43.98$
               & $-57.27$     & $\times$        &   0            & $-101.02$      \\
\specialrule{0.5pt}{3pt}{3pt}
               &$g_{pB_{\bar{3}}B_{\bar{3}}}$ & $g_{\s B_{\bar{3}}B_{\bar{3}}}$
               & $g_{v B_{\bar{3}}B_{\bar{3}}}$& $f_{v B_{\bar{3}}B_{\bar{3}}}$
               &$g_{pB_{\bar{3}}B_{\bar{3}}}$ & $g_{\s B_{\bar{3}}B_{\bar{3}}}$
               & $g_{v B_{\bar{3}}B_{\bar{3}}}$& $f_{v B_{\bar{3}}B_{\bar{3}}}$ \\
\specialrule{0.5pt}{3pt}{3pt}
$\L_Q\L_Q(p,\s,v)$& 0         &  2.82           &  4.59          & $-4.59$
               &    0         &  2.82           &  4.59          & $-4.59$      \\
$\L_Q\Xi_Q(p,\s,v)$&0         & $\times$        &  4.59          & $-4.59$
               &    0         & $\times$        &  4.59          & $-4.59$       \\
$\Xi_Q\Xi_Q(p,\s,v)$&0        &  2.82           &  4.59          & $-4.59$
               &    0         &  2.82           &  4.59          & $-4.59$      \\ [3pt]
\bottomrule[1.pt]
\end{tabular*} \label{re:coupling}
\end{table}

\begin{table}[htp]
\renewcommand{\arraystretch}{1.5}
\caption{The masses of the relevant mesons and baryons
\cite{pdg2010}. The bottomed baryons $\S_b^0$, $\Xi_b^{'0}$ and
$\Xi_b^{'-}$ (marked with $*$) have not been observed
experimentally. Their masses are fixed to be:
$m_{\S_b^0}=[m_{\S_b^+}+m_{\S_b^-}]/2$ and
$m_{\Xi_b^{'0}}=m_{\Xi_b^{'-}}= [m_{\S_b}+m_{\O_b}]/2$. }
\begin{tabular*}{15cm}{@{\extracolsep{\fill}}cccccc}
\toprule[1pt] \addlinespace[3pt]
  meson~     &~mass(MeV)~ & ~baryon~  &~mass(MeV)~&~baryon~ &~mass(MeV)\\
\specialrule{0.8pt}{3pt}{3pt}
$\pi^{\pm}$    &$139.57$ &~$\L_c^+$    & $2286.5$&~$\L_b^{0}$&$5620.2$ \\
$\pi^0$        &$134.98$ &~$\Xi_c^+$   & $2467.9$ &~$\Xi_b^0$ &$5790.5$ \\
$\eta$         &$547.85$ &~$\Xi_c^0$   & $2471.0$ &~$\Xi_b^-$ &$5790.5$ \\
$\rho$         &$775.49$ &~$\S_c^{++}$ & $2454.02$&~$\S_b^+$  &$5807.8$ \\
$\o$           &$782.65$ &~$\S_c^{+}$  & $2452.9$ &~$^*\S_b^0$  &$5811.5$ \\
$\phi$         &$1019.46$&~$\S_c^0$   & $2453.8$&~$\S_b^-$  &$5815.2$ \\
$K^{\pm}$      &$493.68$&~$\Xi_c^{'+}$& $2575.7$ &~$^*\Xi_b^{'0}$&$5941.3$ \\
$K^0$          &$497.61$&~$\Xi_c^{'0}$& $2578.0$ &~$^*\Xi_b^{'-}$&$5941.3$ \\
$K^{*\pm}$     &$891.66$ &~$\O_c^0$    & $2697.5$ &~$\O_b^-$  &$6071.0$   \\
$K^{*0}$       &$895.94$&  ~     &     ~    &        ~  &  ~      \\
$\s$           &$600$   &        &          &           &         \\
\bottomrule[1.0pt]
\end{tabular*} \label{hadron:masses}
\end{table}

\subsection{Effective Potential}

\begin{figure}[htp]
\centering
\begin{tabular}{ccccc}
\includegraphics[width=0.30\textwidth]{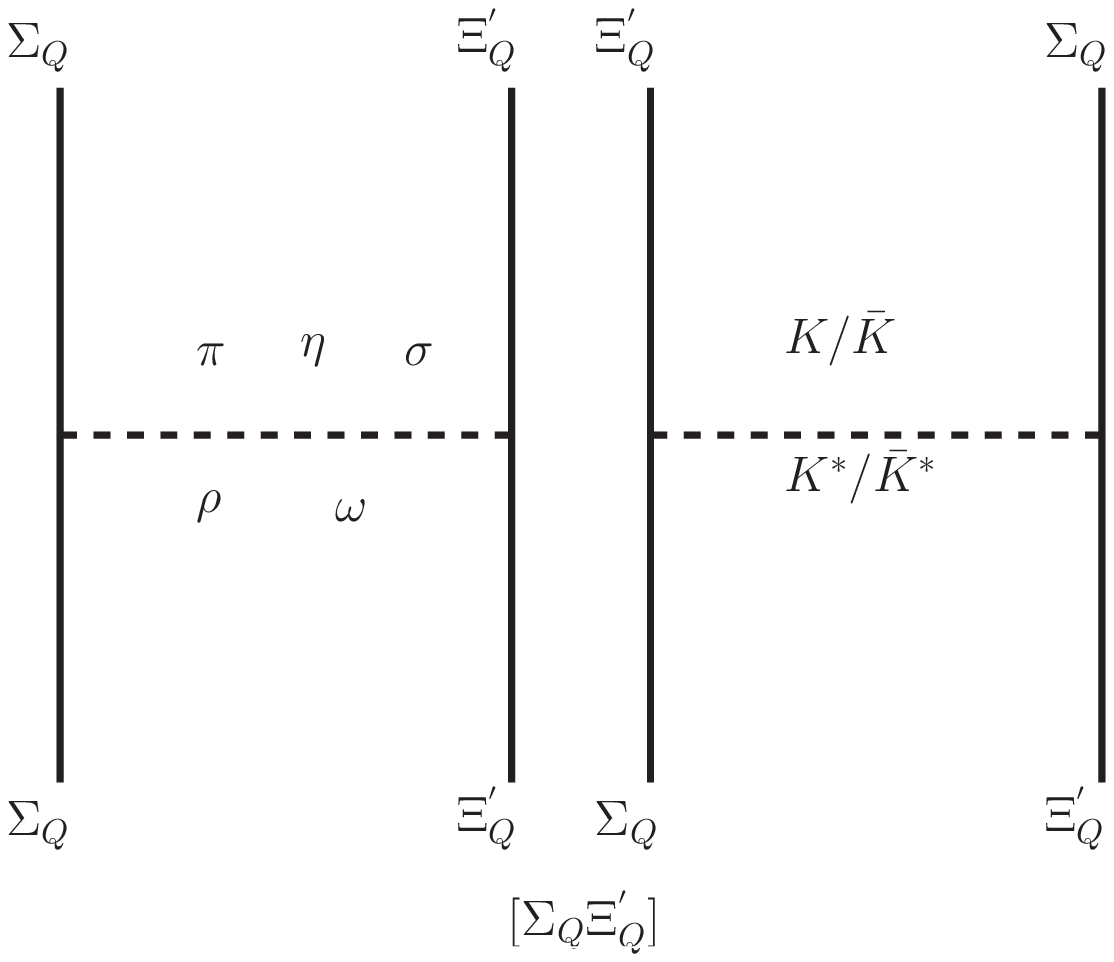}~&~
\includegraphics[width=0.30\textwidth]{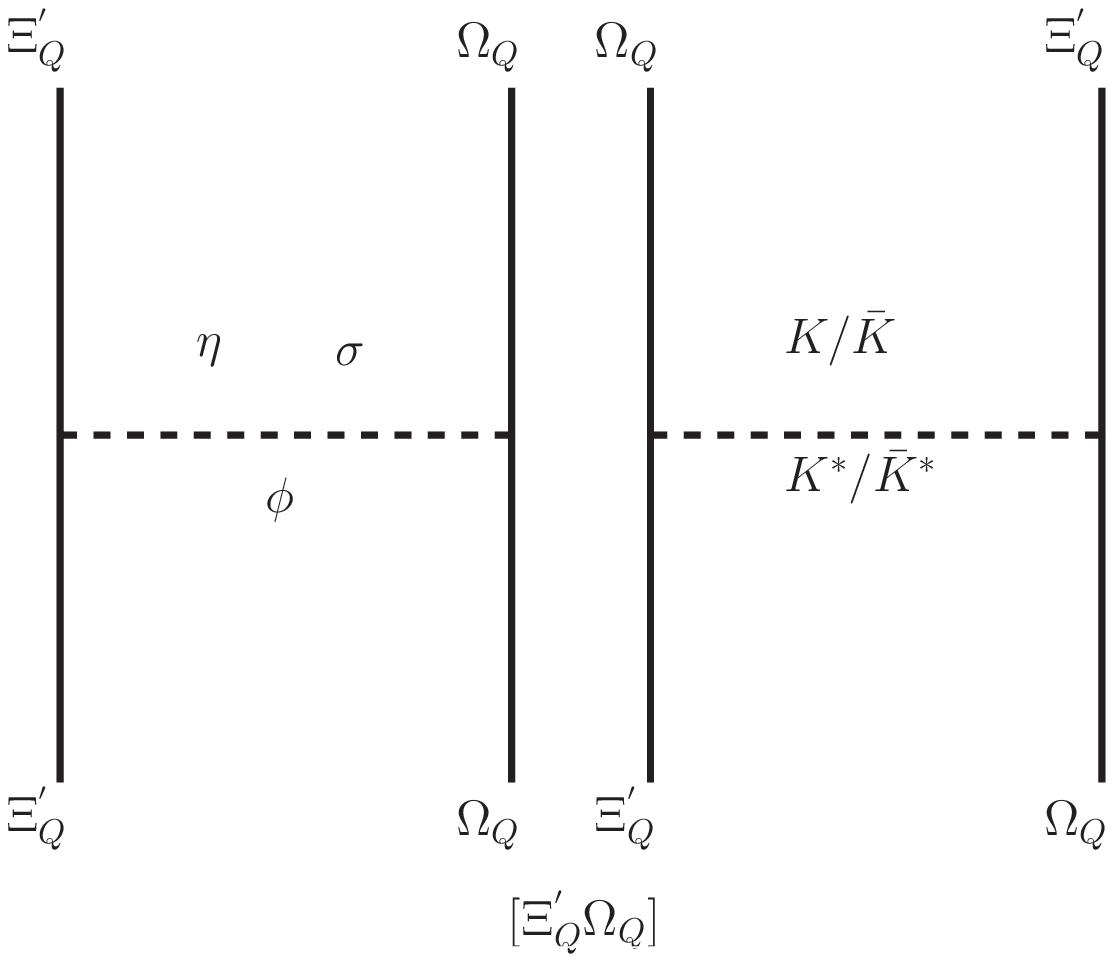}~&~
\includegraphics[width=0.30\textwidth]{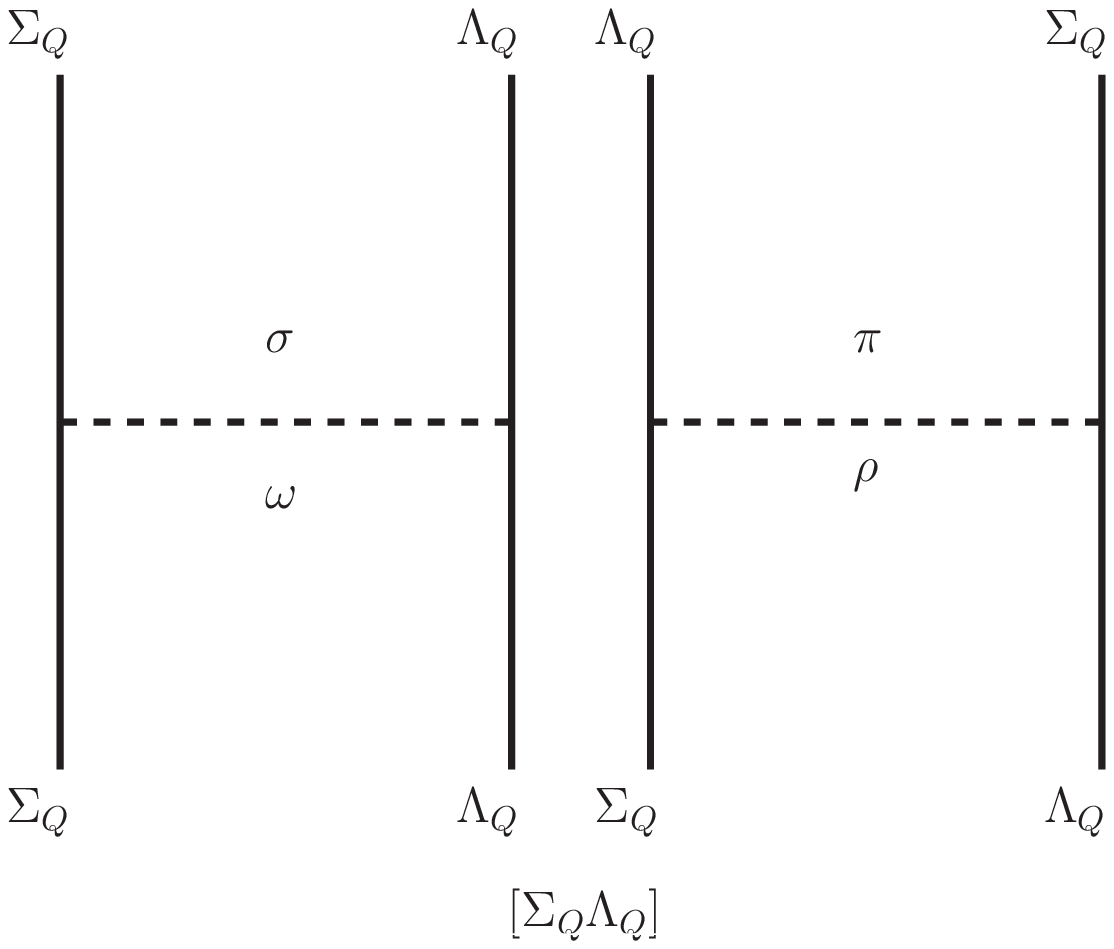}\\
\includegraphics[width=0.30\textwidth]{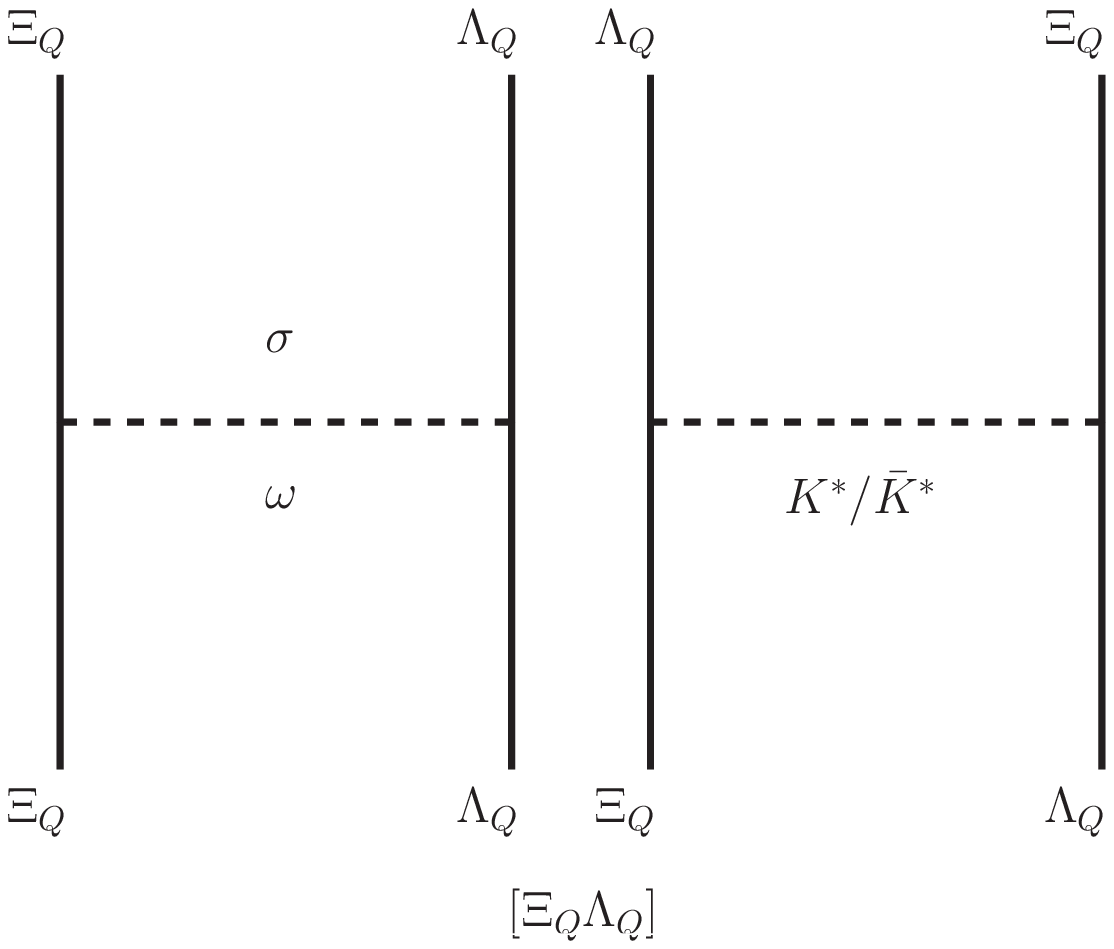}~&~
\includegraphics[width=0.30\textwidth]{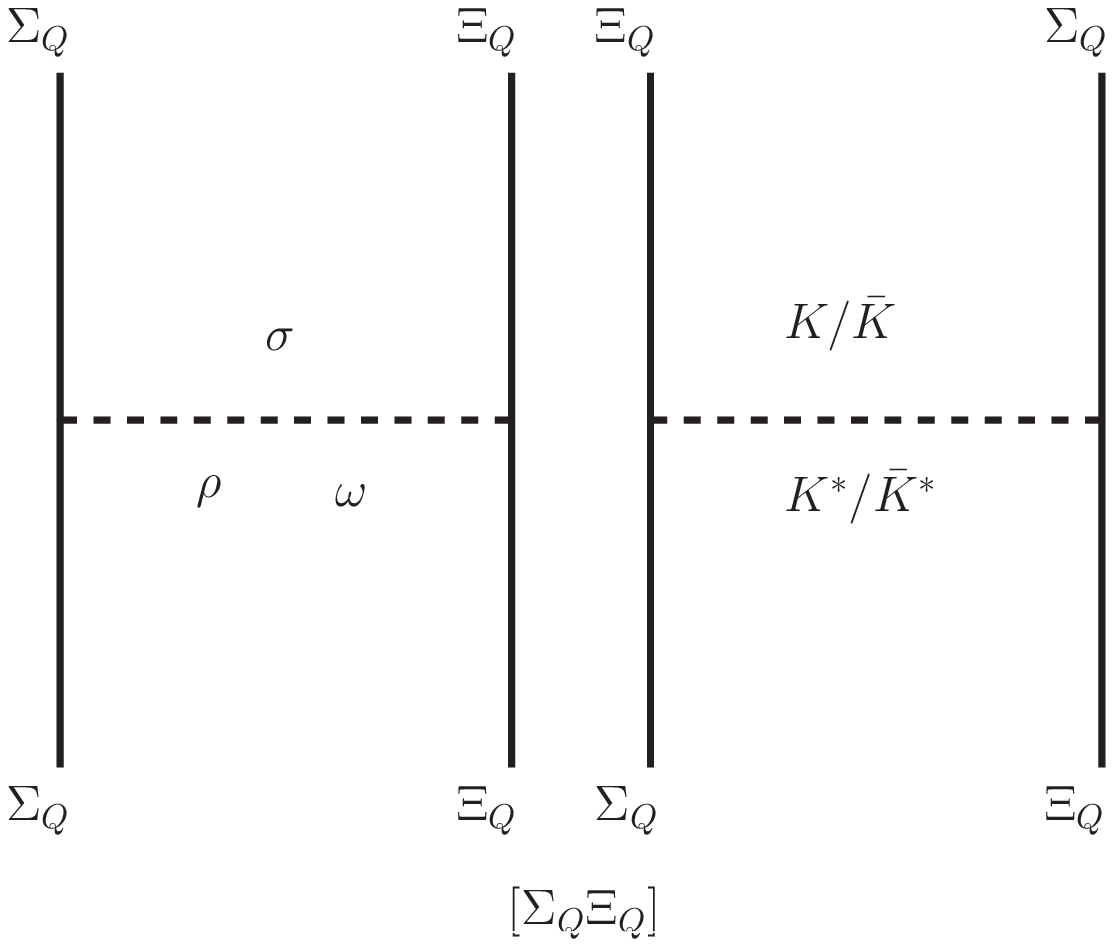}~&~
\includegraphics[width=0.30\textwidth]{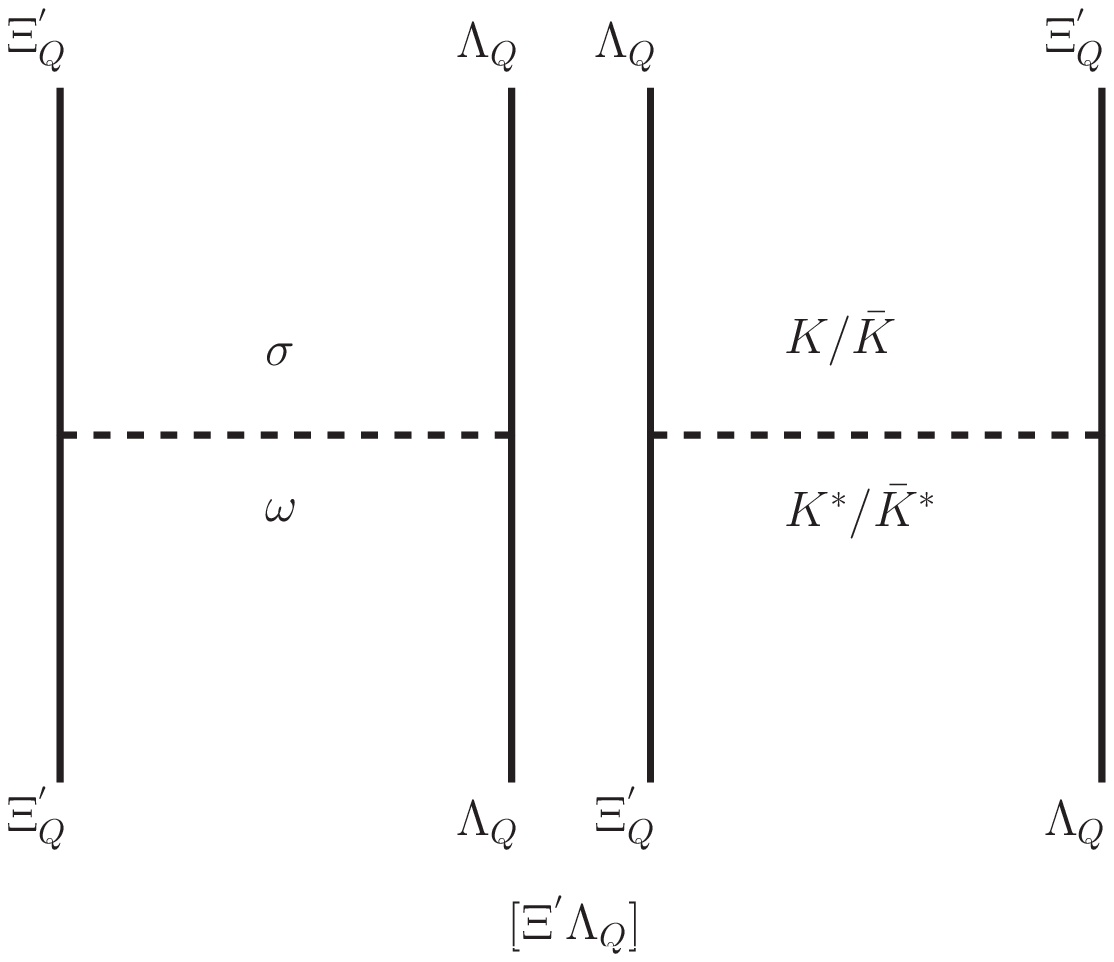}\\
\includegraphics[width=0.30\textwidth]{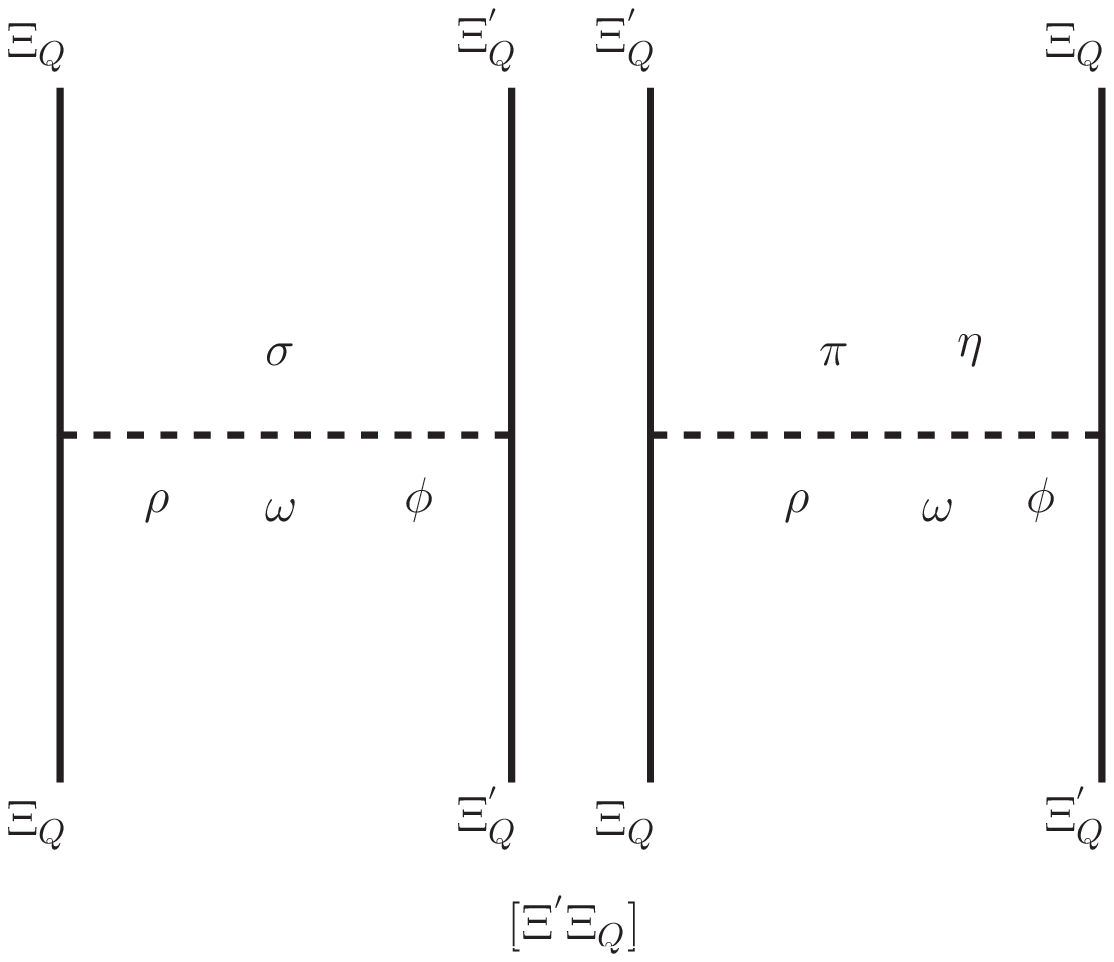}~&~
\includegraphics[width=0.30\textwidth]{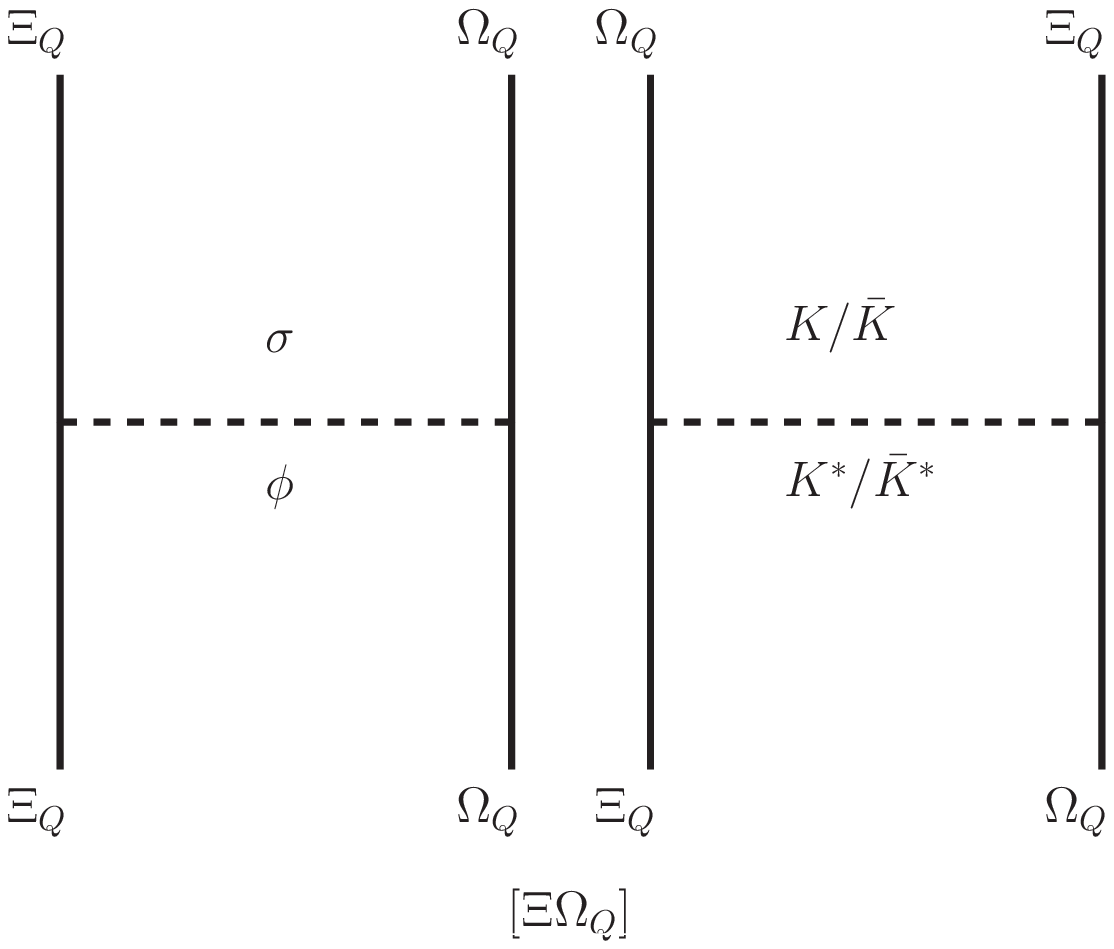}~&~
\includegraphics[width=0.30\textwidth]{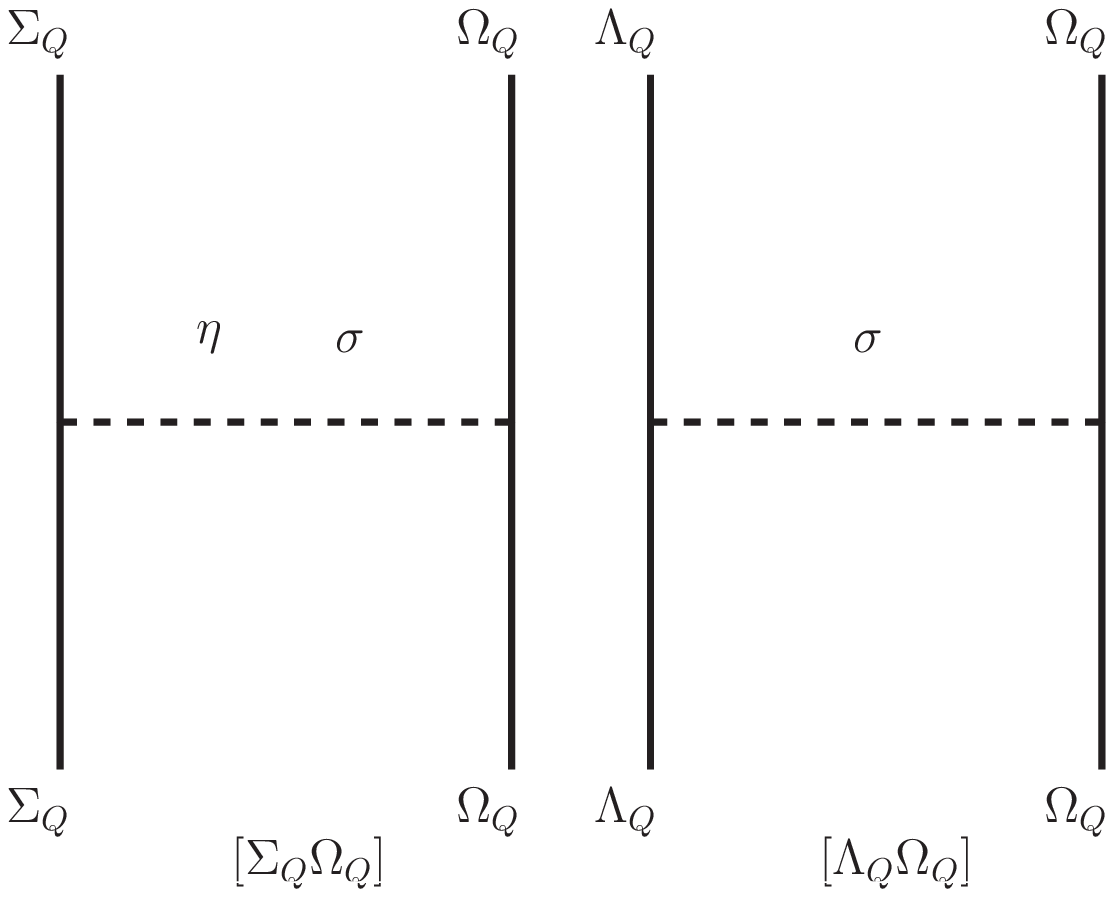}\\
\end{tabular}
\caption{The exchanged mesons for different systems.}\label{Exmeson}
\end{figure}

Applying the lagrangians in
Eqs.(\ref{lagrangian:p}-\ref{lagrangian:s}) one can derive the
effective interaction potential in the momentum space. Given the
hadrons are not fundamental particles, we employ a monopole form
factor at each vertex to roughly describe the structure effect of
the baryon
\begin{eqnarray}
\m{F}(Q)=\f{\L^2-m_{ex}^2}{\L^2-Q^2}=\f{\L^2-m_{ex}^2}{\l^2+\bm{Q}^2},\label{FF}
\end{eqnarray}
where $\L$ is the cutoff parameter by which we can regulate the
exchanged momentum. $m_{ex}$ and $Q$ are the mass and four
momentum of the exchanged meson, respectively, and
$\l^2={\L}^2-Q_0^2$. Making the Fourier transformation,
\begin{eqnarray}
\m{V}(r)=\f{1}{(2\pi)^3}\int dQ^3 e^{i\bm{Q}\cdot\bm{r}}\m{V}(Q)\mathcal{F}^2(Q),
\end{eqnarray}
one obtains the effective interaction potential in the coordinate
space, which are given below. One can refer to Appendix
\ref{FUNCTION} for some formulae. Since the hadronic
molecule is a loosely bound state, the hadrons are not expected to
be very close to each other. We neglect the contact interaction
piece $\delta (r)$ in the potential. The detailed information of
the delta term is given in the Appendix.

We expand the effective interaction potential in terms of
$\f{1}{m}$ (m is the heavy baryon mass) up to order of
$\f{1}{m^2}$. We also adopt the approximation $1/m_A^2\sim 1/m_B^2
\sim 1/(m_Am_B)$ due to the large masses ($m_A$, $m_B$) of the
heavy baryons. The effective potentials have four terms: the
central potential term $\m{V}_C$, the spin-spin term $\m{V}_{SS}$,
the spin-orbit term $\m{V}_{LS}$ and the tensor term $\m{V}_T$.
\begin{eqnarray}
\m{V}^p(r;n,\a)&=&\m{V}_{SS}^p(r;n,\a)+\m{V}_{T}^p(r;n,\a), \n \\
\m{V}^s(r;n=1,\s)&=&\m{V}_{C}^s(r;n=1,\s)+\m{V}_{LS}^s(r;n=1,\s), \n \\
\m{V}^v(r;n,\b)&=&\m{V}_C^v(r;n,\b)+\m{V}_{LS}^v(r;n,\b)
+\m{V}_{SS}^v(r;n,\b)+\m{V}_{T}^v(r;n,\b), \n
\end{eqnarray}
where the superscripts, $p$, $s$ and $v$, denote the pseudoscalar, scalar and vector meson
exchange, respectively. $\a=\pi$, $\eta$, $K^\pm$, $K^0$, $\bar{K}^0$ and
$\b=\o$, $\rho$, $\phi$, $K^{*\pm}$, $K^{*0}$, $\bar{K}^{*0}$.
$n=1$, $2$ denotes the direct and cross diagrams, respectively. For the scalar exchange,
``$n=1$" means that the $\sigma$ exchange only occurs in the direct diagram.
The specific expressions read
\begin{eqnarray}
{\mbox{pseudoscalar~~exchange}},~~~~
\m{V}_{SS}^p(r;n,\a)&=&C_{n,\a}^p\f{g_{1p}g_{2p}}{4\pi}\f{u_{\a}^3}{12m_Am_B}
H_0(\L,m_{\a},r)\bm{\s}_A\cdot\bm{\s}_B, \n\\
\m{V}_T^p(r;n,\a)&=&C_{n,\a}^p\f{g_{1p}g_{2p}}{4\pi}\f{u_{\a}^3}{12m_Am_B}
H_3(\L,m_{\a},r)S_{AB}(\hat{r}),
\label{potential:p1}
\end{eqnarray}
when $u_{\a}^2=m_{\a}^2-(m_f-m_i)^2<0$, they change into,
\begin{eqnarray}
\m{V}_{SS}^p(r;n,\a)&=&C_{n,\a}^p\f{g_{1p}g_{2p}}{4\pi}\f{\theta_{\a}^3}{12m_Am_B}
M_0(\L,m_{\a},r)\bm{\s}_A\cdot\bm{\s}_B, \n\\
\m{V}_T^p(r;n,\a)&=&C_{n,\a}^p\f{g_{1p}g_{2p}}{4\pi}\f{\theta_{\a}^3}{12m_Am_B}
M_3(\L,m_{\a},r)S_{AB}(\hat{r}),\label{potential:p2}
\end{eqnarray}
with $\theta_{\a}^2=-\left[m_{\a}^2-(m_f-m_i)^2\right]$,\\
\begin{eqnarray}
{\mbox{vector~~exchange}},~~~~
\m{V}_C^v(r;n,\b)&=&C_{n,\b}^v\f{u_{\b}}{4\pi}\left[ g_{1v}g_{2v}H_0(\L,m_{\b},r)
+\f{u_{\b}^2}{8m_Am_B}(g_{1v}g_{2v}+2g_{1v}f_{2v}+2g_{2v}f_{1v})H_0(\L,m_{\b},r)\right], \n\\
\m{V}_{SS}^v(r;n,\b)&=&C_{n,\b}^v\left[g_{1v}g_{2v}+g_{1v}f_{2v}+g_{2v}f_{1v}+f_{1v}f_{2v}\right]
\f{1}{4\pi}\f{u_{\b}^3}{6m_Am_B}H_0(\L,m_{\b},r)\bm{\s}_A\cdot\bm{\s}_B, \n\\
\m{V}_T^v(r;n,\b)&=&-C_{n,\b}^v\left[g_{1v}g_{2v}+g_{1v}f_{2v}+g_{2v}f_{1v}+f_{1v}f_{2v}\right]
\f{1}{4\pi}\f{u_{\b}^3}{12m_Am_B}H_3(\L,m_{\b},r)S_{AB}(\hat{r}), \n\\
\m{V}_{LS}^v(r;n,\b)&=&-C_{n,\b}^v\f{1}{4\pi}\f{u_{\b}^3}{2m_Am_B}H_2(\L,m_{\b},r)
\left[3g_{1v}g_{2v}\bm{L}
\cdot\bm{S}+4g_{1v}f_{2v}\bm{L}\cdot\bm{S}_A
+4g_{v2}f_{1v}\bm{L}\cdot\bm{S}_B\right],
\label{potential:v}
\end{eqnarray}
\begin{eqnarray}
{\mbox{scalar~~exchange}},~~~~
\m{V}_C^s(r;n=1,\s)&=&-C_{n,\s}^su_{\s}\f{g_{1s}g_{2s}}{4\pi}\left[H_0(\L,m_{\s},r)
-\f{u_{\s}^2}{8m_Am_B}H_0(\L,m_{\s},r)\right],\n\\
\m{V}_{LS}^s(r;n=1,\s)&=&-C_{n,\s}^s\f{g_{1s}g_{2s}}{4\pi}\f{u_{\s}^3}{2m_Am_B}
H_2(\L,m_{\s},r)\bm{L}\cdot\bm{S}.\label{potential:s}
\end{eqnarray}
$C_{n,\a}^p$, $C_{n,\b}^v$ and $C_{n,\s}^s$ are the isospin
factors. Their numerical values are given in Table
\ref{coefficients}, and the exchanged mesons are shown in Fig.~\ref{Exmeson}.
$\bm{L}$ is the relative orbit momentum
operator between the two baryons ``$A_Q$" and ``$B_Q$". $\bm{S}_A$,
$\bm{S}_B$ are the spin operators of the two baryons while
$\bm{S}=\bm{S}_A+\bm{S}_B$ is the total spin operator.
$S_{AB}(\hat{r})=3\bm{\s}_A\cdot\bm{r}
\bm{\s}_B\cdot\bm{r}/r^2-\bm{\s}_A\cdot\bm{\s}_B$ is the tensor
operator. $g_{1p}$, $g_{2p}$, $\ldots$ are the coupling constants
given in Eqs. (\ref{coupling:p}-\ref{coupling:v}). The values of $u_i$ read
\begin{eqnarray}
{\mbox{direct~~diagram}},~~~~u_{\a}&=&u_{\s}=u_{\b}=0, \n\\
{\mbox{cross~~diagram}},~~~~~
u_{\a}^2&=&m_{\a}^2-(m_A-m_B)^2,\quad \theta_{\a}^2=-\left[m_{\a}^2-(m_A-m_B)^2\right],\quad
u_{\b}^2=m_{\b}^2-(m_A-m_B)^2.
\end{eqnarray}

Substituting the masses of the corresponding baryons for $m_A$ and
$m_B$ in Eqs. (\ref{potential:p1}-\ref{potential:s}), one obtains
the effective interaction potentials. Besides Eqs.
(\ref{potential:p1}-\ref{potential:p2}) in the one-pion-exchange
(OPE) model, we need include the contributions from the other
heavier exchanged mesons in the the one-boson-exchange (OBE)
model. The potential within the OBE model reads
\begin{eqnarray}
\m{V}(r)&=&\m{V}_C(r)+\m{V}_{SS}(r)+\m{V}_{LS}(r)+\m{V}_T(r) \n\\
&=&\left[\m{V}_C^s(r;n=1,\s)
+\sum_{n,\b}\m{V}_C^v(r;n,\b)\right]
+\left[\sum_{n,\a}\m{V}_{SS}^p(r;n,\a)
+\sum_{n,\b}\m{V}_{SS}^v(r;n,\b)\right] \n\\
&~&+\left[\m{V}_{LS}^s(r;n=1,\s)
+\sum_{n,\b}\m{V}_{LS}^v(r;n,\b)\right]
+\left[\sum_{n,\a}\m{V}_T^p(r;n,\a)
+\sum_{n,\b}\m{V}_T^v(r;n,\b)\right].\label{potential:singlet}
\end{eqnarray}

The systems with two spin-half particles are either spin-singlet
(S=0) or spin-triplet (S=1). For the spin-singlet, we focus on the
ground state $^1S_0$ while for the spin-triplet, we take both
$^3S_1$ and $^3D_1$ into account. The wave functions can be
expressed as
\begin{eqnarray*}
\Psi(r)^S=y_S^S(r)|^1S_0>,\quad
\Psi(r)^T=
\left(\begin{array}{c}
y_S^T(r) \\
 0     \\
\end{array}\right) |^3S_1>+
\left(\begin{array}{c}
0  \\
y_D^T(r)\\
\end{array}\right)|^3D_1>,
\end{eqnarray*}
where $y_S^S(r)$, $y_S^T(r)$ and $y_D^T$ are the radial wave
functions. The operators can be written in the following matrix
form:
\begin{eqnarray}
{\mbox{spin-singlet}},\quad\bm{\s}_A\cdot\bm{\s}_B=-3,\quad \bm{L}\cdot\bm{S}=0,\quad\bm{L}\cdot\bm{S}_A=0,
\quad\bm{L}\cdot\bm{S}_B=0,\quad S_{AB}(\hat{r})=0, \label{operator:singlet}
\end{eqnarray}
\begin{eqnarray}
{\mbox{spin-triplet}},\quad\bm{\s}_A\cdot\bm{\s}_B=
\left(\begin{array}{cc}
 1  &   0  \\
 0  &   1  \\
 \end{array} \right),~~
 S_{AB}(\hat{r})=
 \left(\begin{array}{ccc}
0   &  \sqrt{8}  \\
\sqrt{8}  &  -2   \\
\end{array}\right),~~
\bm{L}\cdot\bm{S}=
\left(\begin{array}{cc}
0  &  0 \\
0  & -3 \\
\end{array}\right),~~
 \bm{L}\cdot \bm{S}_A=
 \left(\begin{array}{cc}
 0  &   0  \\
 0  &  -\f{3}{2} \\
 \end{array}\right),~~
 \bm{L}\cdot \bm{S}_B=
 \left( \begin{array}{ccc}
 0   &    0 \\
 0   &  -\f{3}{2} \\
\end{array}\right). \label{operator:triplet}
\end{eqnarray}

\begin{table}[htp]
\renewcommand{\arraystretch}{1.5}
\caption{The isospin factors. The superscript in the first column
is the isospin of the system. The values outside the square
brackets are for the direct diagram while the ones inside the
square brackets are for the crossed diagram.  }
\begin{tabular*}{16cm}{@{\extracolsep{\fill}}ccccccccc}
\toprule[1pt] \addlinespace[3pt]
  ~States~  &~$C_{n,\pi}^p$~~&~~$C_{n,\eta}^p$~~&~$C_{n,K}^p$~&~~$C_{n,\rho}^v$~~&
  ~~$C_{n,\o}^v$~~&~$C_{n,\phi}^v$~&~$C_{n,K^*}^p~$~&~$C_{n,\s}^s$ \\
\specialrule{0.8pt}{3pt}{3pt}
$[\S_Q\Xi_Q^{'}]^{I=1/2}$     &   $-1/2$      & $-1/12$     & $[-1/4]$ &$-1/2$       &  $1/4$   &
                              & $[-1/4]$      &   $1$   \\
$[\S_Q\Xi_Q^{'}]^{I=3/2}$     &$1/4$          & $-1/12$     & $[1/2]$  &     $1/4$   &   $1/4$  &
                              & $[1/2]$       &    $1$  \\
$[\Xi_Q^{'}\O_Q]^{I=1/2}$     &               &   $1/6$     & $[1/2]$  &             &           &
                     $ 1/2$   & $[1/2]$       &    $1$  \\
$[\S_Q\O_Q]^{I=1/2}$          &               &   $-1/3$    &          &             &           &
                              &               &   $1$  \\
$[\Xi_Q\L_Q]^{I=1/2}$         &               &             &          &             &   $1$     &
                              &  $[1]$        &    $4$  \\
$[\S_Q\L_Q]^{I=1}$            &  $[1]$        &             &          &   $[1]$     &   $1$    &
                              &               &   $2$  \\
$[\S_Q\Xi_Q]^{I=1/2}$         &               &             & $[-1/2]$ &    $-1$     &   $1/2$  &
                              & $[-1/2]$      &   $2$  \\
$[\S_Q\Xi_Q]^{I=3/2}$         &               &             & $[1]$    &   $1/2$     &   $1/2$  &
                              &  $[1]$        &   $2$  \\
$[\Xi_Q^{'}\L_Q]^{I=1/2}$     &               &             & $[1/2]$  &             &   $1/2$  &
                              & $[1/2]$       &   $2$  \\
$[\Xi_Q\Xi_Q^{'}]^{I=0}$      & $[3/4]$       &     $[-3/4]$&          & $-3/4[3/4]$  &$1/4[-1/4]$&
$ 1/2[-1/2]$                  &               &   $2$  \\
$[\Xi_Q\Xi_Q^{'}]^{I=1}$      &    $[1/4]$    &$     [3/4]$ &          &$1/4[1/4]$   &$1/4[1/4]$&
$ 1/2[1/2]$                   &               &   $2$  \\
$[\Xi_Q\O_Q]^{I=1/2}$         &               &             &  $[1]$     &             &           &
     $1$                     &   $[1]$        &   $2$  \\
$[\L_Q\O_Q]^{I=0}$            &               &             &          &             &           &
                              &               &   $2$  \\ [5pt]
\bottomrule[1pt]
\end{tabular*} \label{coefficients}
\end{table}

%%%%%%%%%%%%%%%%%%%%%%%%%%%%%%%%%%%%%%%%%%%%%%%%%%%%%%%%%%%%
\section{The Coupled channel analysis of the $\L_Q\L_Q$
system}\label{lambdaq}
%%%%%%%%%%%%%%%%%%%%%%%%%%%%%%%%%%%%%%%%%%%%%%%%%%%%%%%%%%%%

The $\L_Q\L_Q\left[0(0^+)\right]$ system is very interesting,
which can be viewed as the heavy analogue of the H dibaryon. The
heavy quark mass $m_{b,c}$, the S-D wave mixing and the coupled
channel effect in the flavor space all may play an important role
in the formation of the possible loosely bound states.
Investigation and comparison of the $\L_b\L_b$, $\L_c\L_c$, and
$\L\L$ systems may reveal which underlying mechanism is dominant.

In the present work we shall perform an extensive analysis of
$\L_Q\L_Q$ with quantum numbers $I(J^P)=0(0^+)$, $0(0^-)$ and
$0(1^-)$. We list the flavor channels which we take into account
in Table \ref{LL}. Besides the lagrangians given in Eqs.
(\ref{lagrangian:p}-\ref{lagrangian:s}), we also need the
following effective Lagrangians:
\begin{eqnarray}
  \mbox{pseudoscalar exchange},\quad \mathcal{L}_{p}&=&
  \left[-\f{g_3}{\sqrt{2}f_{\pi}}\mbox{Tr}\left(\bar{B}_{6\mu}^*\partial^{\mu}MB_6\right)+h.c.\right]
  +\left[-\f{g_4}{\sqrt{2}f_{\pi}}\mbox{Tr}\left(\bar{B}_{6\mu}^*\partial^{\mu}MB_{\bar{3}}\right)+h.c.\right]\\
  &~&-\f{g_5}{\sqrt{2}f_{\pi}}
  \mbox{Tr}\left(\bar{B}_{6\mu}^*\g_{\nu}\g_5\partial^{\nu}MB_6^{*\mu}\right)\\
  \mbox{scalar exchange},\quad \mathcal{L}_{s}&=&l_s\mbox{Tr}
  \left(\bar{B}_{6\mu}^*\sigma B_6^{*\mu}\right),\\
  \mbox{vector exchange},\quad \mathcal{L}_{v}&=&\f{\beta_s g_v}{\sqrt{2}}
  \mbox{Tr}\left(\bar{B}_{6\mu}^*\g_{\nu}V^{\nu}B_6^{*\mu}\right)
  +\f{i\l_sg_v}{\sqrt{2}}\mbox{Tr}\left[\bar{B}_{6\mu}^*
  (\partial^{\mu}V^{\nu}-\partial^{\nu}V^{\mu})B_{6\nu}^*\right] \nonumber\\
  &~&+\left\{-\f{i\l_sg_v}{\sqrt{6}}\mbox{Tr}
  \left[\bar{B}_{6\mu}^*\left(\partial^{\mu}V^{\nu}-\partial^{\nu}V^{\mu}\right)
  \g_{\nu}\g_5B_6\right]+h.c.\right\} \nonumber \\
  &~&+\left\{-i\sqrt{2}\l_Ig_v\mbox{Tr}\left[B_{6\mu}^{*}
  \left(\partial^{\mu}V^{\nu}-\partial^{\nu}V^{\mu}\right)\g_{\nu}\g_5B_{\bar{3}}\right]+h.c.\right\}.
\end{eqnarray}
The coupling constants are $g_4=0.999$, $g_3=\sqrt{6}g_4$,
$g_5=-\sqrt{2}g_4$, $l_{s}=6.2$, $f_{\pi}=92.3$ MeV,
$(\beta_sg_v)=12$, $(\l_sg_v)=19.2~\mbox{GeV}^{-1}$ and
$(\l_Ig_v)=-(\l_sg_v)/\sqrt{8}$ \cite{Meguro:2011nr}. Besides the
potentials in Eqs. (\ref{potential:p1}-\ref{potential:s}), we also
need the following potentials
\begin{eqnarray}
\mathcal{V}^p(r)&=&C^p(i,j)\f{u^3}{24\pi}\left[H_3(\L,m_{ex},r)\Delta_{ten}
+H_0(\L,m_{ex},r)\Delta_{SS}\right],\\
\mathcal{V}^v(r)&=&C_1^v(i,j)\f{u}{4\pi}H_0(\L,m_{ex},r)+C_2^v(i,j)\f{u^3}{12\pi}
\left[-H_3(\L,m_{ex},r)\Delta_{ten}+2H_0(\L,m_{ex},r)\Delta_{SS}\right],\\
\mathcal{V}^s(r)&=&C^s(i,j)\f{m_{\sigma}}{4\pi}H_0(\L,m_{\sigma},r).
\end{eqnarray}
where $\Delta_{ten}$ and $\Delta_{SS}$ denote the tensor and
spin-spin operators respectively. They are channel-dependent.
Their specific expressions are given in Table \ref{operator}.

For the baryon masses, we use $m_{\S_c^*}=2518.0$ MeV and $m_{\S_b^*}=5832.5$ MeV \cite{pdg2010}.
Due to the conservation of the energy and momentum, we keep the
non-vanishing zeroth component of the exchanged four momentum
$Q_0$ and define $u$ as the following
\begin{eqnarray*}
\S_Q\S_Q^*\leftrightarrow\S_Q^*\S_Q,&~~& u^2=m_{ex}^2-\left(m_{\S_Q^*}-m_{\S_Q}\right)^2,~~
\L_Q\L_Q\leftrightarrow\S_Q\S_Q^*,~~ u^2=m_{ex}^2-\left(\f{m_{\S_Q^*}^2-m_{\S_Q}^2}{4m_{\L_Q}}\right)^2,\\
\S_Q\S_Q\leftrightarrow\S_Q\S_Q^*,&~~& u^2=m_{ex}^2-\left(\f{m_{\S_Q^*}^2-m_{\S_Q}^2}{4m_{\S_Q}}\right)^2,~~
\S_Q^*\S_Q^*\leftrightarrow\S_Q\S_Q^*,~~ u^2=m_{ex}^2-\left(\f{m_{\S_Q^*}^2-m_{\S_Q}^2}{4m_{\S_Q}^*}\right)^2,~~
\mbox{Other channels},~~ u^2=m_{ex}^2.
\end{eqnarray*}

\begin{table}[htp]
\renewcommand{\arraystretch}{1.5}
\caption{The flavor channels for the $\L_Q\L_Q$ system, where
$Q=b$ or $c$.}\label{LL}
\begin{tabular*}{18cm}{@{\extracolsep{\fill}}ccccccccc}
\toprule[1pt]\addlinespace[3pt]
Channels             &      1              &         2          &   3
&      4             &      5              &        6           &   7       \\
\specialrule{0.8pt}{3pt}{3pt}
$I(J^P)=0(0^+)$      &$\L_Q\L_Q(^1S_0)$    &$\S_Q\S_Q(^1S_0)$   &$\S_Q^*\S_Q^*(^1S_0)$
&$\S_Q\S_Q^*(^5D_0)$ &$\S_Q^*\S_Q^*(^5D_0)$&                    &           \\
$I(J^P)=0(0^-)$      &$\L_Q\L_Q(^3P_0)$    &$\S_Q\S_Q(^3P_0)$   &$\S_Q^*\S_Q^*(^3P_0)$
&$\S_Q\S_Q^*(^3P_0)$  &$\S_Q^*\S_Q^*(^7F_0)$ &                    &           \\
$I(J^P)=0(1^-)$      &$\L_Q\L_Q(^3P_1)$    &$\S_Q\S_Q(^3P_1)$   &$\S_Q^*\S_Q^*(^3P_1)$
&$\S_Q^*\S_Q^*(^7F_1)$&$\S_Q\S_Q^*(^3P_1)$ &$\S_Q\S_Q^*(^5P_1)$ &$\S_Q\S_Q^*(^5F_1)$ \\
\addlinespace[3pt]
\bottomrule[1pt]
\end{tabular*} \label{coupled}
\end{table}

\begin{table}
\caption{The specific expressions of operators $\Delta_{ten}$ and
$\Delta_{SS}$ for the individual channels. $S_t$ and
$\f{3}{2}\s_{rs}\equiv -\f{3}{2}S_{t\mu}\s S^{t\mu}$ are the
transition matrix and the spin operator of the spin-$\f{3}{2}$
baryons, respectively. One can refer to Ref. \cite{Meguro:2011nr}
for their definitions.}\label{operator}
\begin{tabular*}{10cm}{@{\extracolsep{\fill}}ccc}
\toprule[1.0pt] \addlinespace[5pt]
Channels             &  $\Delta_{ten}$      &$\Delta_{SS}$   \\
\specialrule{0.8pt}{3pt}{3pt}
$\L_Q\L_Q\leftrightarrow\S_Q^*\S_Q^*$ &  \multirow{2}*{$3\mathbf{S}_{t1}^{\dag}\cdot\mathbf{\hat{r}}
\mathbf{S}_{t2}^{\dag}\cdot\mathbf{\hat{r}}-\mathbf{S}_{t1}^{\dag}\cdot\mathbf{S}_{t2}^{\dag}$}&
\multirow{2}*{$\mathbf{S}_{t1}^{\dag}\cdot\mathbf{S}_{t2}^{\dag}$}\\
$\S_Q\S_Q\leftrightarrow\S_Q^*\S_Q^*$ &                  &                       \\[10pt]
$\L_Q\L_Q\leftrightarrow\S_Q\S_Q^*$   &  \multirow{2}*{$3\mathbf{\sigma}_{1}\cdot\mathbf{\hat{r}}
\mathbf{S}_{t2}^{\dag}\cdot\mathbf{\hat{r}}-\mathbf{\sigma}_{1}\cdot\mathbf{S}_{t2}^{\dag}$}&
\multirow{2}*{$\mathbf{\sigma}_{1}\cdot\mathbf{S}_{t2}^{\dag}$} \\
$\S_Q\S_Q\leftrightarrow\S_Q\S_Q^*$                           &                      \\[10pt]
$\S_Q^*\S_Q^*\leftrightarrow\S_Q^*\S_Q^*$   & $3\mathbf{\sigma}_{rs1}\cdot\mathbf{\hat{r}}
\mathbf{\sigma}_{rs2}\cdot\mathbf{\hat{r}}-
\mathbf{\sigma}_{rs1}\cdot\mathbf{\sigma}_{rs2}$&
$\mathbf{\sigma}_{rs1}\cdot\mathbf{\sigma}_{rs2}$ \\[10pt]
$\S_Q\S_Q^*\leftrightarrow\S_Q\S_Q^*$   & $3\mathbf{\sigma}_{1}\cdot\mathbf{\hat{r}}
\mathbf{\sigma}_{rs2}\cdot\mathbf{\hat{r}}-
\mathbf{\sigma}_{1}\cdot\mathbf{\sigma}_{rs2}$&
$\mathbf{\sigma}_{1}\cdot\mathbf{\sigma}_{rs2}$ \\[10pt]
$\S_Q\S_Q^*\leftrightarrow\S_Q^*\S_Q$   & $3\mathbf{S}_{t1}^{\dag}\cdot\mathbf{\hat{r}}
\mathbf{S}_{t2}\cdot\mathbf{\hat{r}}-
\mathbf{S}_{t1}^{\dag}\cdot\mathbf{S}_{t2}$&
$\mathbf{S}_{t1}^{\dag}\cdot\mathbf{S}_{t2}$ \\[3pt]
\bottomrule[1.0pt]
\end{tabular*}
\end{table}
%%%%%%%%%%%%%%%%%%%%%%%%%%%%%%%%%%%%%%%%%%%%%%%%%%%%%%%%%%%%%%%%%
\section{Numerical Results for the $A_QB_Q$ systems}\label{NUMERICAL}
%%%%%%%%%%%%%%%%%%%%%%%%%%%%%%%%%%%%%%%%%%%%%%%%%%%%%%%%%%%%%%%%

In our numerical analysis, we apply the Fortran program
FESSDE~\cite{fessde} to solve the multichannel Schr{\"o}dinger
equation. Solving the Sch{\"o}dinger equations with the potentials
derived in the previous sections, we obtain the numerical results
including the binding energy (B.E.), the root-mean-square radius
($r_{rms}$) and the probabilities of the individual channels. We
also plot the dependence of the binding energy on the cutoff
parameter in the Appendix.

The hadronic molecule is a loosely bound state. Its constituents
are expected to be well separated. One expects that the size of
the molecules should be much larger than that of the conventional
$q\bar{q}$ and $qqq$ hadrons. Recall that the size of the deuteron
is about $1.96$ fm \cite{Simon:1981br}. T\"ornqvist argued that
the size of the meson-meson molecule is even up to 3 fm
\cite{Tornqvist:1991ks}. We expect the size of the hadronic
molecules composed of two heavy baryons should be comparable to
the size of the deuteron. The size of the molecular system may
tell us whether the present framework and numerical results are
self-consistent or not. To be more specific, the size of the
molecular states composed of two charmed (or bottomed) baryons is
expected to be larger than that of $J/\psi$ (or Upsilon).

Generally the value of the cutoff parameter is determined through
fit to experimental data. In our case, there is almost no
information on the heavy baryon-baryon interaction through which
we can extract the cutoff parameter. Fortunately, the
one-boson-exchange model is very successful in explaining the
deuteron with the cutoff parameter 0.80 GeV$<{\mbox{cutoff}}<$1.5
GeV, which provides us a good benchmark.

\begin{figure}
  \centering
  \begin{tabular}{ccc}
  \includegraphics[width=0.33\textwidth]{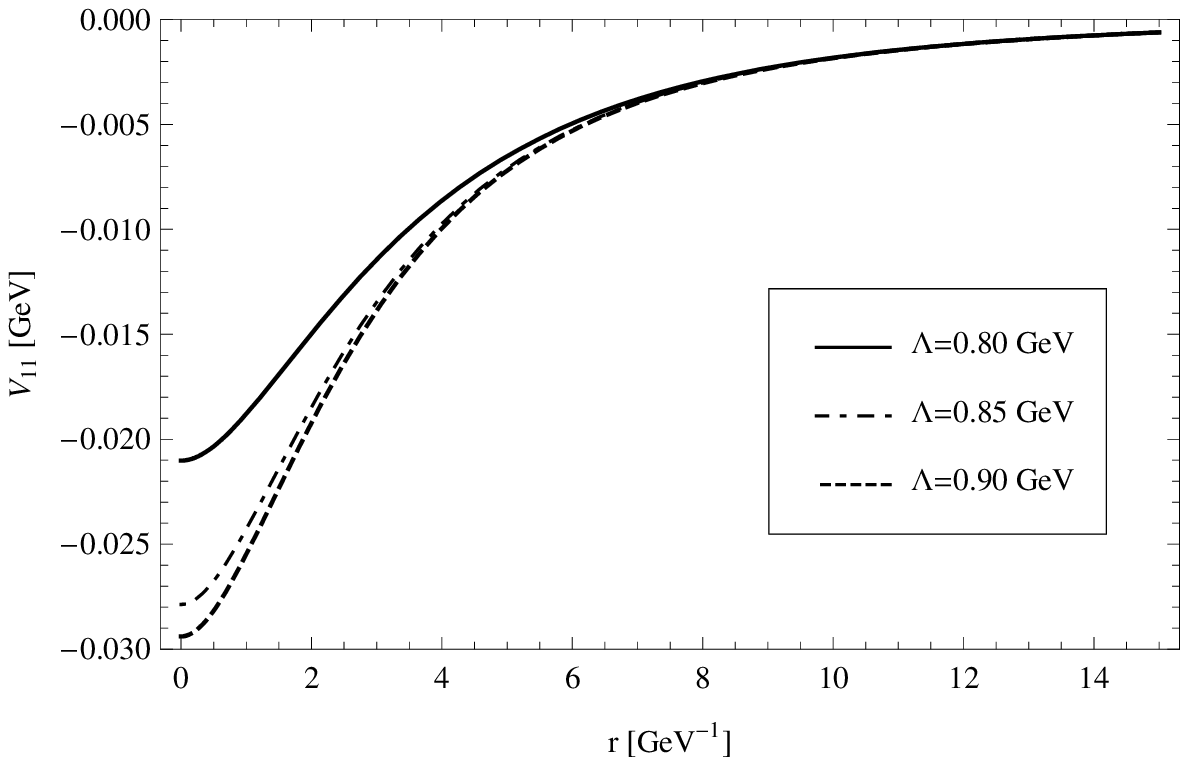}&
  \includegraphics[width=0.33\textwidth]{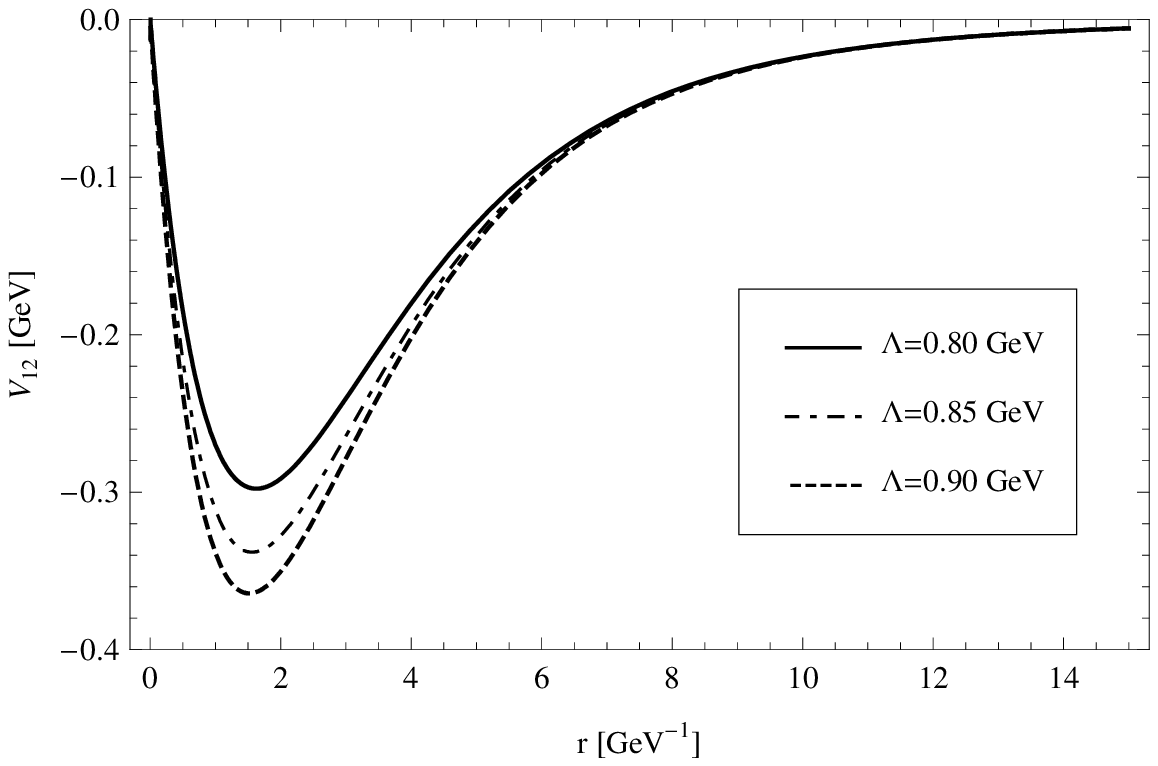}&
  \includegraphics[width=0.33\textwidth]{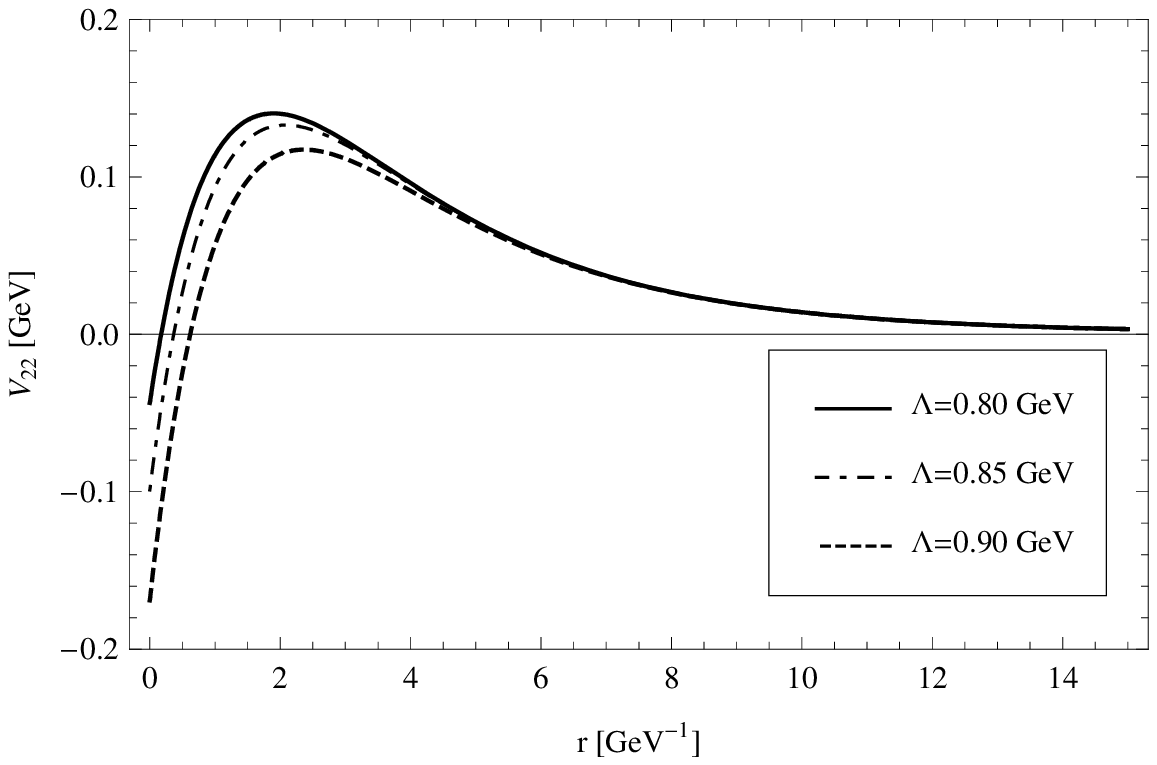}
  \end{tabular}
  \caption{The interaction potentials of the deuteron within the OBE model.
  $V_{11}$, $V_{12}$ and $V_{22}$ are for the transitions
  $^3S_1$$\leftrightarrow$$^3S_1$, $^3S_1$$\leftrightarrow$$^3D_1$ and
  $^3D_1$$\leftrightarrow$$^3D_1$, respectively.}\label{deu:potential}
\end{figure}

The present OBE model is rather crude. For example, we adopt the same cutoff
parameter for all the meson exchange. We plot the interaction potentials of the
deuteron with the OBE model in Fig.~\ref{deu:potential}. We present the
numerical results in Table~\ref{deu:numerical}. In order to study the effect of
the contact interaction, we compare the results (1) when the $\delta (r)$
function is omitted and (2) when the $\delta (r)$ function is explicitly kept.
Without the contact interaction piece, the binding energy of the deuteron is
9.31 MeV and the root-mean-square radius is 2.14 fm when the cutoff is 0.8 GeV.
If we include the contact interaction, the binding energy decreases to 1.87 MeV
and the root-mean-square radius increases to 4.06 fm. In other words, both
approaches roughly reproduce the qualitative feature of the loosely bound
deuteron. In the following, we present the numerical results without the
contact interaction. For comparison, we collect the results with the contact
interaction in Appendix \ref{numerical:contact}.

\begin{figure}
\centering
\begin{tabular}{cc}
\includegraphics[width=0.40\textwidth]{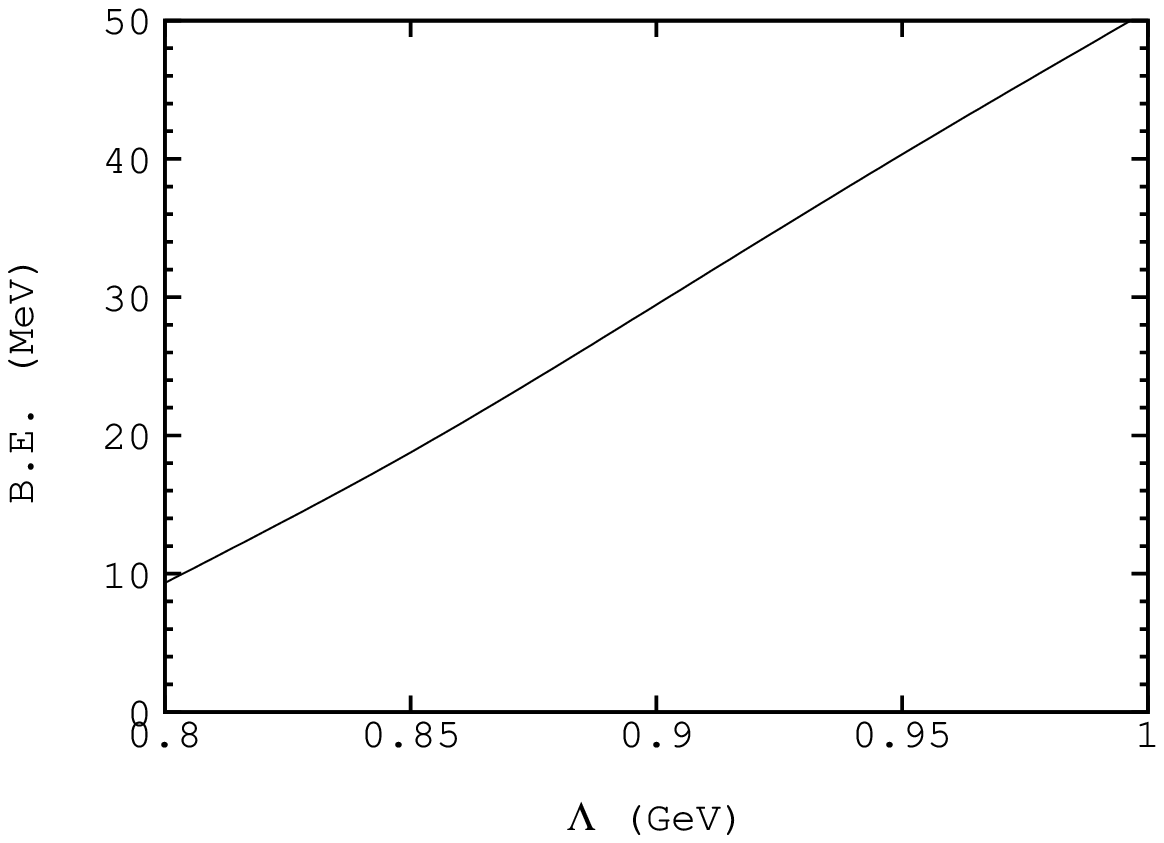}&
\includegraphics[width=0.40\textwidth]{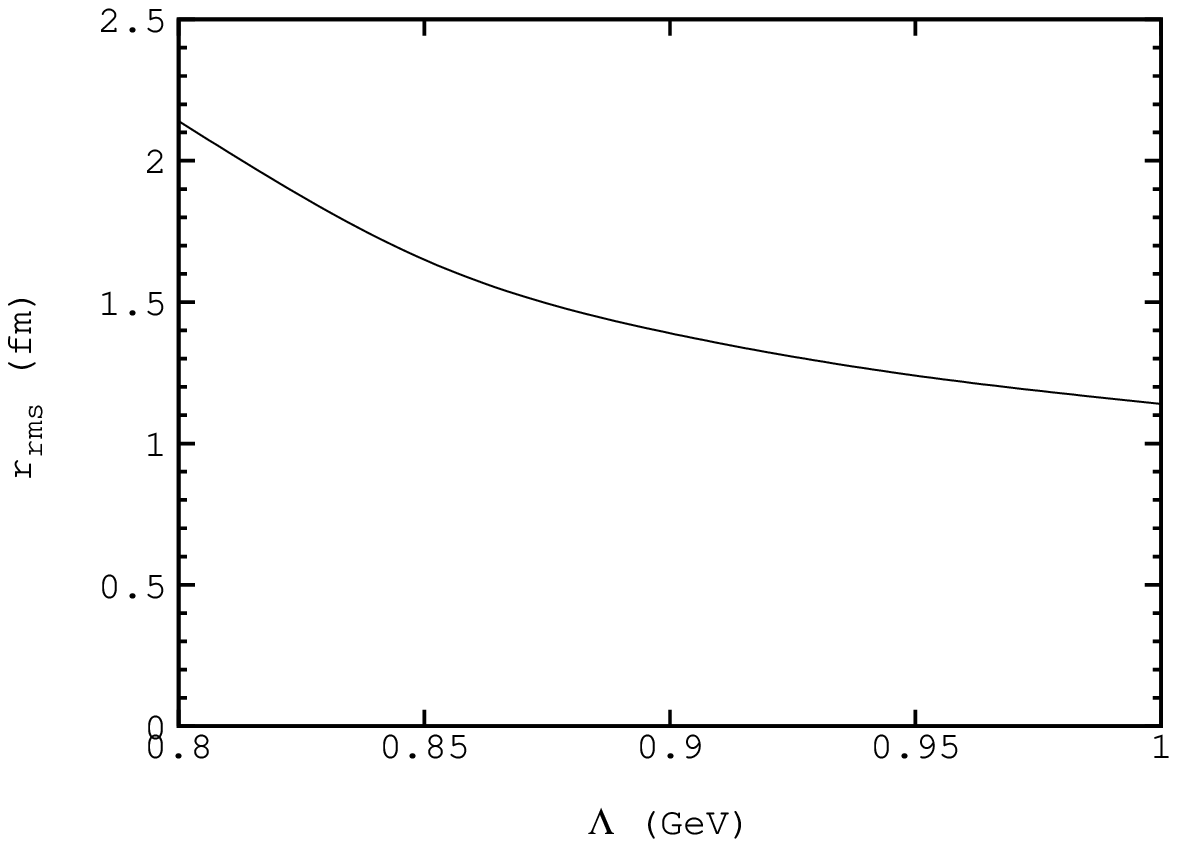}
\end{tabular}
\caption{The variations of the binding energy and the
root-mean-square radius of the deuteron with the cutoff
parameter.}\label{deu:BR}
\end{figure}

However, if we shut down the $^3D_1$ channel, we can not find the binding
solutions, which means that the S-D mixing effect is very important in the
formation of the loosely deuteron bound state although the probability of the D
wave is as small as $\sim 6\%$. One can refer to Fig.~\ref{deu:BR} for the
variations of the binding energy and the root-mean-square radius of the
deuteron with the cutoff parameter.

\begin{table}
\renewcommand{\arraystretch}{1.5}
\centering \caption{The numerical result of the deuteron with the
OBE potential. ``$\L$" is the cutoff parameter. ``B.E." means the
binding energy. ``$P_S$" and ``$P_D$" are the probabilities of the
S wave and D wave respectively.}
\begin{tabular*}{16cm}{@{\extracolsep{\fill}}ccccccccccc}
\toprule[1.0pt] \addlinespace[3pt]
 \multicolumn{5}{c}{without contact term}      & &
             \multicolumn{5}{c}{with contact term} \\
$\L$ (GeV)   &  B.E. (MeV)  & $r_{rms}$ (fm)  &  $P_S(\%)$   &  $P_D(\%)$&
&$\L$ (GeV)  &  B.E. (MeV)  & $r_{rms}$ (fm)  &  $P_S(\%)$   &  $P_D(\%)$ \\
\specialrule{0.6pt}{3pt}{3pt}
0.80         & 9.31         &  2.14           &   93.19      &  6.81 &
&0.80        &   1.87       & 4.06            &  95.00       &  5.00   \\
0.85         & 18.77        &  1.65           &   92.40      &  7.60 &
&0.85        & 2.58         &  3.59           &   94.14      &  5.86   \\
0.90         & 29.45        &  1.39           &   92.14      &  7.86 &
&0.90        & 2.88         &  3.37           &  93.82       &  6.18   \\ [3pt]
\bottomrule[1pt]\label{deu:numerical}
  \end{tabular*}
\end{table}

It's interesting to investigate whether the long-range pion-exchange
interaction plays a dominant role in forming the hadronic molecules. Therefore,
in the first part we give the numerical results with the one-pion-exchange
(OPE) potential for the systems where the pion exchange is allowed. In the
second part we take into account the scalar and vector boson exchanges, which
account for the medium- and short-range interactions, as well as the $\pi$
exchange.

%%%%%%%%%%%%%%%%%%%%%%%%%%%%%%%%%%%%%%%%%%%%%%%%%%%%%%%%%%%%%%%%%%
\subsection{The Results of The ``$A_QB_Q$" Systems with The OPE Potential}

From Fig.~\ref{Exmeson} and Table~\ref{coefficients}, we can see
that there exists the one-pion-exchange force for five systems:
$[\S_Q\Xi_Q^{'}]^{(I=1/2,3/2)}_S$, $[\S_Q\L_Q]^{I=1}_S$ and
$[\Xi_Q\Xi_Q^{'}]^{(I=0,1)}_S$. For $[\S_Q\Xi_Q^{'}]^{I=1/2}_S$ and
$[\S_Q\Xi_Q^{'}]^{3/2}_S$, the pion exchange exists in the direct
channel while the pion exchange occurs in the cross channel for
the other three systems.

\begin{figure}
  \begin{tabular}{ccc}
  \includegraphics[width=0.245\textwidth]{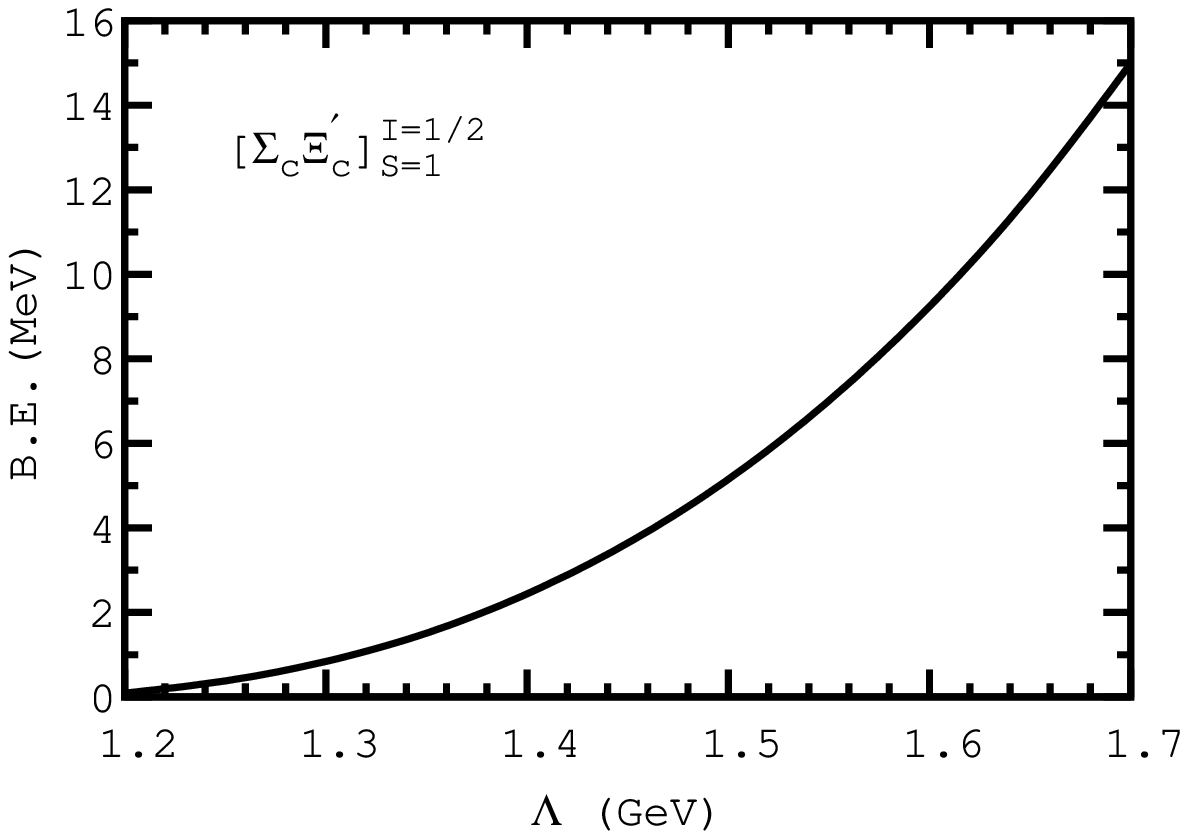}&
\includegraphics[width=0.245\textwidth]{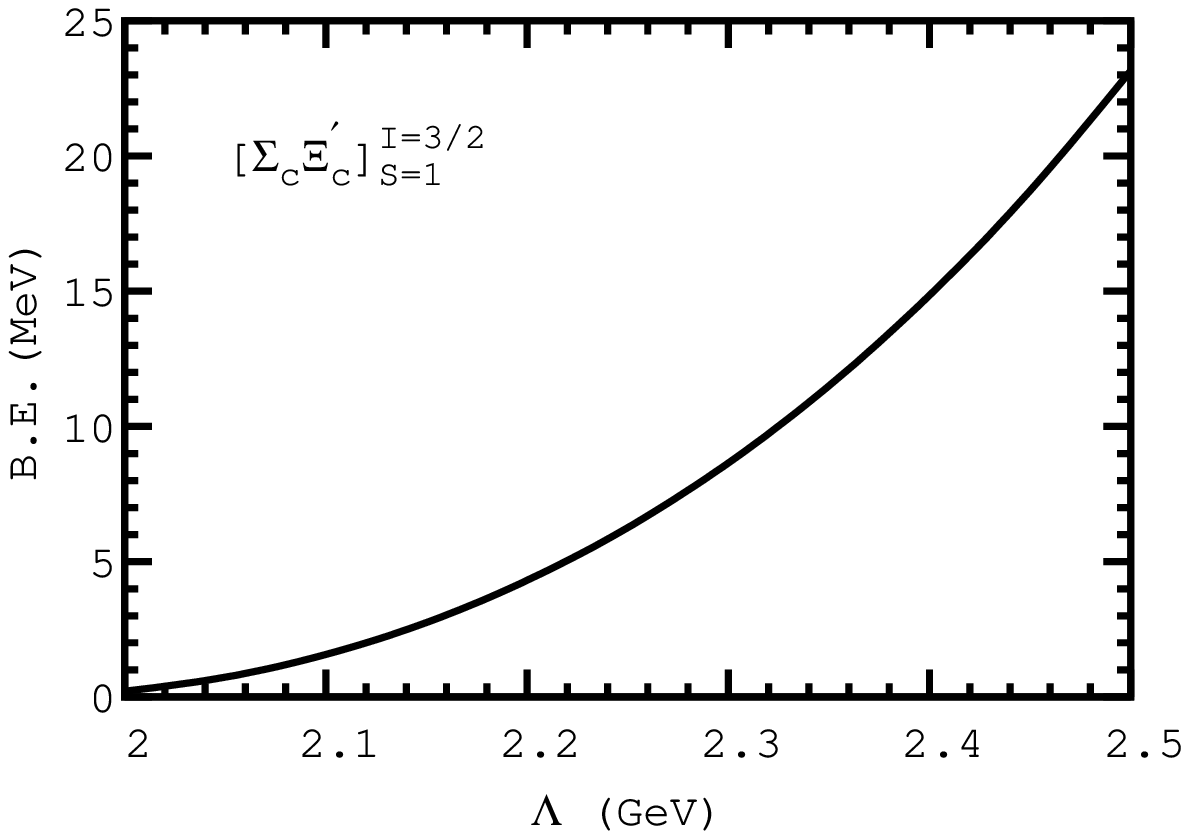}&
\includegraphics[width=0.245\textwidth]{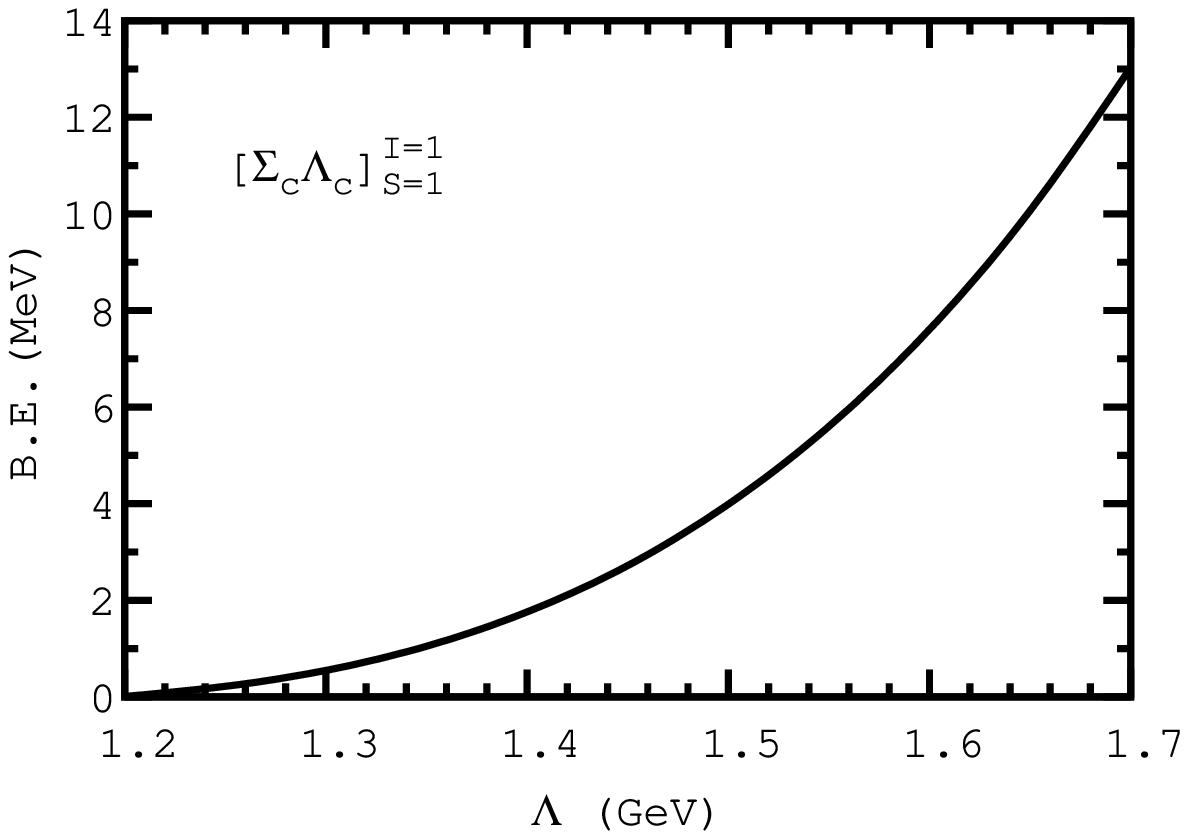}\\
\includegraphics[width=0.245\textwidth]{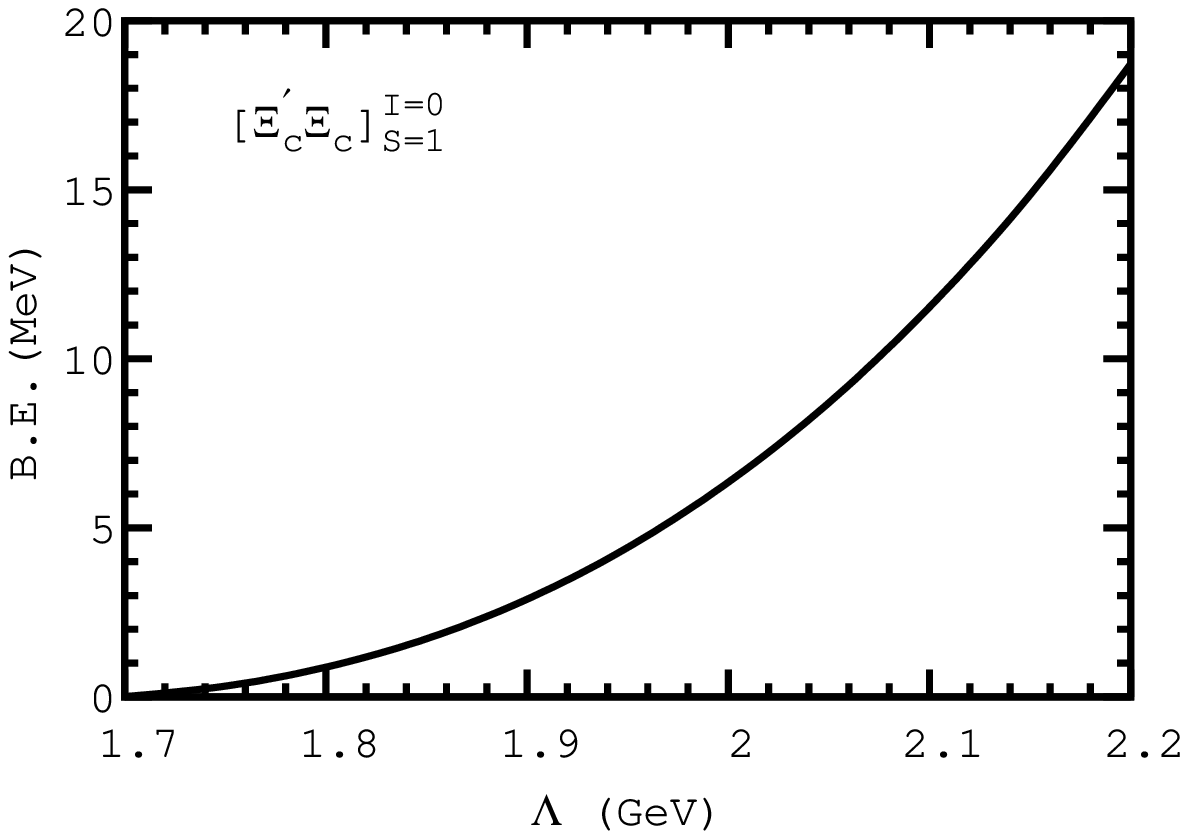}&
    \includegraphics[width=0.245\textwidth]{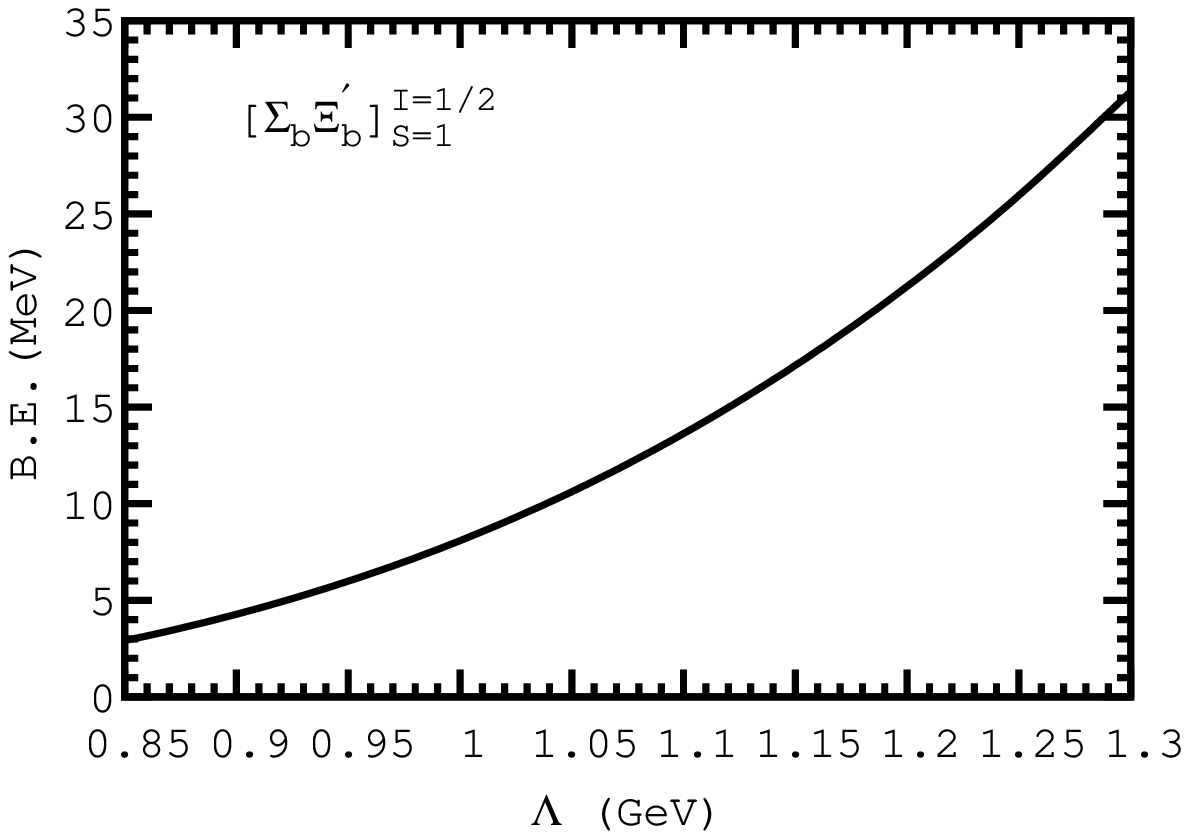}&
\includegraphics[width=0.245\textwidth]{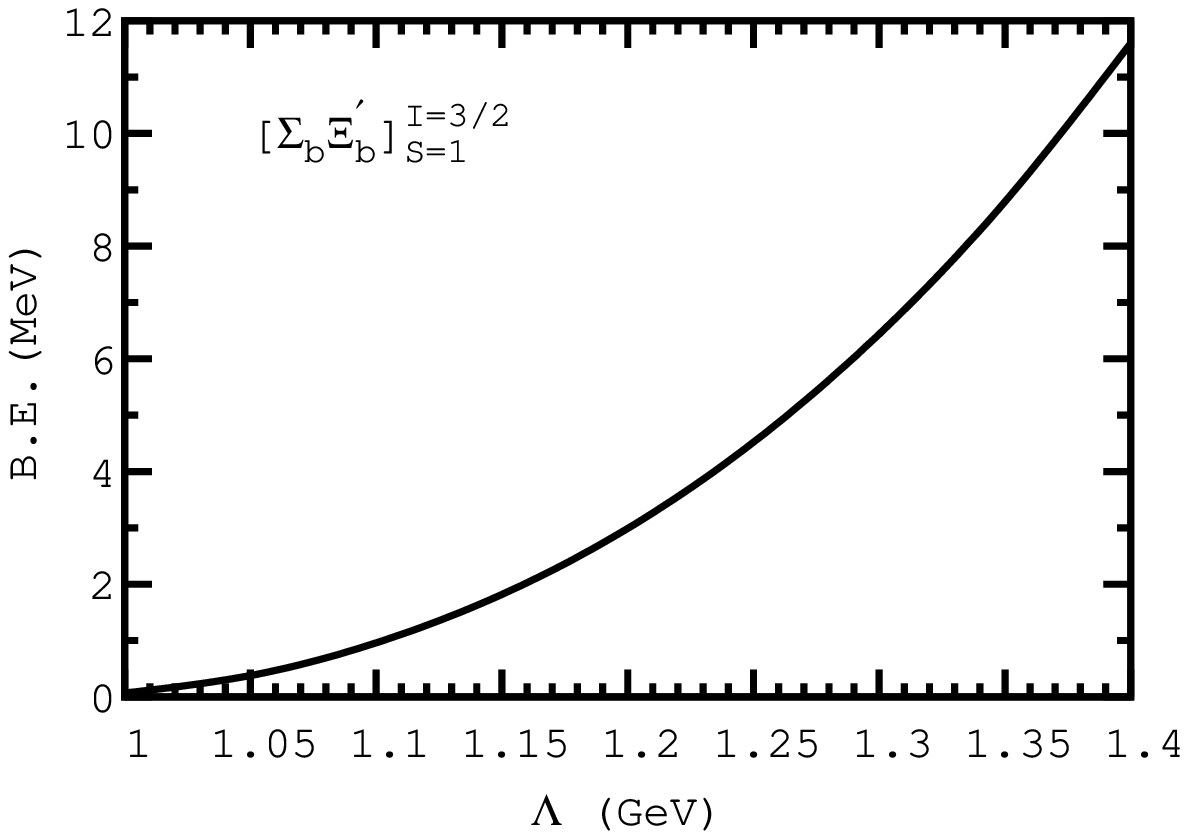}\\
\includegraphics[width=0.245\textwidth]{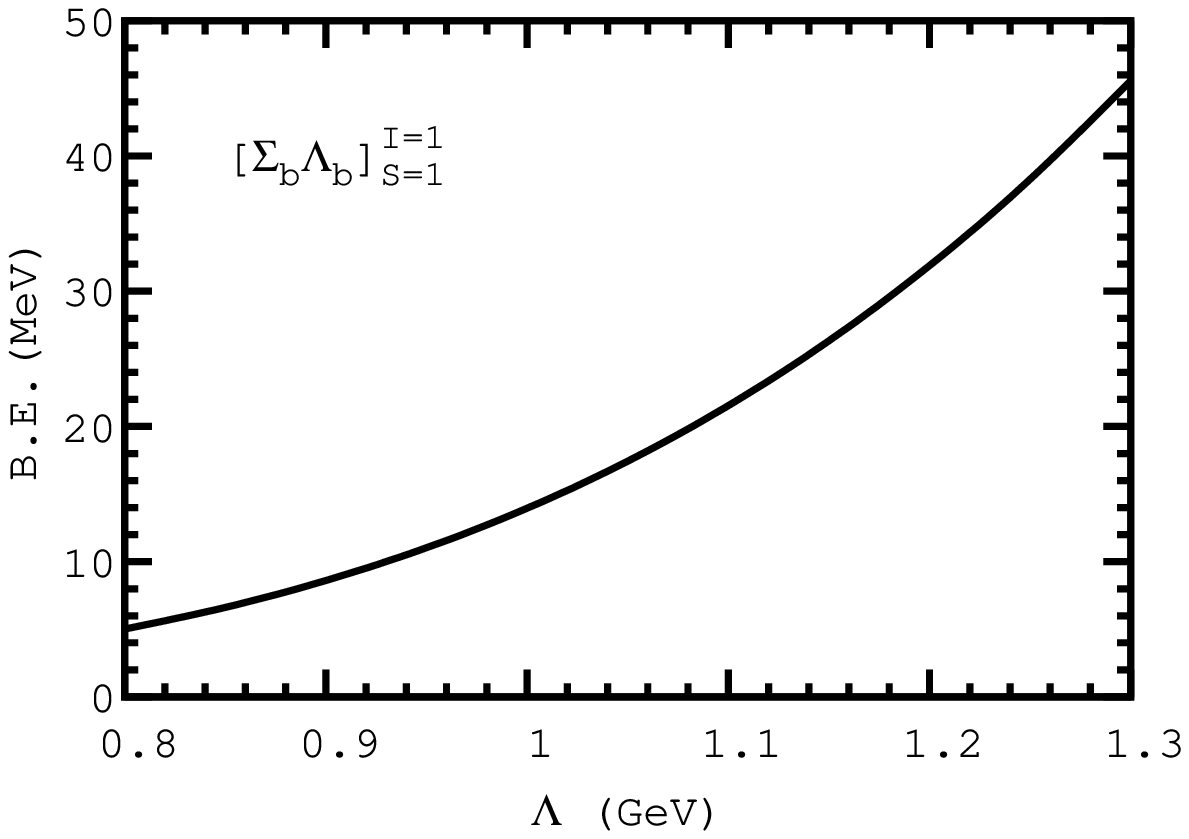}&
\includegraphics[width=0.245\textwidth]{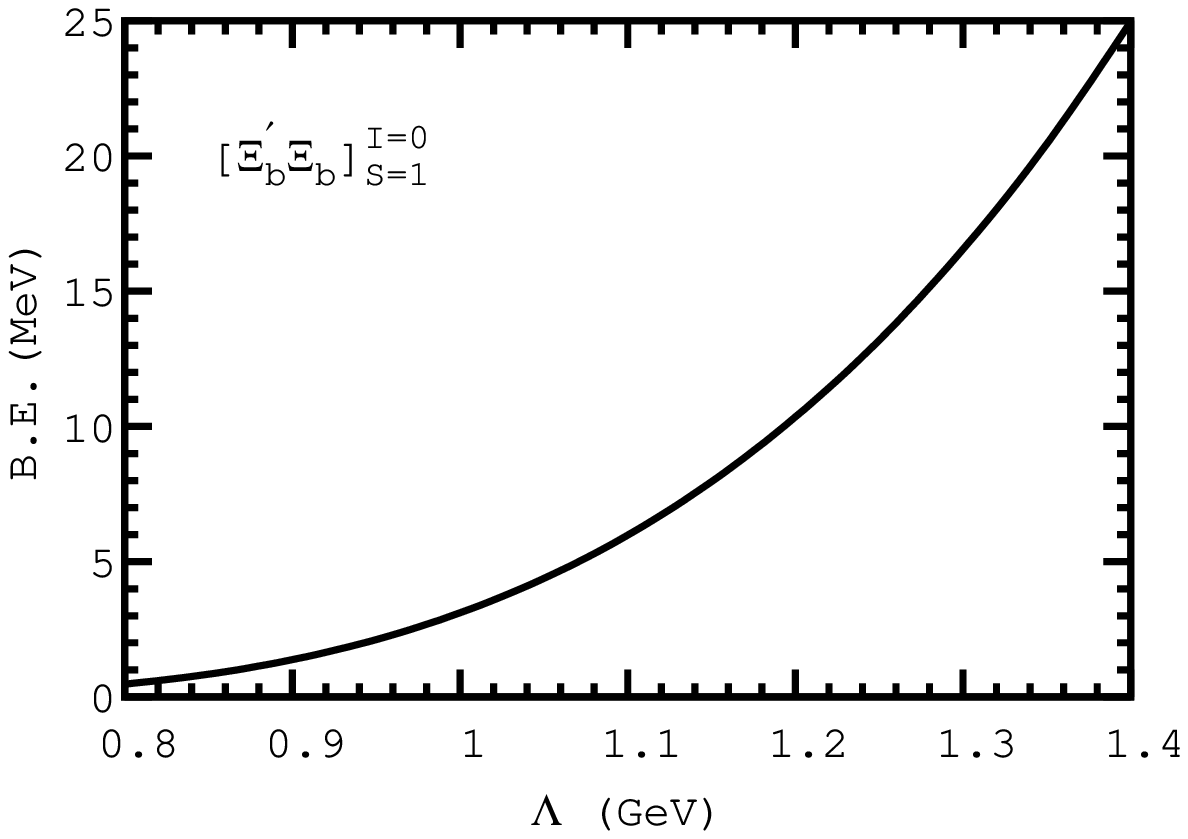}&
    \includegraphics[width=0.245\textwidth]{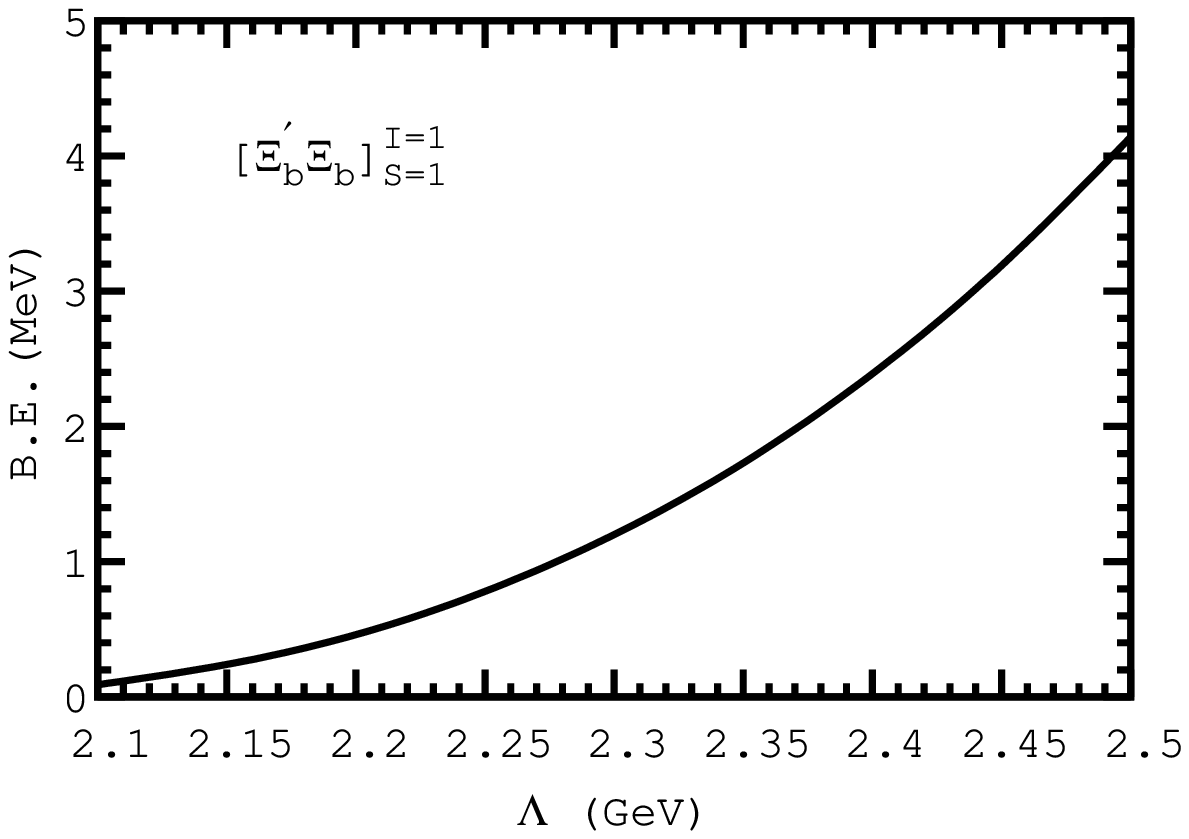}
  \end{tabular}
  \caption{The dependence of the binding energy on the cutoff parameter
for the ``$A_QB_Q$" system with the OPE potential.}\label{PLOTP}
\end{figure}

There are no binding solutions for the five spin-singlet (S=0)
systems. For the spin-triplet (S=1) case, we list the numerical
results in Table~\ref{STrOP} and plot the dependence of the
binding energy on the cutoff parameter in Fig.~\ref{PLOTP}. We
obtain binding solutions for all the states except for
$[\Xi_c\Xi_c^{'}]^{I=1}_{S=1}$, see Table~\ref{STrOP}. In the
charmed sector, loosely bound states
$[\S_c\Xi_c^{'}]^{I=1/2}_{S=1}$ and $[\S_c\L_c]^{I=1}_{S=1}$ have
small binding energy around a few MeV for a reasonable cutoff
parameter about $1.20\sim1.50$ GeV. They are good candidates of
molecules. For these two states, the D wave contribution is less
than $13\%$. For the states $[\S_c\Xi_c^{'}]^{I=3/2}_{S=1}$ and
$[\Xi_c\Xi^{'}_c]^{I=0}_{S=1}$, the binding solutions exist with the
cutoff parameter around 2.0 GeV and 1.80 GeV, respectively. And,
the D wave probabilities for the two states are less than $15\%$.

The bottomed case is similar to the charmed case except that the
binding energy of the bottomed bound states is deeper. This is mainly
because that the larger mass of the bottomed baryon reduces the
kinetic energy. For the states
$[\S_b\Xi_b^{'}]^{(I=1/2,3/2)}_{S=1}$, $[\S_b\L_b]^{I=1}_{S=1}$ and
$[\Xi_b\Xi_b^{'}]^{I=0}_{S=1}$, we obtain loosely bound states
with binding energy less than 30 MeV and the root-mean-square
radius larger than 0.7 fm, when the cutoff parameter is about
$0.9\sim1.30$ GeV. The D wave contribution is larger for the
bottomed systems than for the charmed systems, see Table
\ref{STrOP}. In~\cite{Gerasyuta:2011zx}, the authors performed a
study of the $\S_Q\L_Q$ system at the quark level and obtained
larger binding energy. The $[\Xi_b\Xi_b^{'}]^{I=1}_{S=1}$ bound
state appears when the cutoff parameter is 2.20 GeV. When
2.20 GeV$<{\mbox {cutoff}}<$2.40 GeV, the binding energy is
$0.46\sim2.39$ MeV.

By comparing the numerical results of the two pairs of isospin
multiplets, $[\S_Q\Xi_Q^{'}]^{I=1/2}_{S}$ and
$[\S_Q\Xi_Q^{'}]^{I=3/2}_{S}$ and $[\Xi_Q\Xi_Q^{'}]^{I=0}_{S}$ and
$[\Xi_Q\Xi_Q^{'}]^{I=1}_{S}$, we can see that the results are
different for different isospin multiplets of the same flavor
system since the potentials are isospin-dependent. Comparing the
results of the charmed systems with those of the bottomed systems,
it is obvious that the large heavy quark mass is salient in the
formation of the molecular states.

\begin{table}
\renewcommand{\arraystretch}{1.3}
\centering \caption{The numerical results for the spin-triplet
system (S=1) with the OPE potential. ``$\times$" indicates no
binding solutions. $\L$ is the cutoff parameter. $B.E.$ is the
binding energy while $r_{rms}$ is the root-mean-square radius.
$P_S$ and $P_D$ indicate the the probabilities of the S wave and
the D-wave, respectively. $Q=c$ or $b$ denotes the charmed or the
bottomed systems.}\label{STrOP}
\begin{tabular*}{18cm}{@{\extracolsep{\fill}}cccccccccccc}
\toprule[1.0pt] \addlinespace[5pt]
&\multicolumn{10}{c}{S=1}\\
             &\multicolumn{5}{c}{Q=c}   & \multicolumn{5}{c}{Q=b}\\
Systems      & $\L$(GeV)   &  B.E.(MeV)    & $r_{rms}$(fm) & $P_S(\%)$  & $P_D(\%)$
&  $\L$(GeV) & B.E.(MeV)      & $r_{rms}$(fm)& $P_S(\%)$     & $P_D(\%)$   \\
\specialrule{0.8pt}{3pt}{3pt}
\multirow{3}*{$[\S_Q\Xi_Q^{'}]^{I=1/2}_S$}
& 1.20  & 0.09   & 5.99    &  97.82     &  2.18
& 0.90  & 4.28   & 1.53    &  88.01     &  11.99  \\
& 1.30  & 0.84   & 3.51    &  95.69     &  4.31
& 1.10  & 13.63  & 1.01    &  85.20     &  14.80  \\
& 1.50  & 5.15   & 1.67    &  92.15     &  7.85
& 1.30  & 31.33  & 0.74    &  83.37     &  16.63  \\ [7pt]
\multirow{3}*{$[\S_Q\Xi_Q^{'}]^{I=3/2}_{S}$}
& 2.00  & 0.21   & 5.07    &  96.66     &  3.34
& 1.10  & 0.96   & 2.34    &  86.98     &  13.02  \\
& 2.20  & 4.31   & 1.61    &  89.98     &  10.02
& 1.20  & 2.99   & 1.53    &  82.01     &  17.99  \\
& 2.40  & 14.85  & 0.98    &  85.41     &  14.59
& 1.30  & 6.44   & 1.17    &  78.41     &  21.59  \\ [7pt]
\multirow{3}*{$[\S_Q\L_Q]^{I=1}_S$}
& 1.30  & 0.55   & 4.20    &  97.46     &  7.54
& 0.90  & 8.61   & 1.40    &  71.76     &  28.24  \\
& 1.40  & 1.76   & 2.71    &  89.71     &  10.29
& 1.00  & 13.95  & 1.16    &  70.04     &  29.96  \\
& 1.50  & 3.99   & 1.94    &  87.29     &  12.71
& 1.10  & 21.54  & 0.98    &  68.35     &  31.65  \\ [7pt]
\multirow{3}*{$[\Xi_Q\Xi_Q^{'}]^{I=0}_{S}$}
& 1.80  & 0.88   & 3.25    &  93.65     &  6.35
& 0.90  & 1.38   & 2.37    &  83.17     & 16.83  \\
& 1.90  & 2.59   & 1.96    &  90.33     & 9.67
& 1.10  & 5.98   & 1.35    &  77.00     & 23.00  \\
& 2.00  & 6.35   & 1.42    &  87.63     & 12.38
& 1.30  & 16.55  & 0.93    &  72.40     & 27.60  \\ [7pt]
\multirow{3}*{$[\Xi_Q\Xi_Q^{'}]^{I=1}_{S}$}
& \multicolumn{5}{c}{\multirow{3}*{$\times$}}
& 2.20  &  0.46  &  3.01   &  94.06     & 5.94 \\
&       &        &         &            &
& 2.30  &  1.20  &  1.99   &  91.76     & 8.24 \\
&       &        &         &            &
& 2.40  &  2.39  &  1.48   &  89.69     & 10.31\\ [3pt]
\bottomrule[1pt]
\end{tabular*}
\end{table}

%%%%%%%%%%%%%%%%%%%%%%%%%%%%%%%%%%%%%%%%%%%%%%%%%%%%%%%%%%%%%%%%%%%%%
\subsection{The Results of The ``$A_QB_Q$" Systems with The OBE Potential}

\begin{figure}[htp]
\hfill
\begin{minipage}{0.245\textwidth}
\centering
\includegraphics[width=\textwidth]{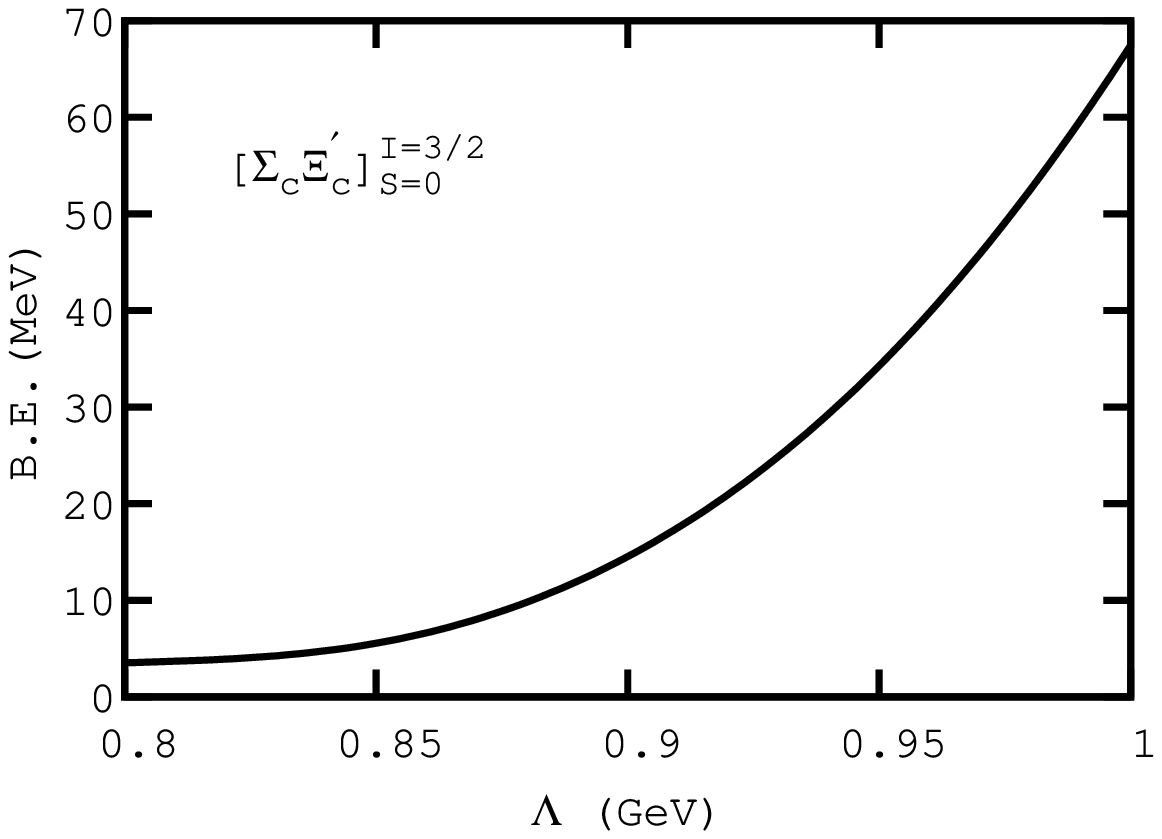}\\
\includegraphics[width=\textwidth]{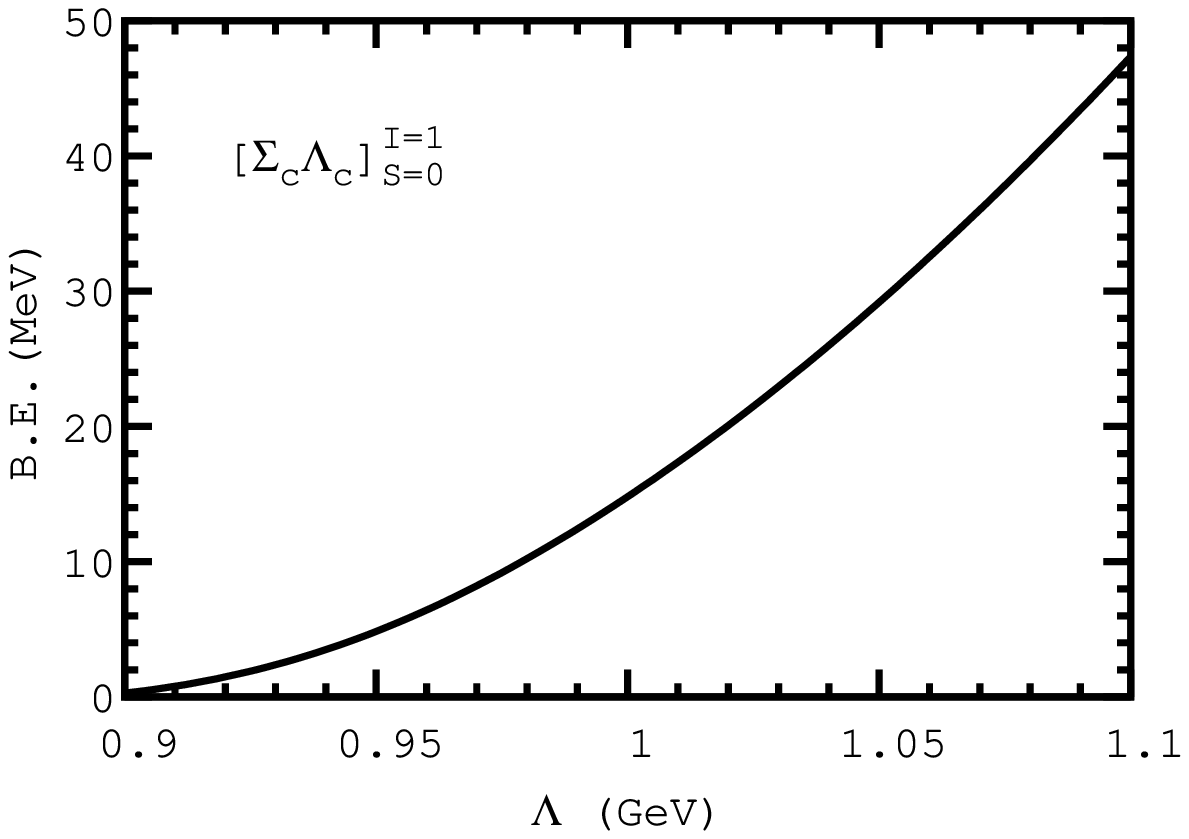}\\
\includegraphics[width=\textwidth]{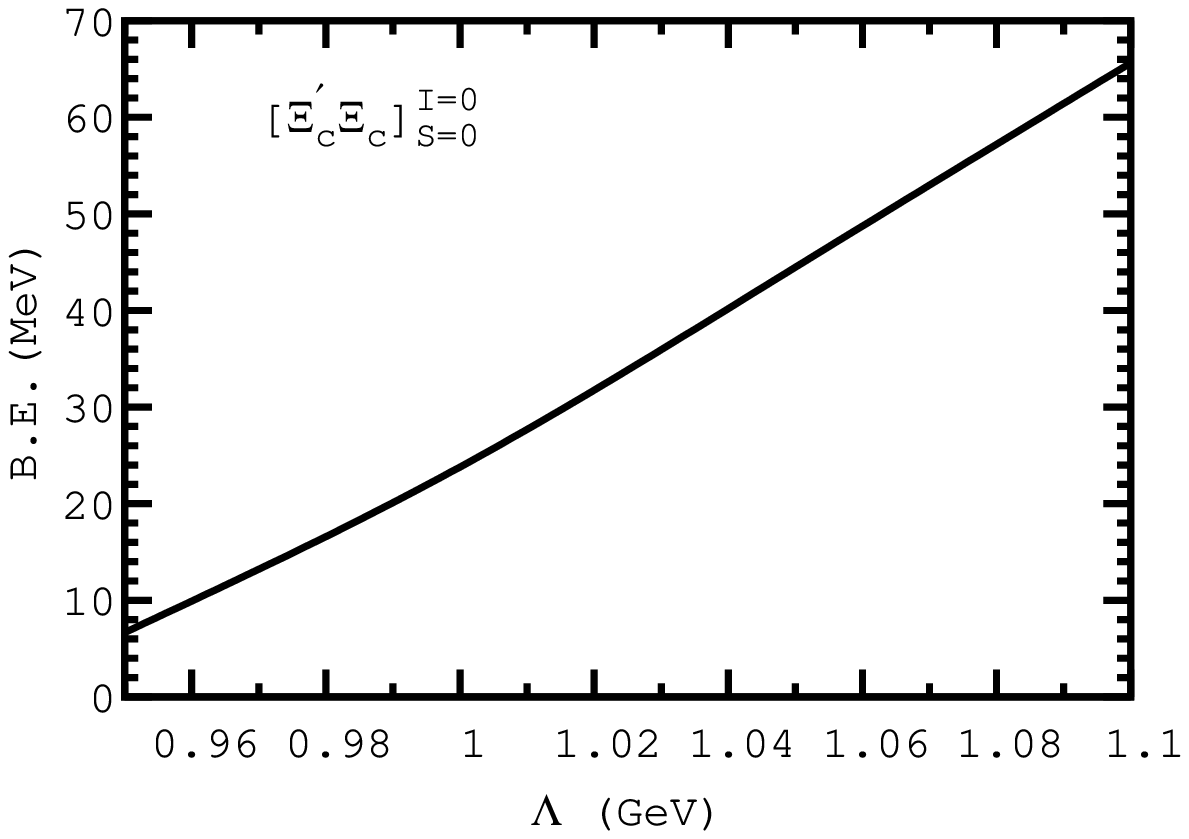}\\
\end{minipage}
\hfill
\begin{minipage}{0.245\textwidth}
\centering
\includegraphics[width=\textwidth]{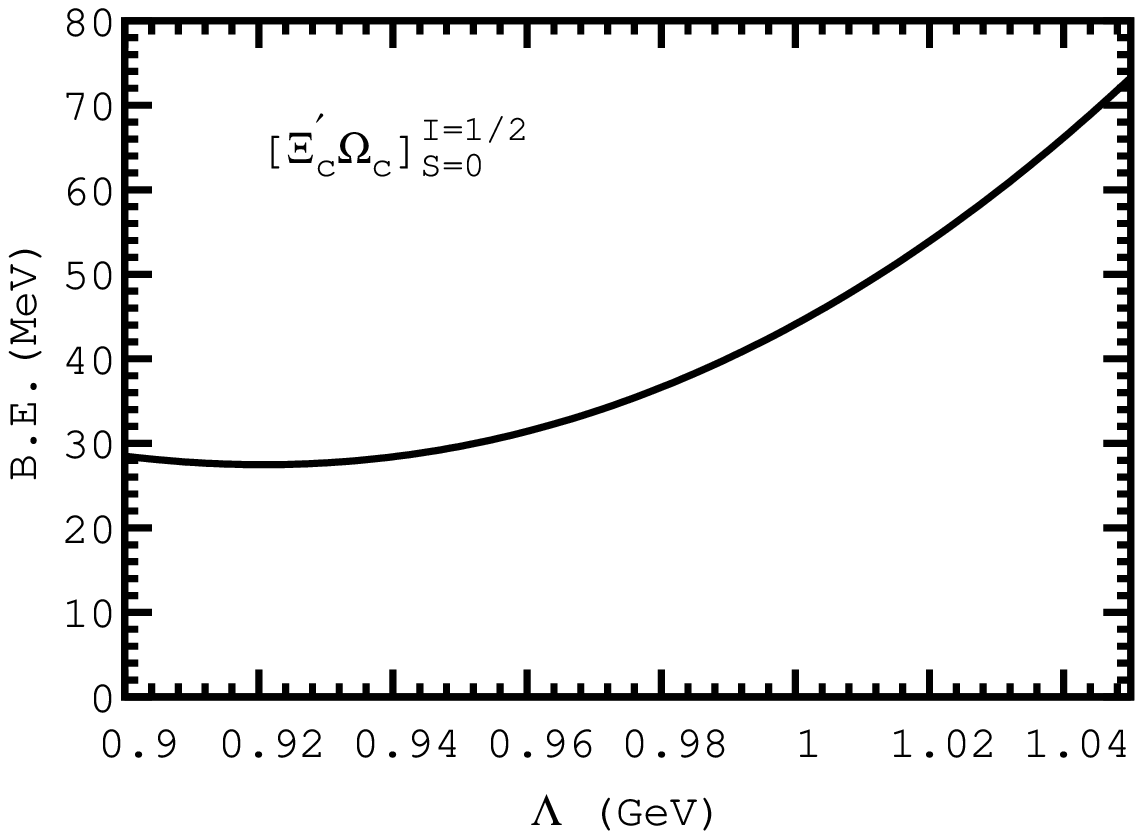}\\
\includegraphics[width=\textwidth]{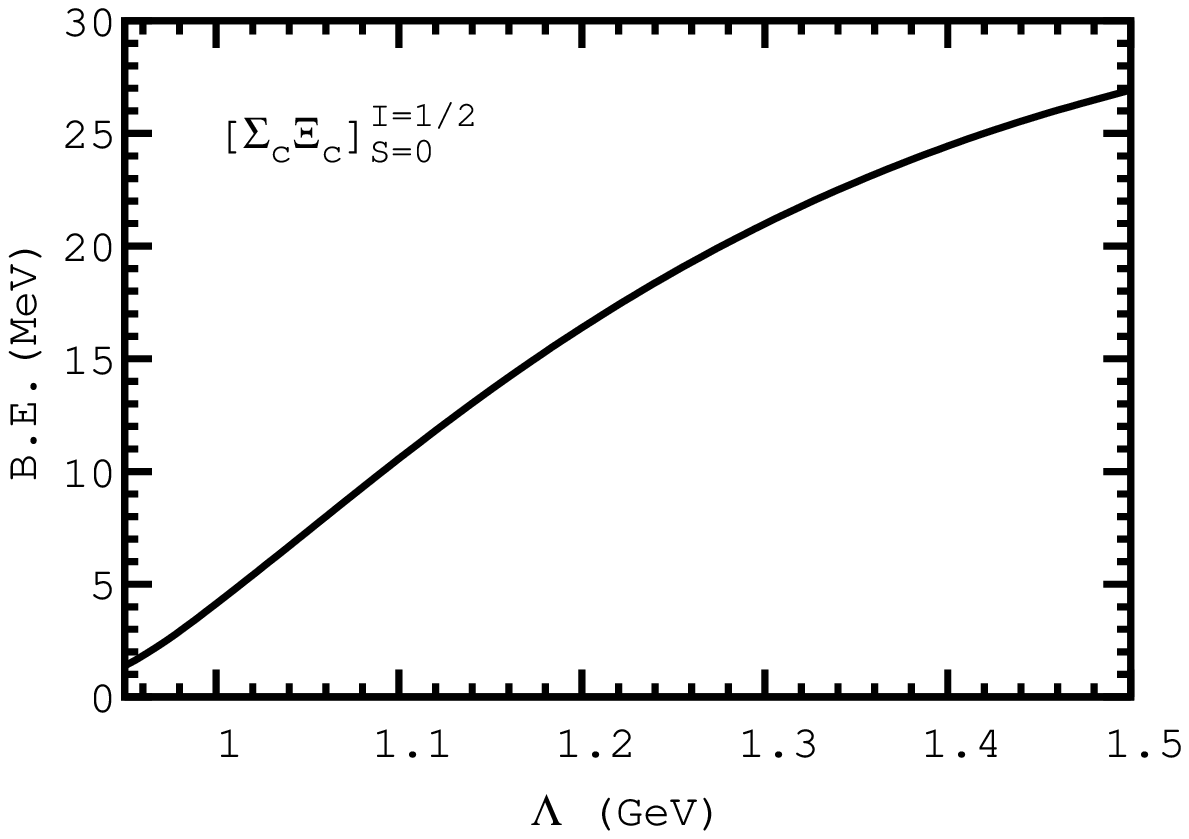}\\
\includegraphics[width=\textwidth]{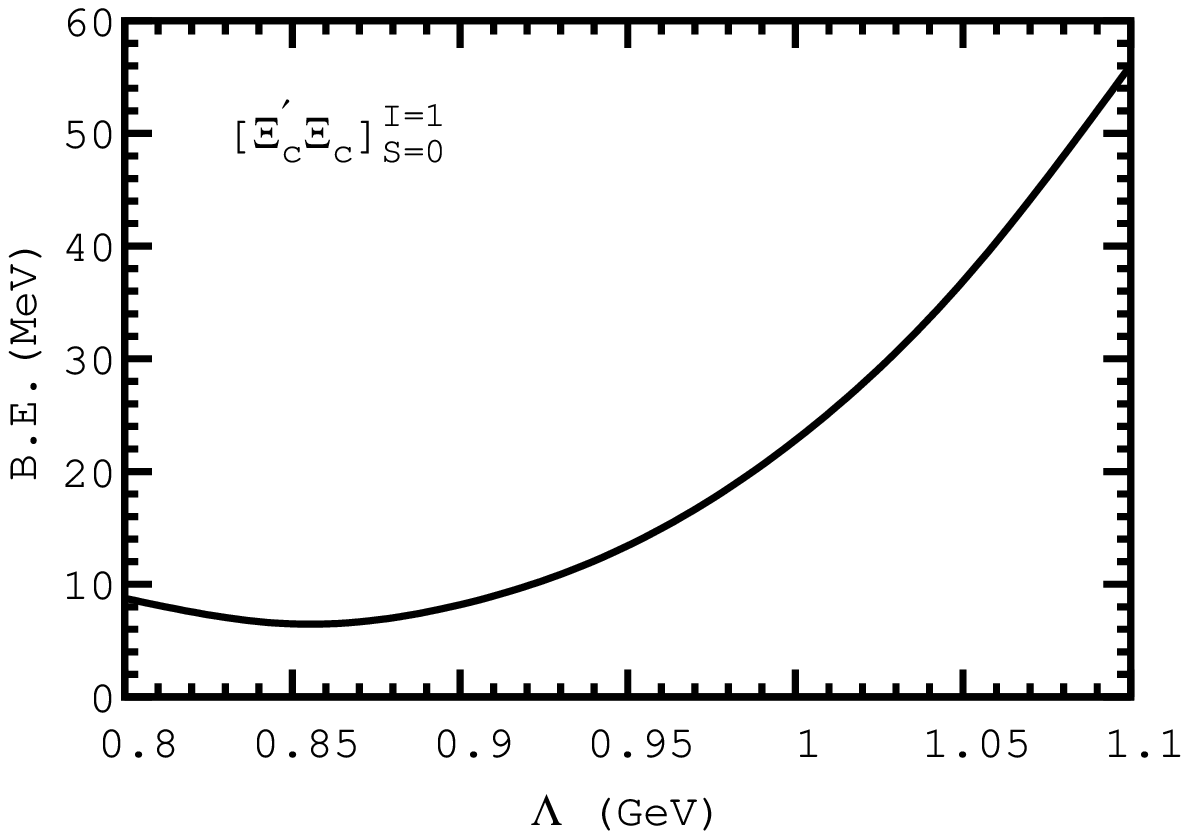}\\
\end{minipage}
\hfill
\begin{minipage}{0.245\textwidth}
\centering
\includegraphics[width=\textwidth]{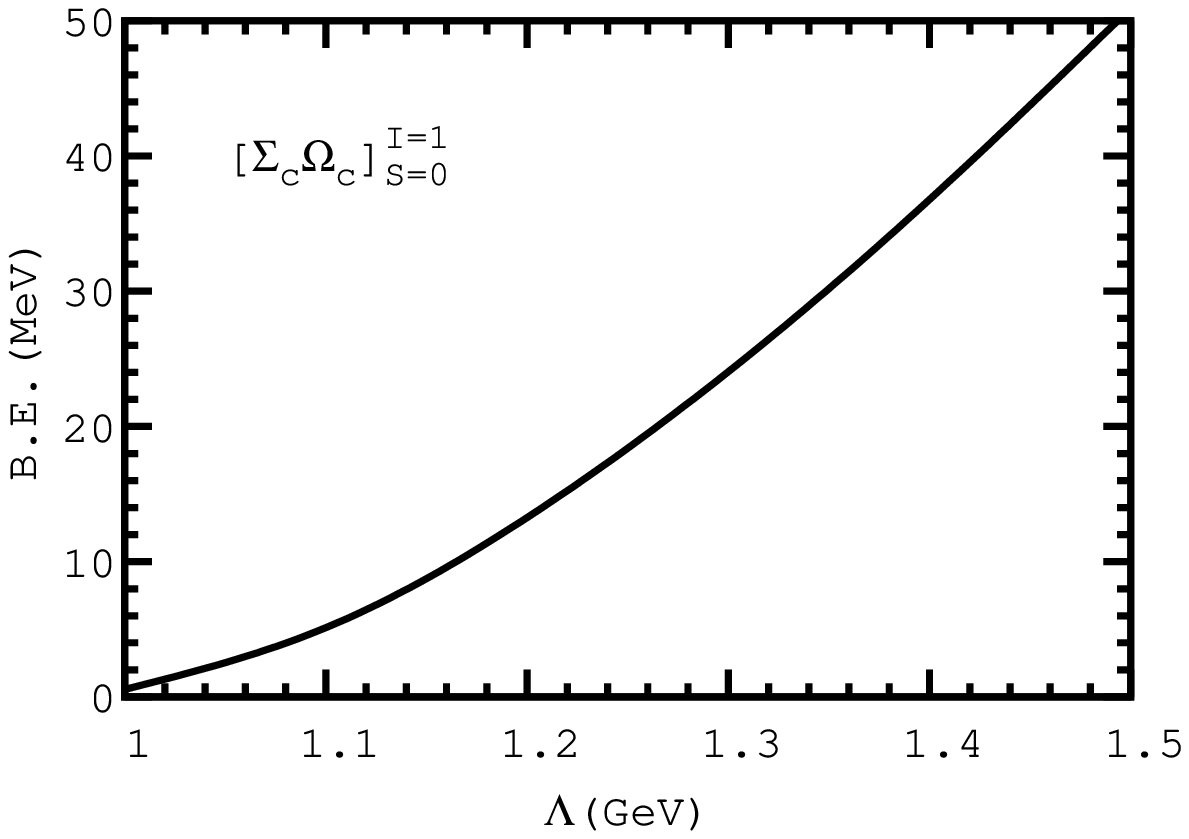}\\
\includegraphics[width=\textwidth]{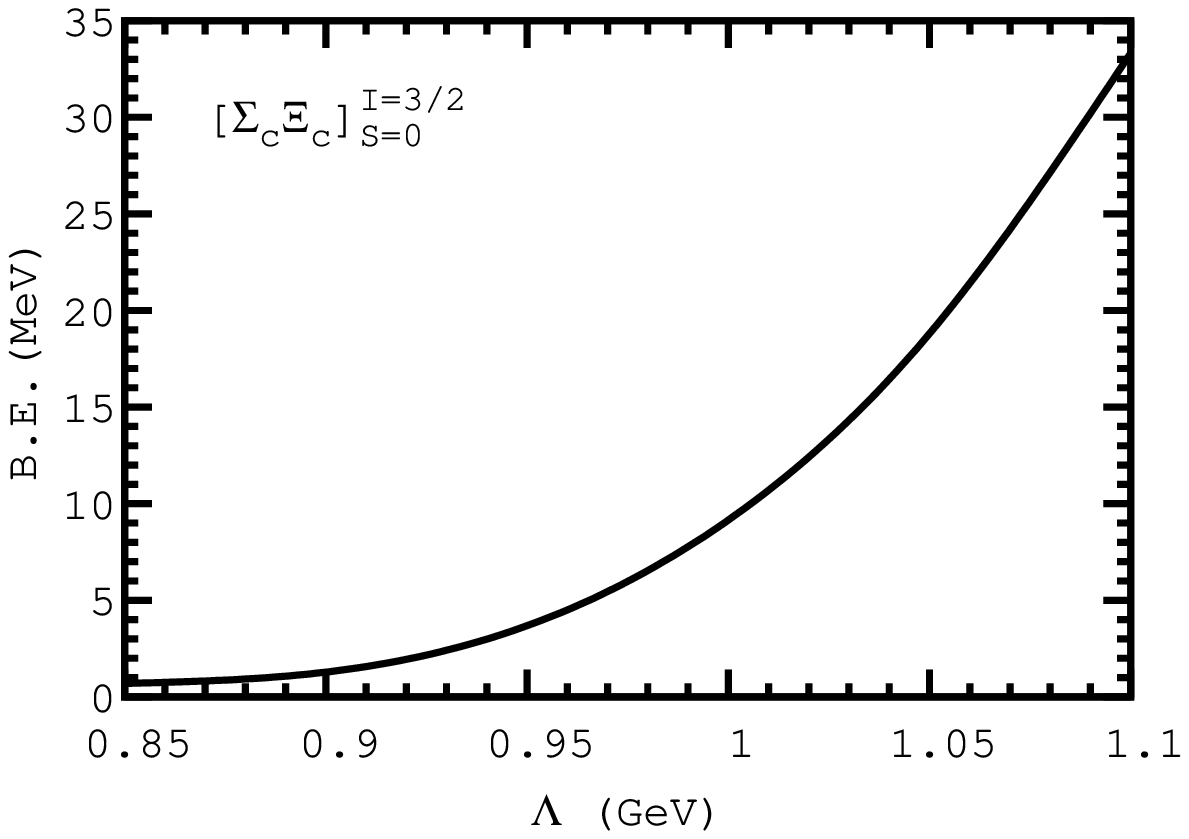}\\
\includegraphics[width=\textwidth]{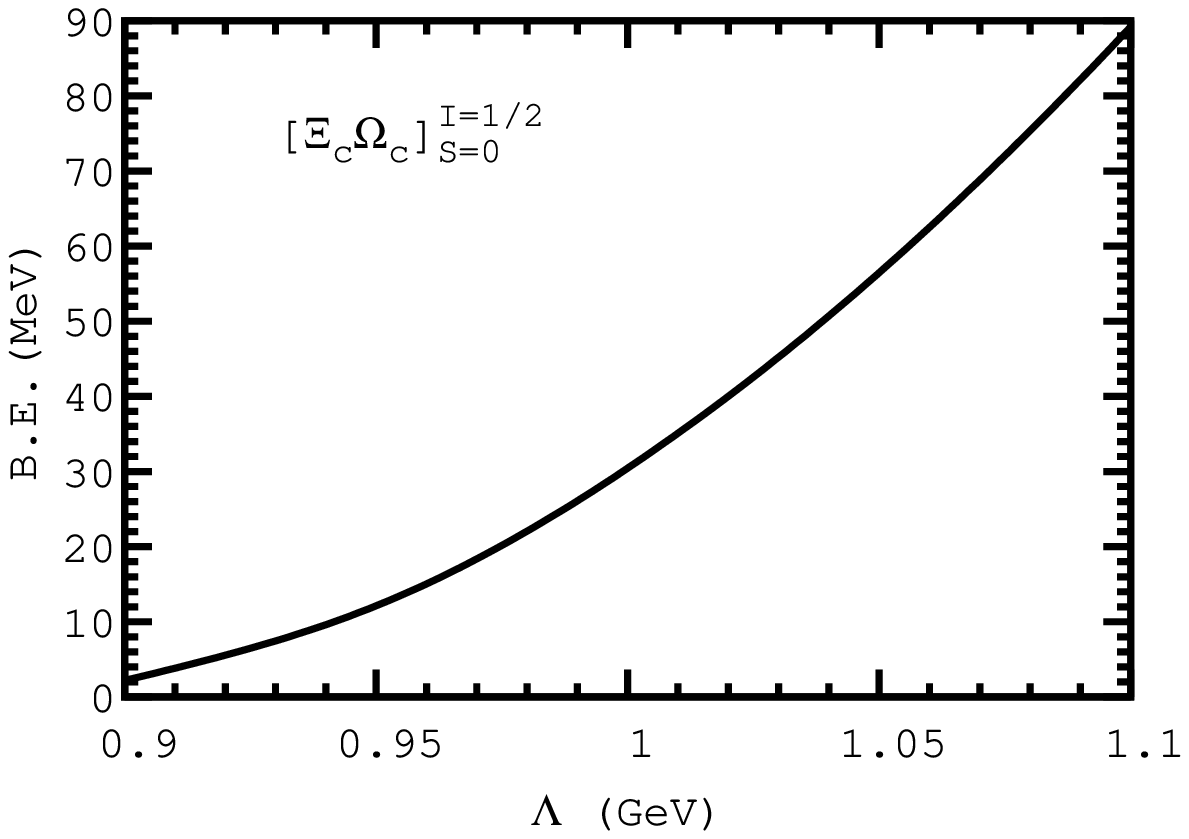}\\
\end{minipage}
\hfill
\begin{minipage}{0.245\textwidth}
\centering
\includegraphics[width=\textwidth]{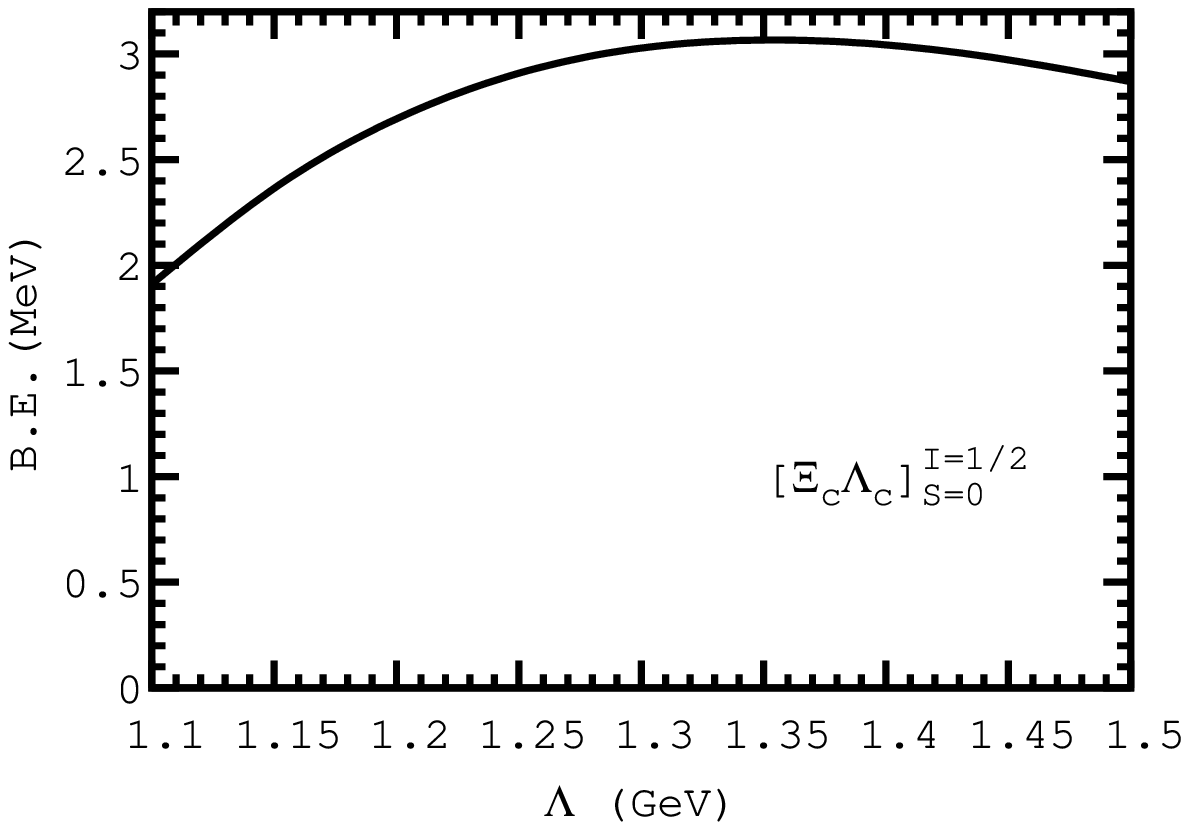}\\
\includegraphics[width=\textwidth]{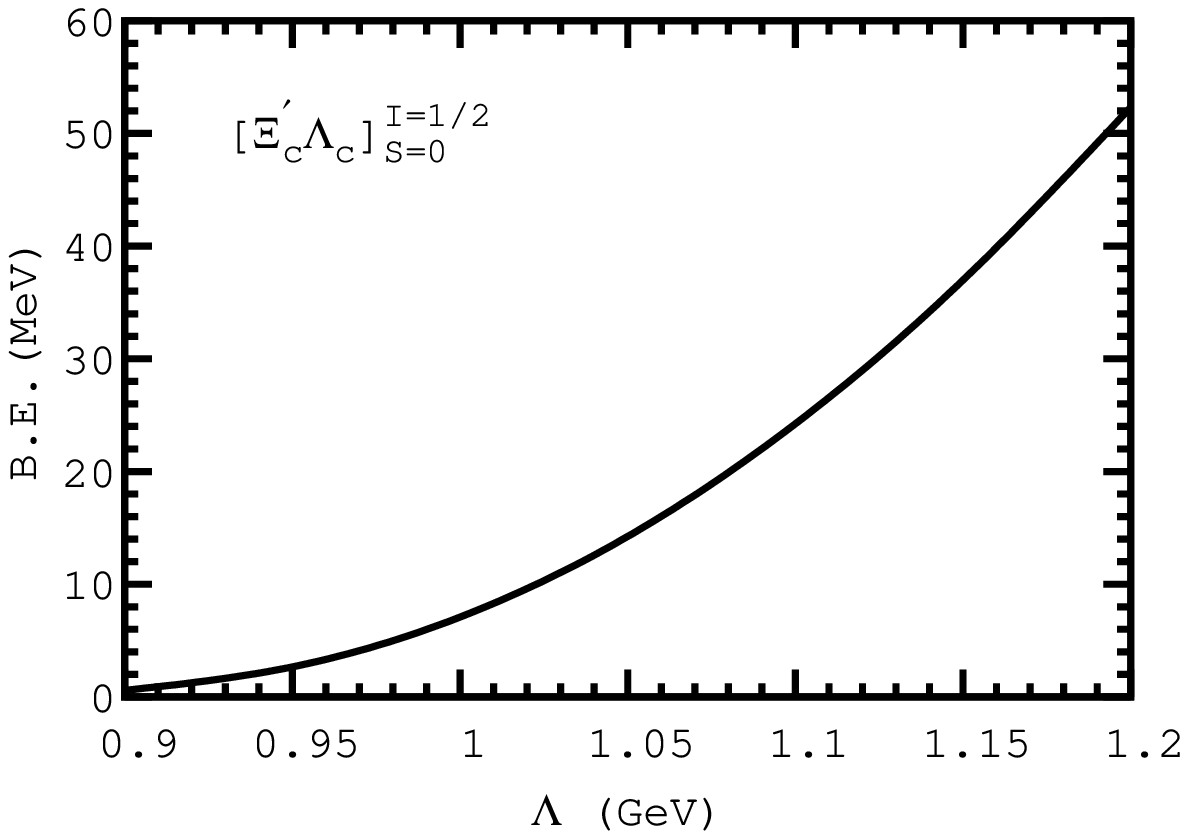}\\
\includegraphics[width=\textwidth]{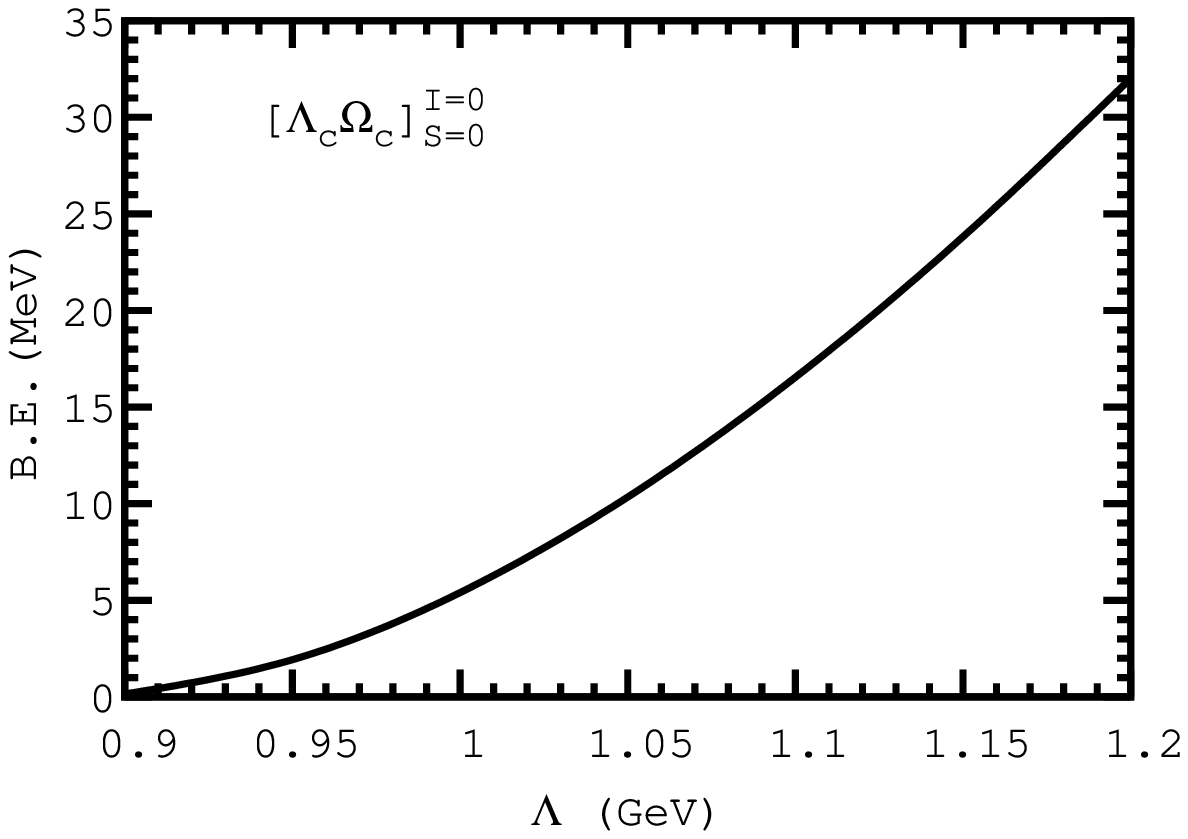}\\
\end{minipage}
\caption{The dependence of the binding energy on the cutoff momentum
for the spin-singlet system ``$A_cB_c$" with the OBE potential.}\label{plotCS}
\end{figure}

\begin{figure}
\centering
  \begin{tabular}{cccc}
    \includegraphics[width=0.245\textwidth]{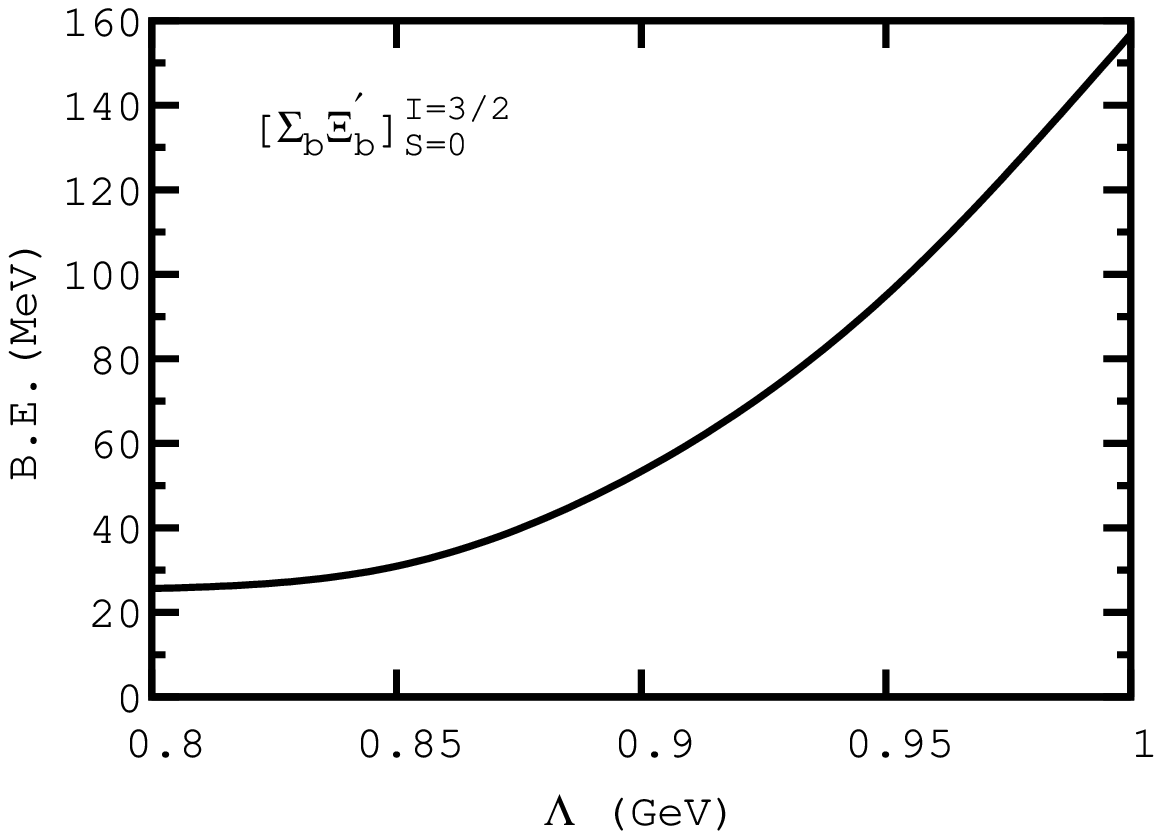}&
\includegraphics[width=0.245\textwidth]{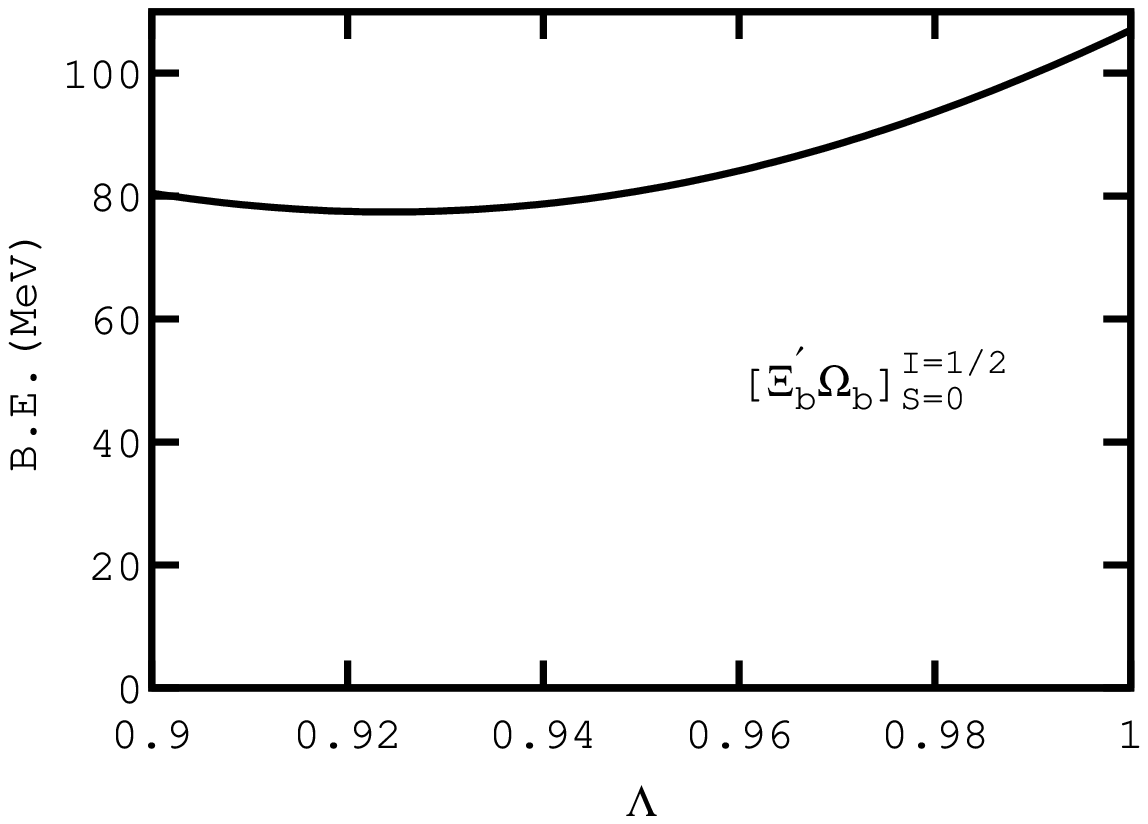}&
\includegraphics[width=0.245\textwidth]{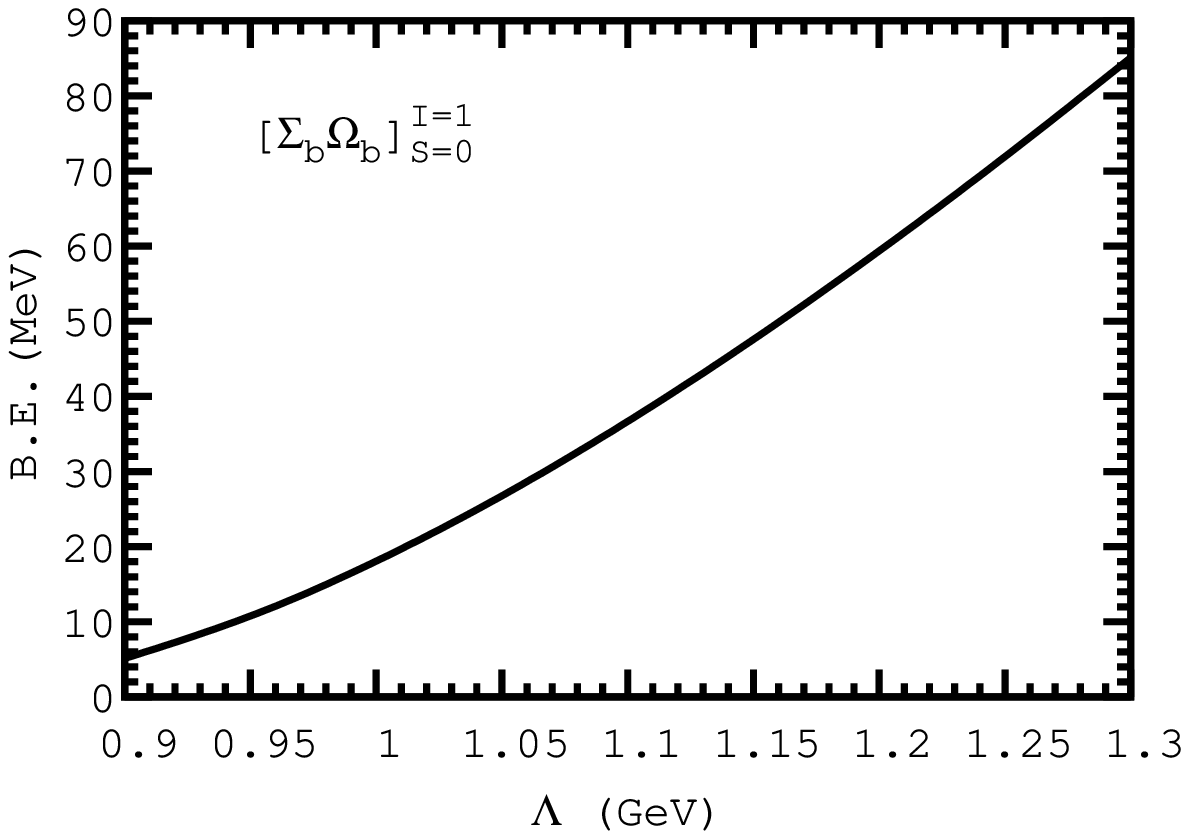}&
\includegraphics[width=0.245\textwidth]{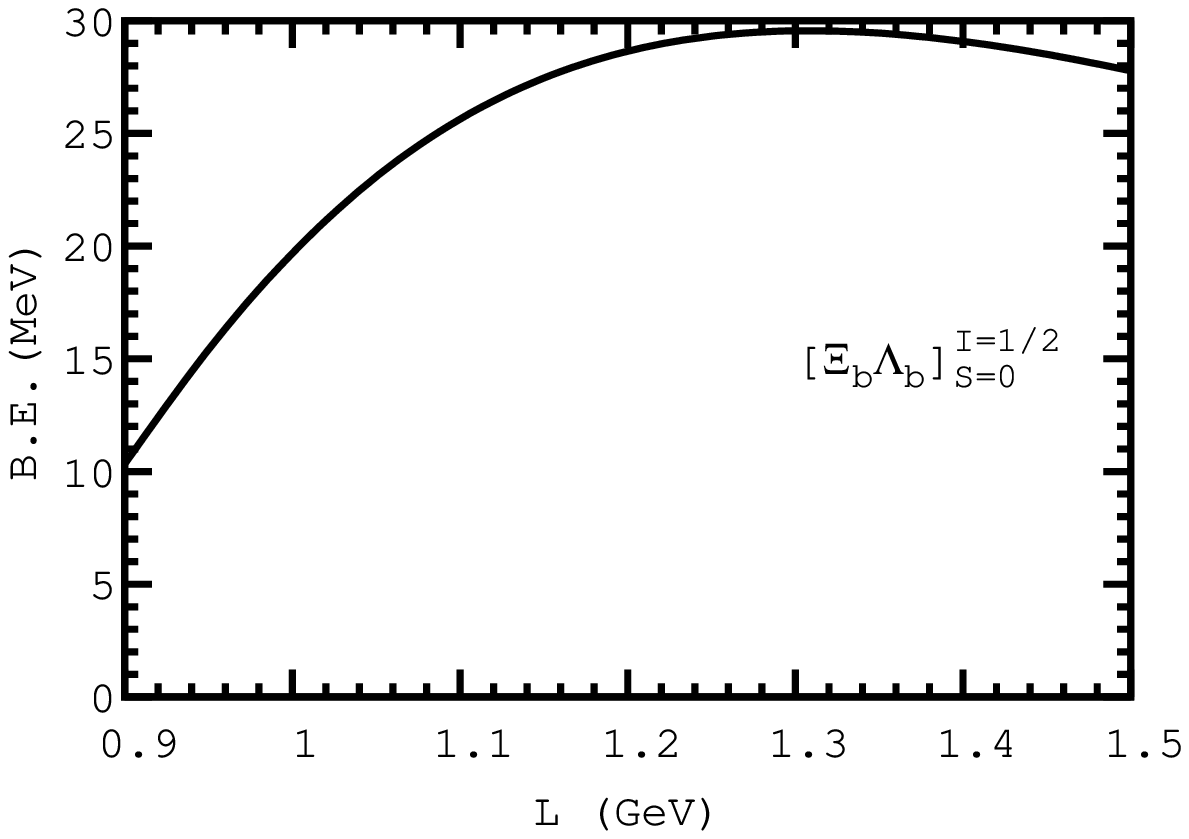}\\
    \includegraphics[width=0.245\textwidth]{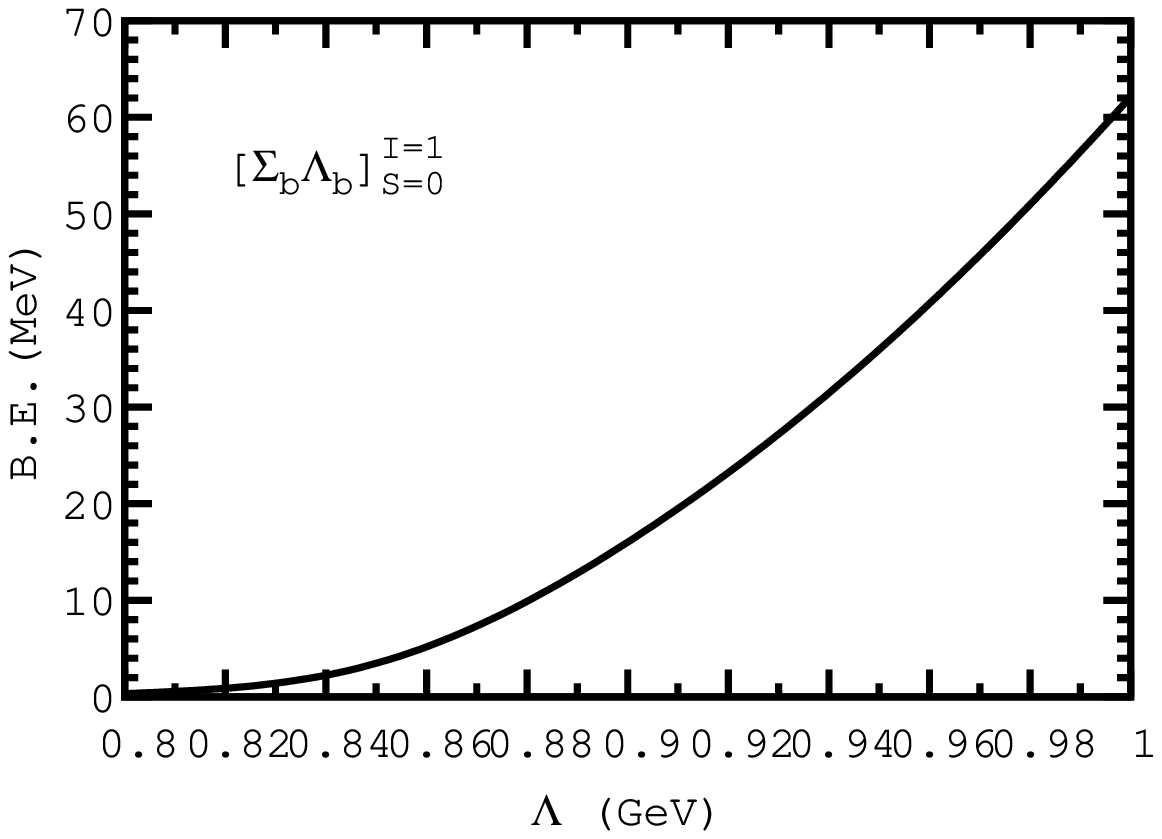}&
\includegraphics[width=0.245\textwidth]{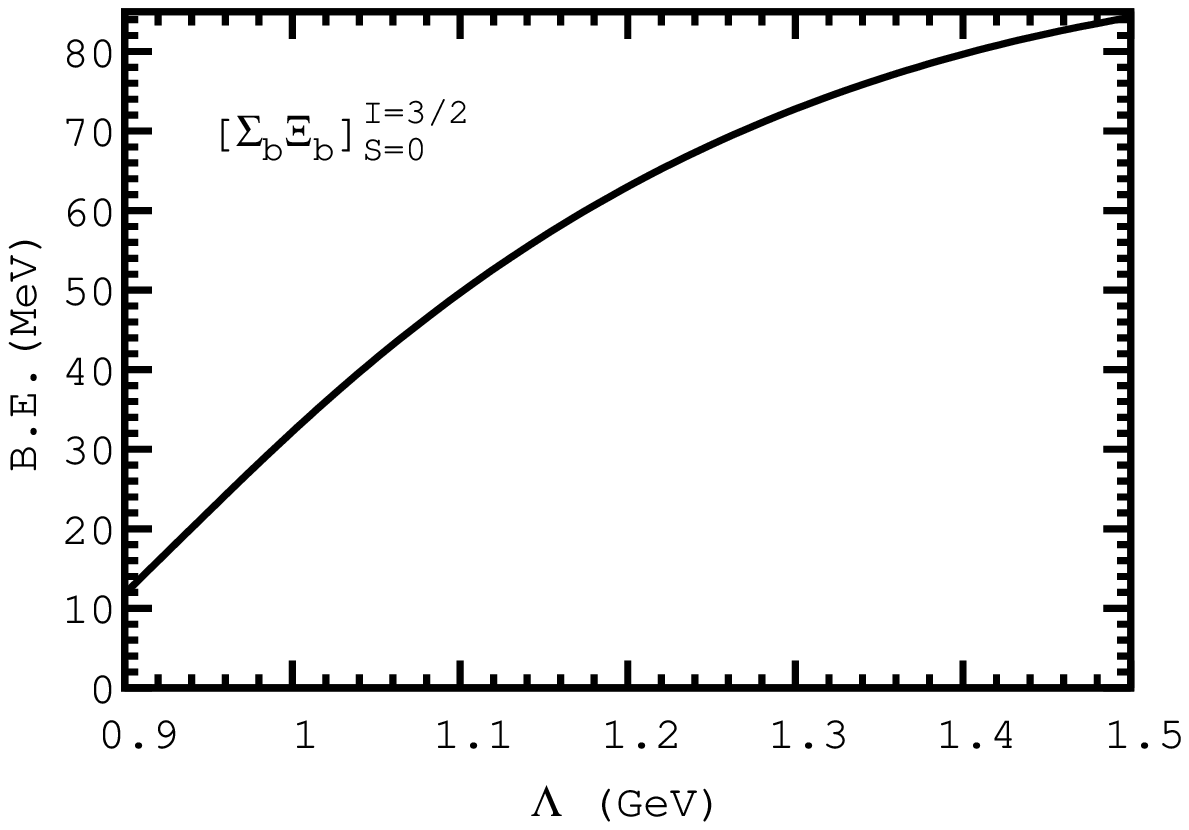}&
\includegraphics[width=0.245\textwidth]{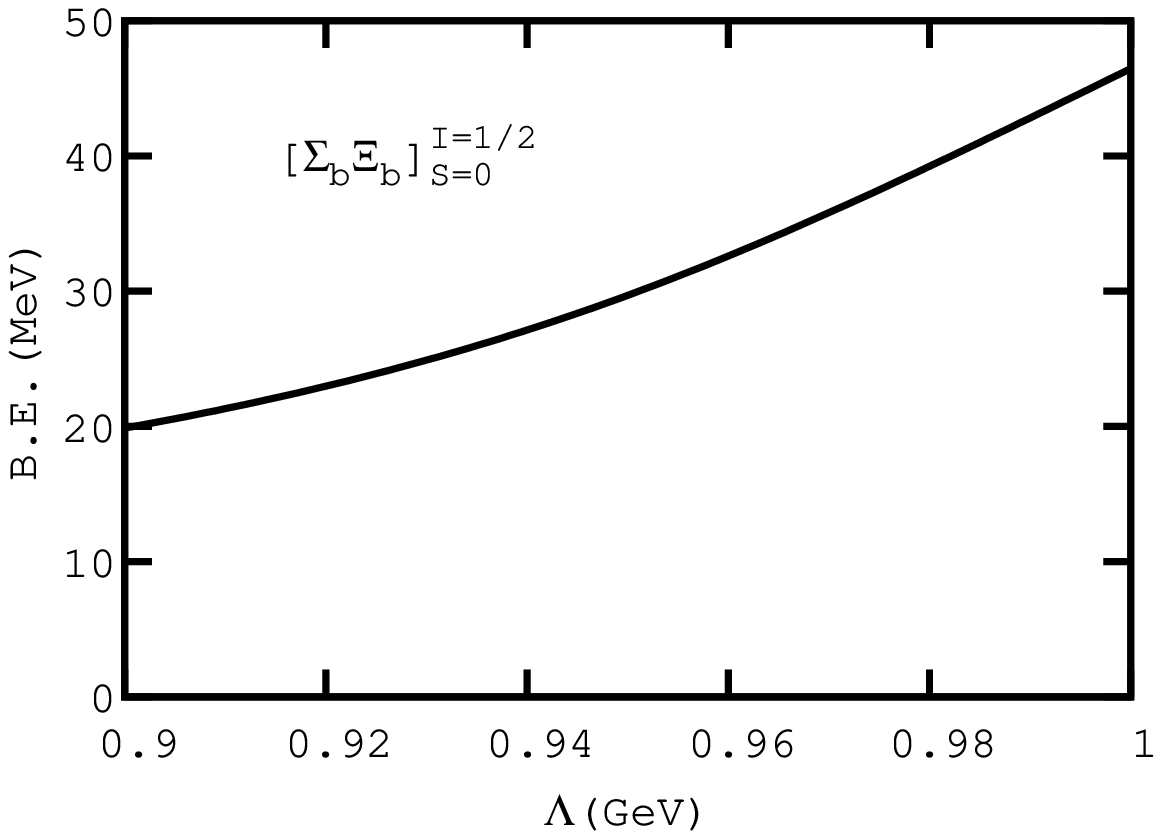}&
\includegraphics[width=0.245\textwidth]{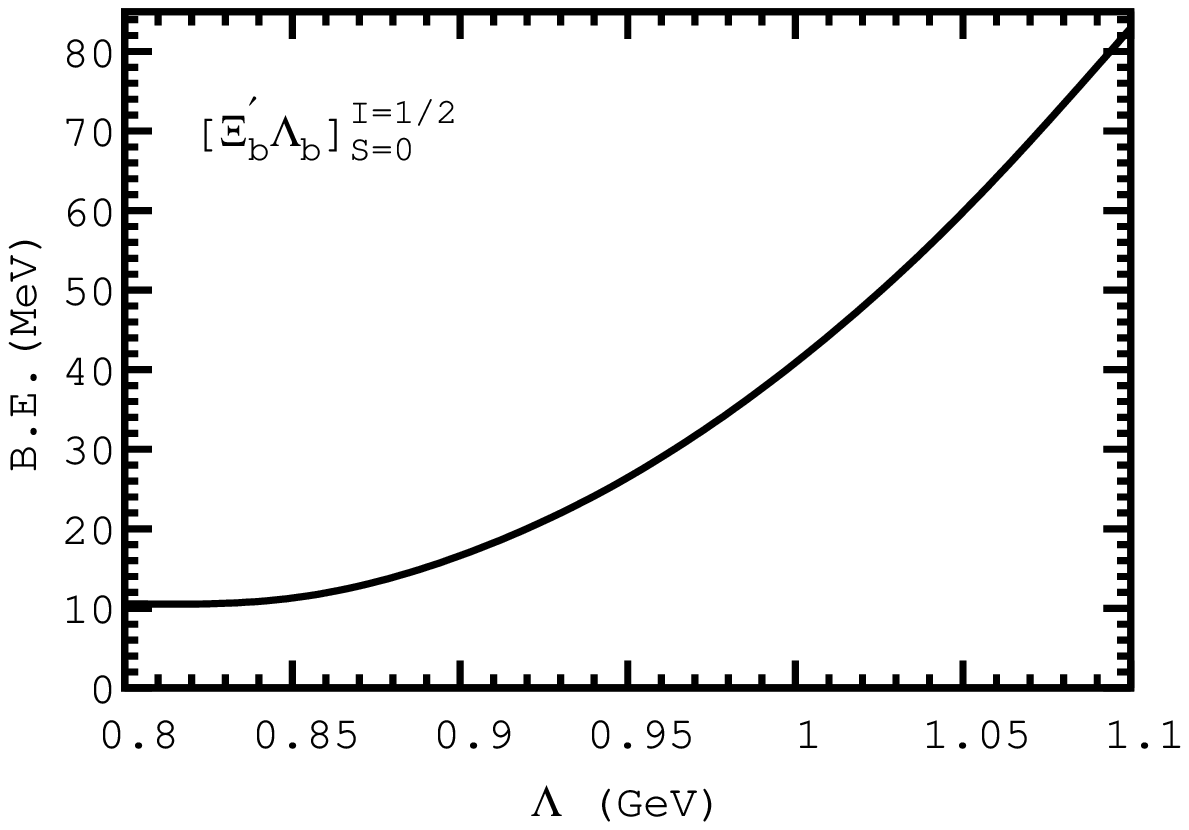}\\
    \includegraphics[width=0.245\textwidth]{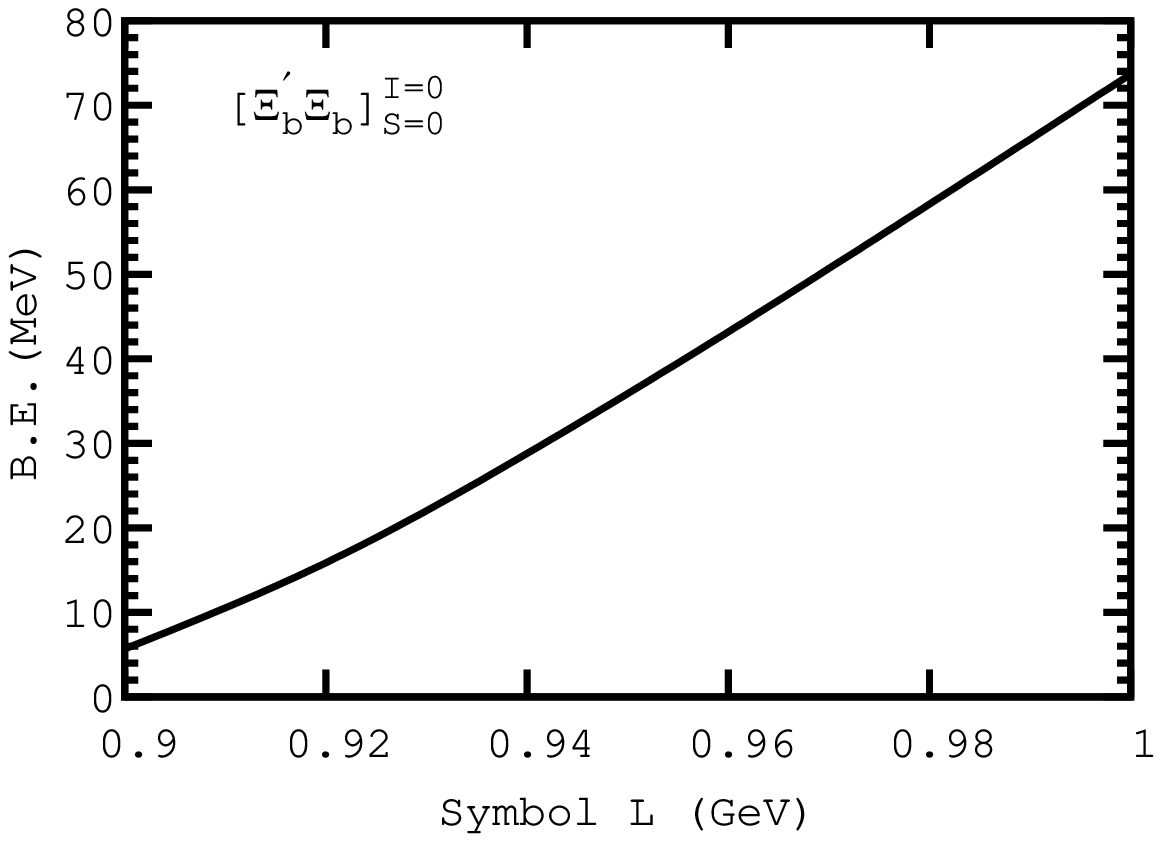}&
\includegraphics[width=0.245\textwidth]{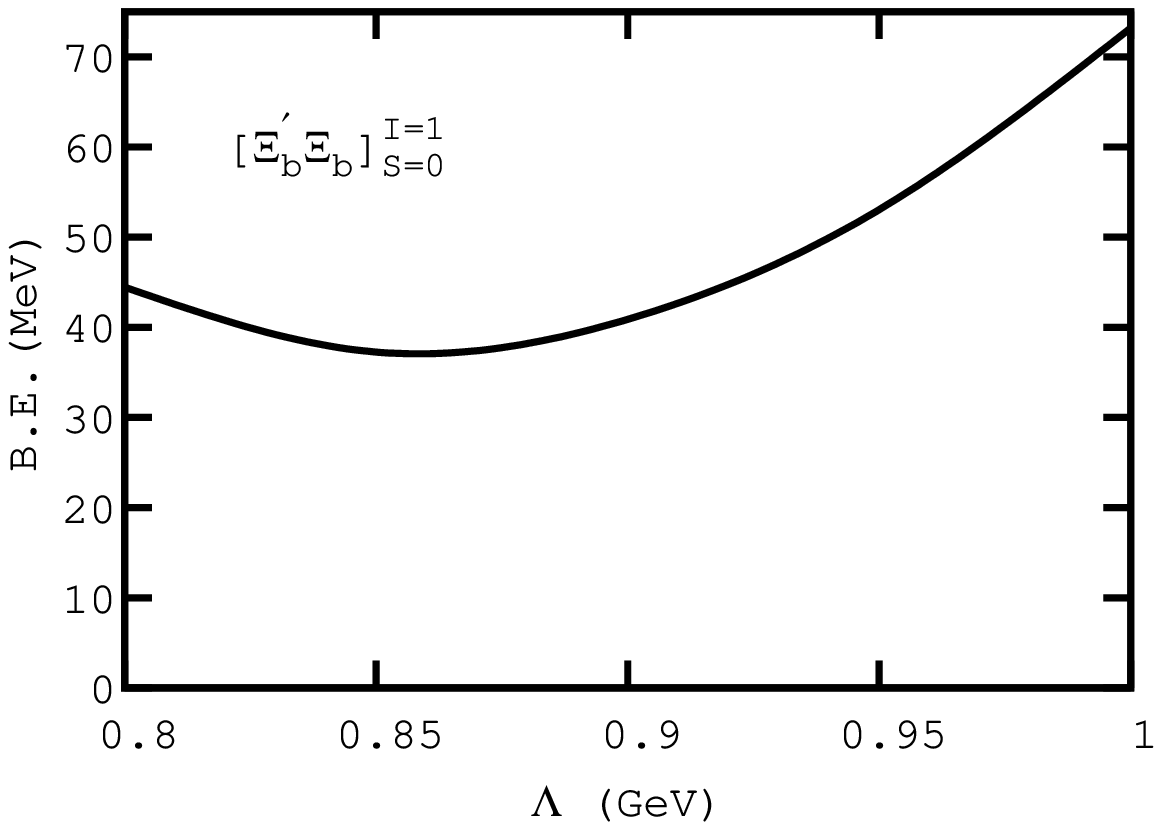}&
\includegraphics[width=0.245\textwidth]{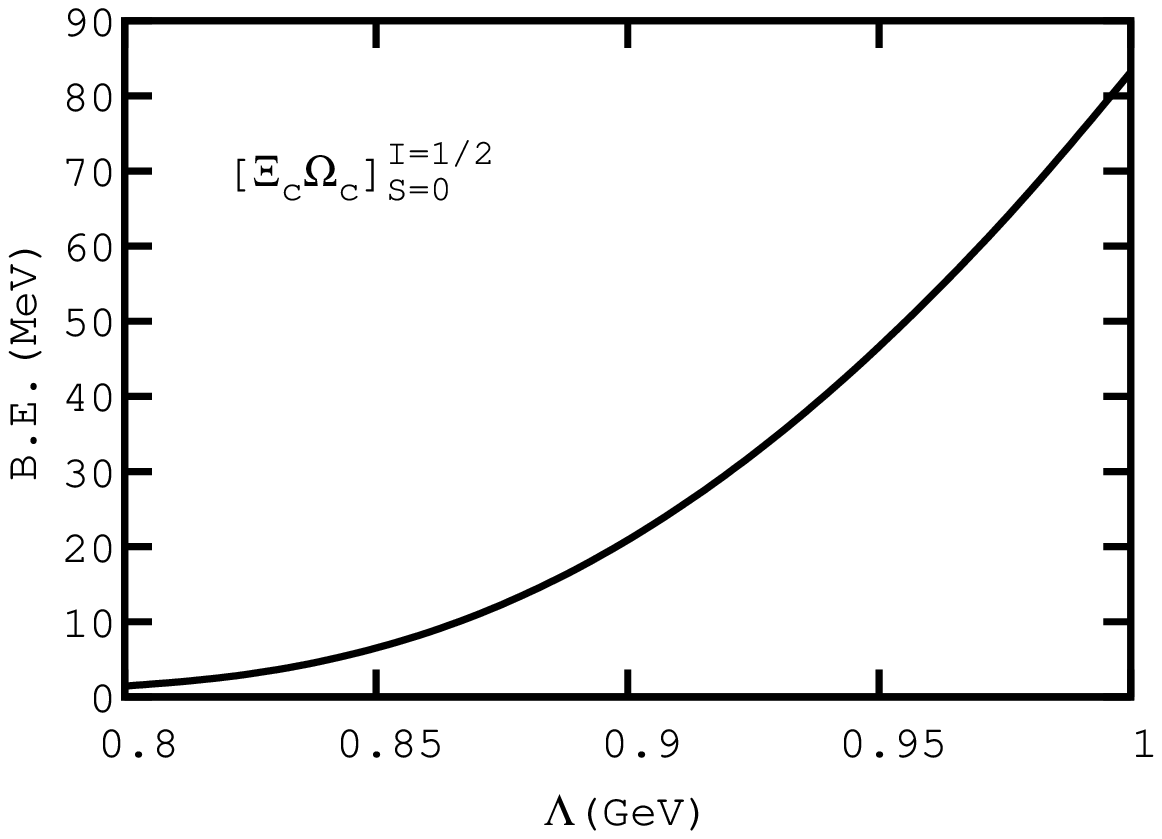}&
\includegraphics[width=0.245\textwidth]{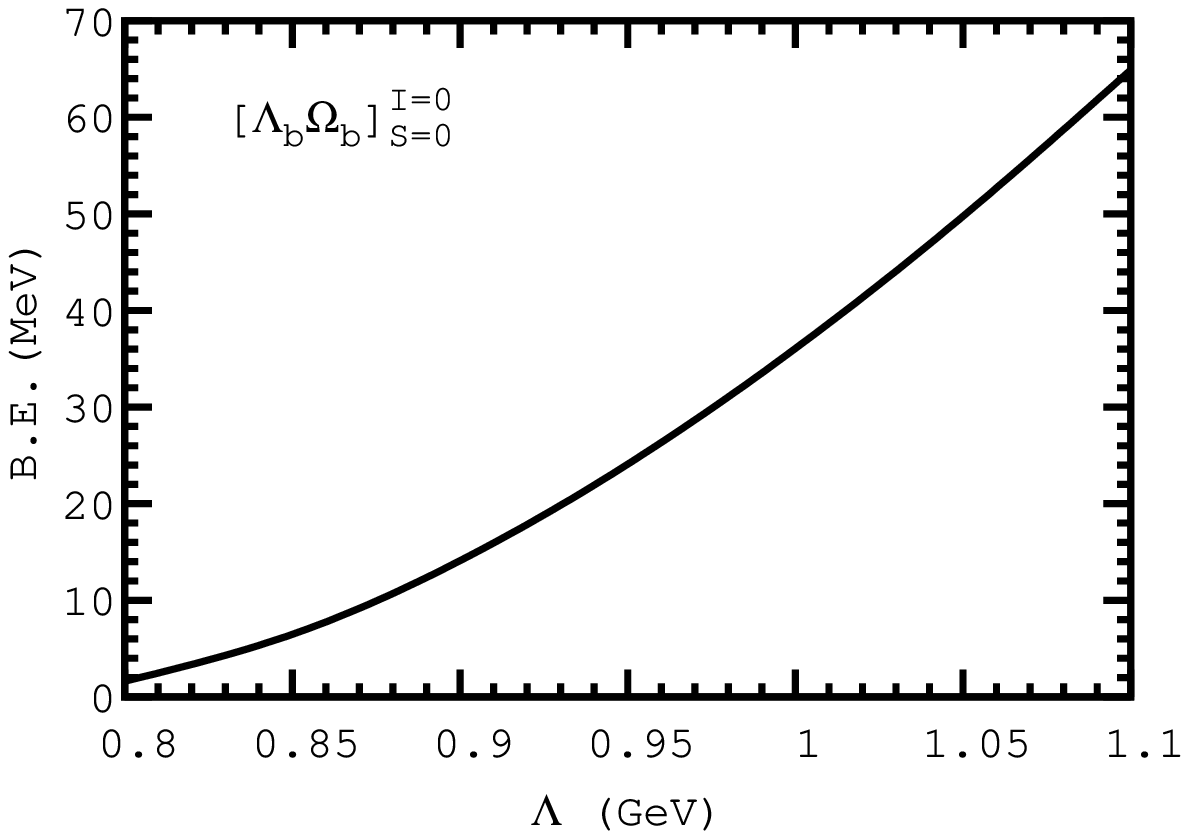}
  \end{tabular}
    \caption{The dependence of the binding energy on the cutoff parameter
    for the spin-singlet ``$A_bB_b$" system with OBE potential.}\label{plotBS}
\end{figure}

\begin{figure}
  \begin{tabular}{cccc}
  \includegraphics[width=0.245\textwidth]{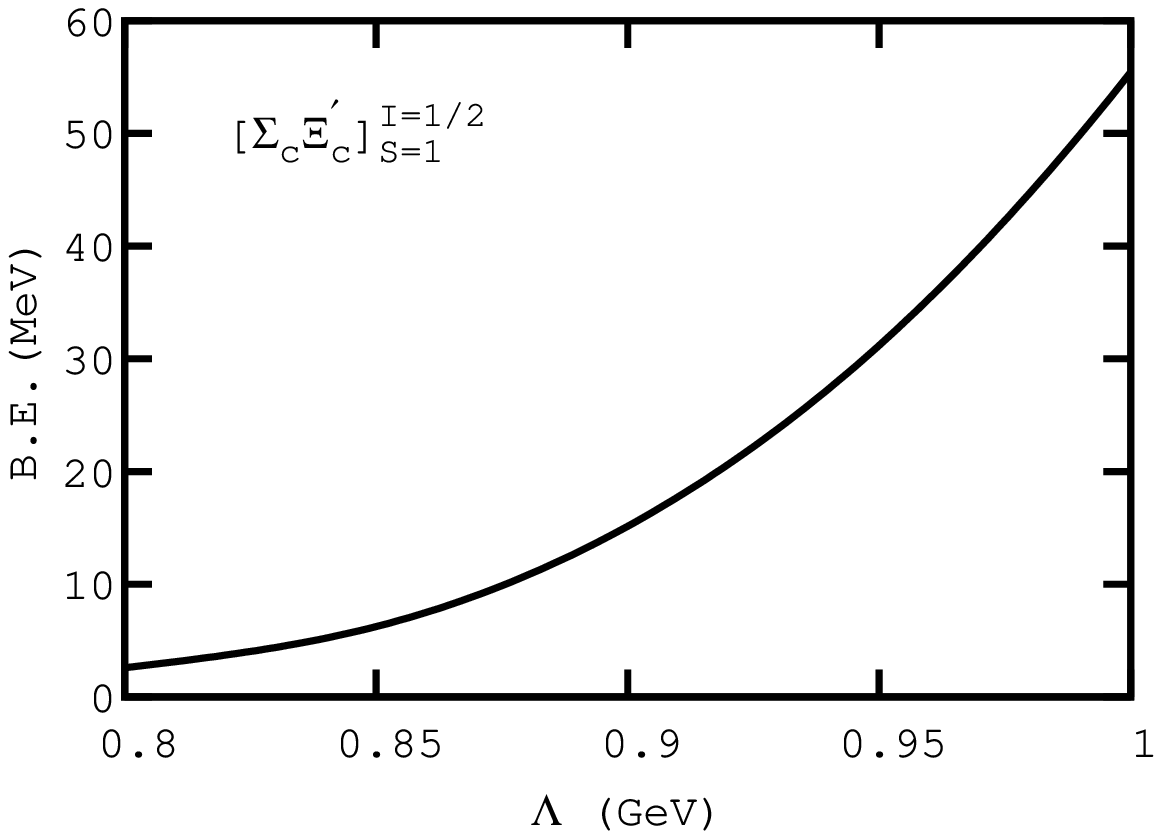}&
\includegraphics[width=0.245\textwidth]{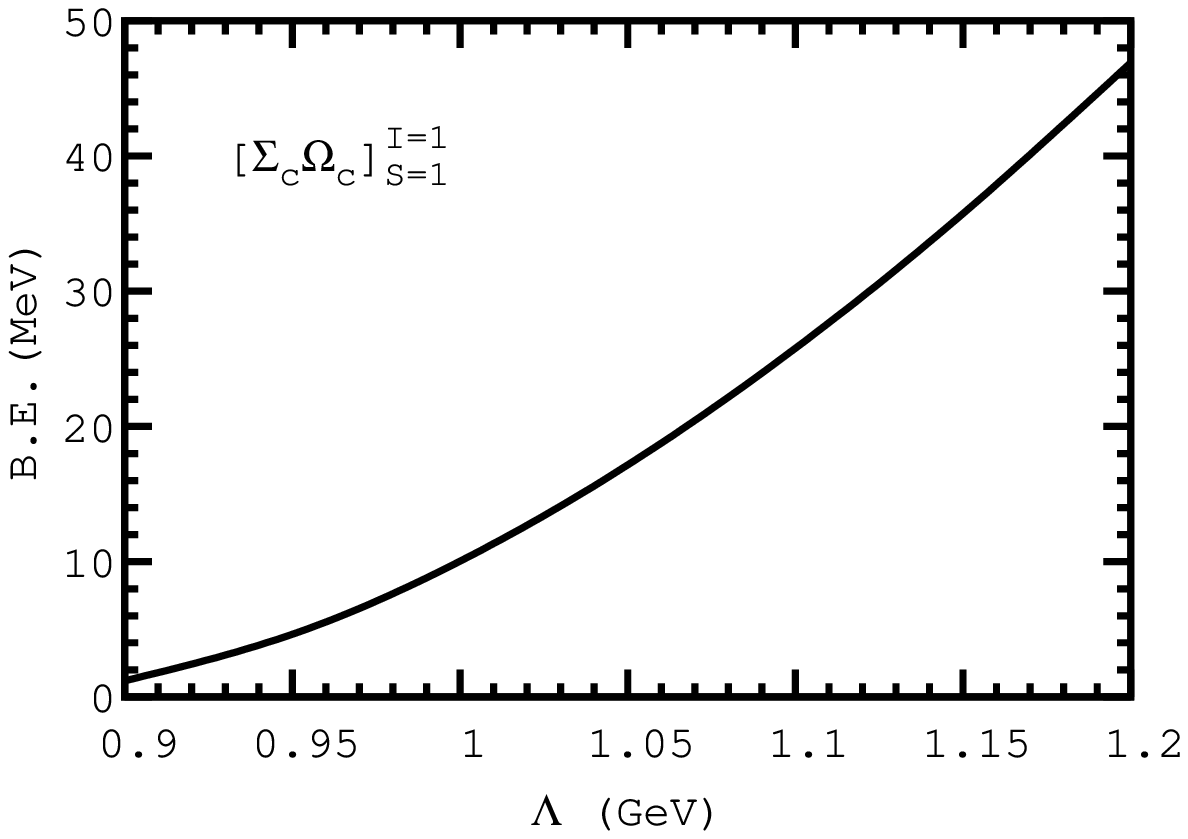}&
\includegraphics[width=0.245\textwidth]{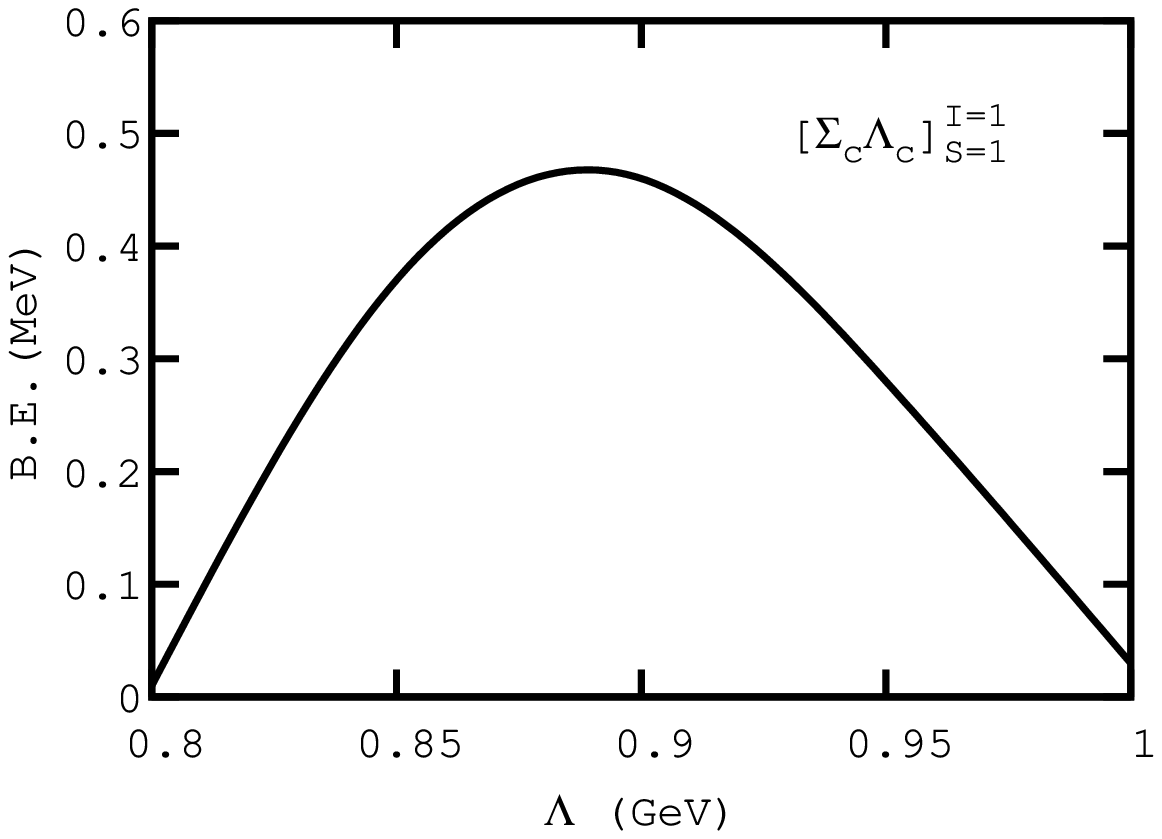}&
\includegraphics[width=0.245\textwidth]{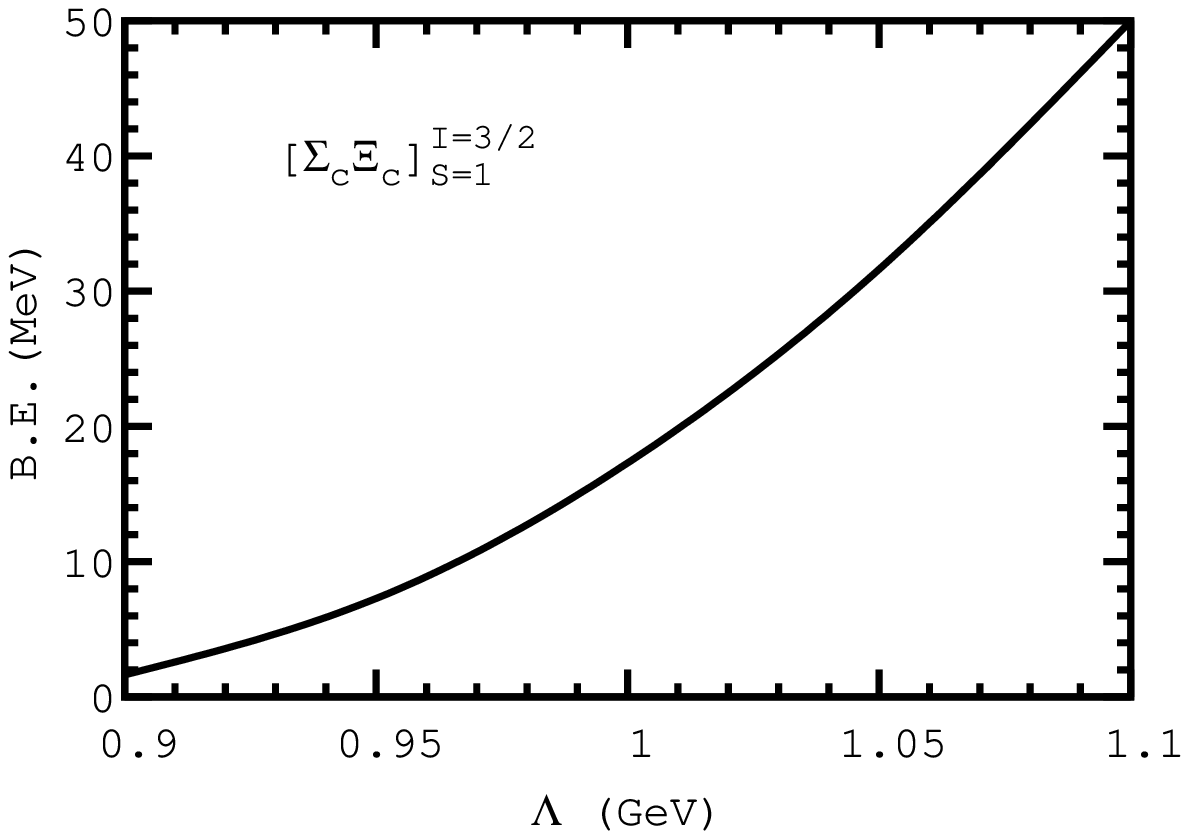}\\
    \includegraphics[width=0.245\textwidth]{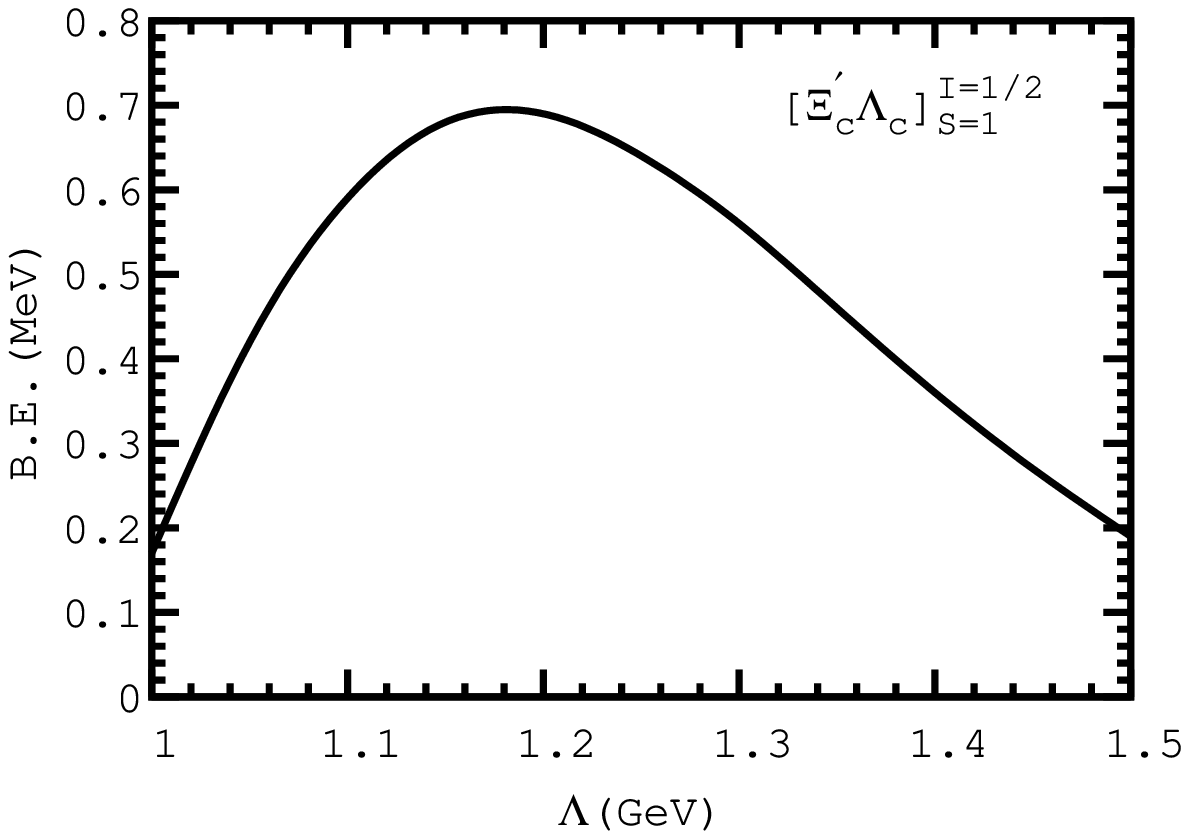}&
\includegraphics[width=0.245\textwidth]{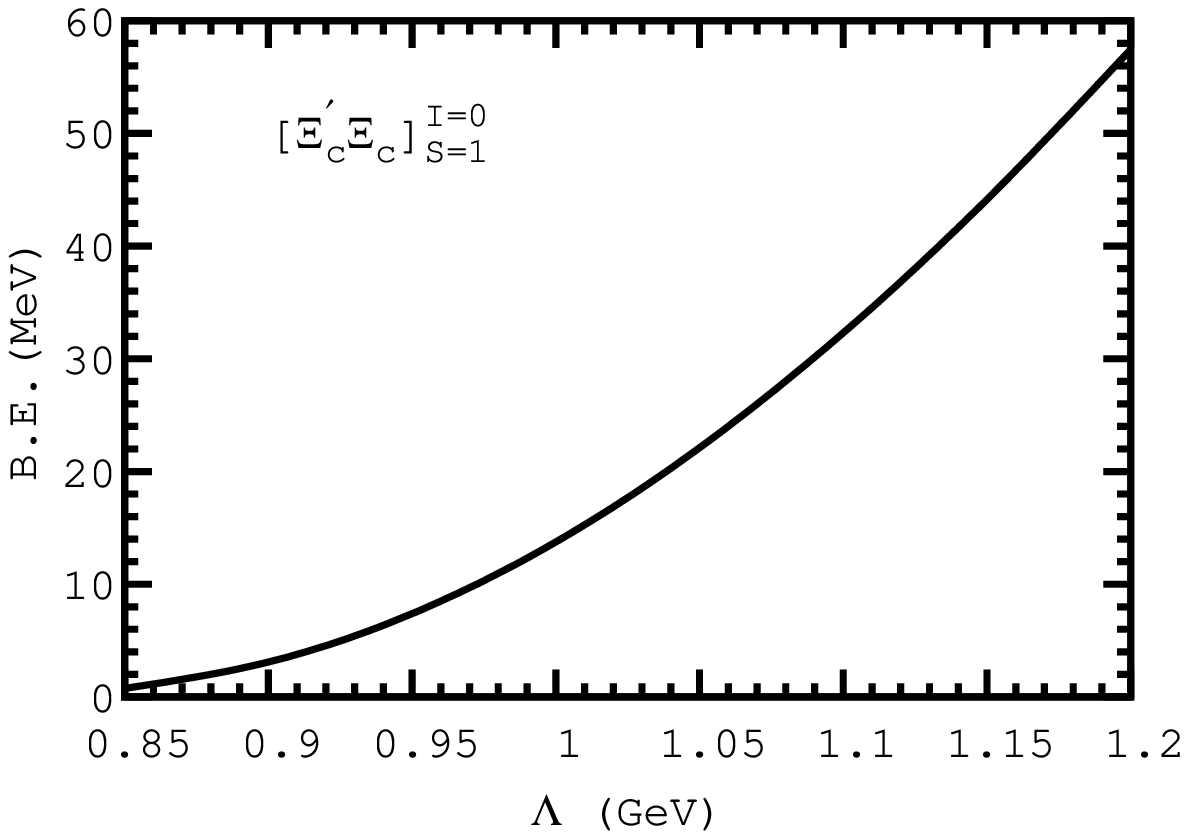}&
\includegraphics[width=0.245\textwidth]{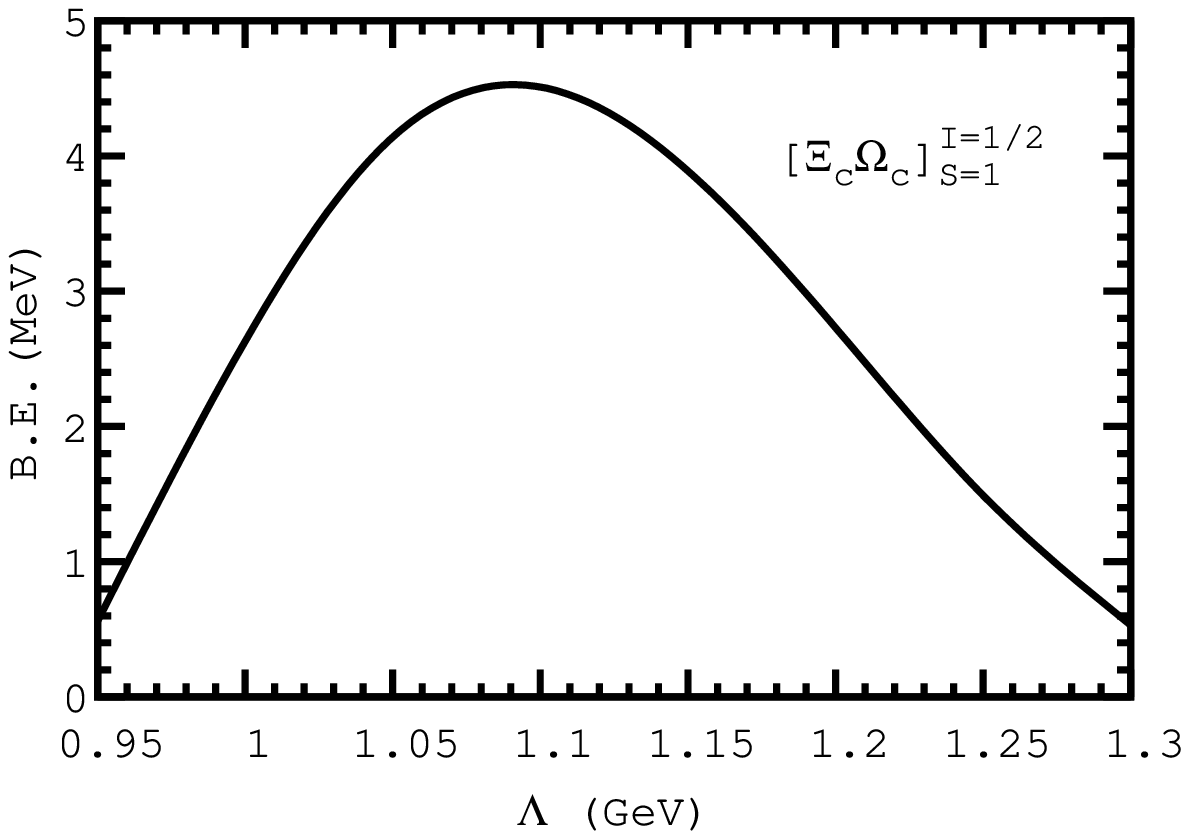}&
\includegraphics[width=0.245\textwidth]{CT9.eps}\\
  \end{tabular}
  \caption{The dependence of the binding energy on the cutoff parameter
  for the spin-triplet ``$A_cB_c$" system with the OBE potential.}\label{plotCT}
\end{figure}

\begin{figure}
\centering
  \begin{tabular}{cccc}
    \includegraphics[width=0.245\textwidth]{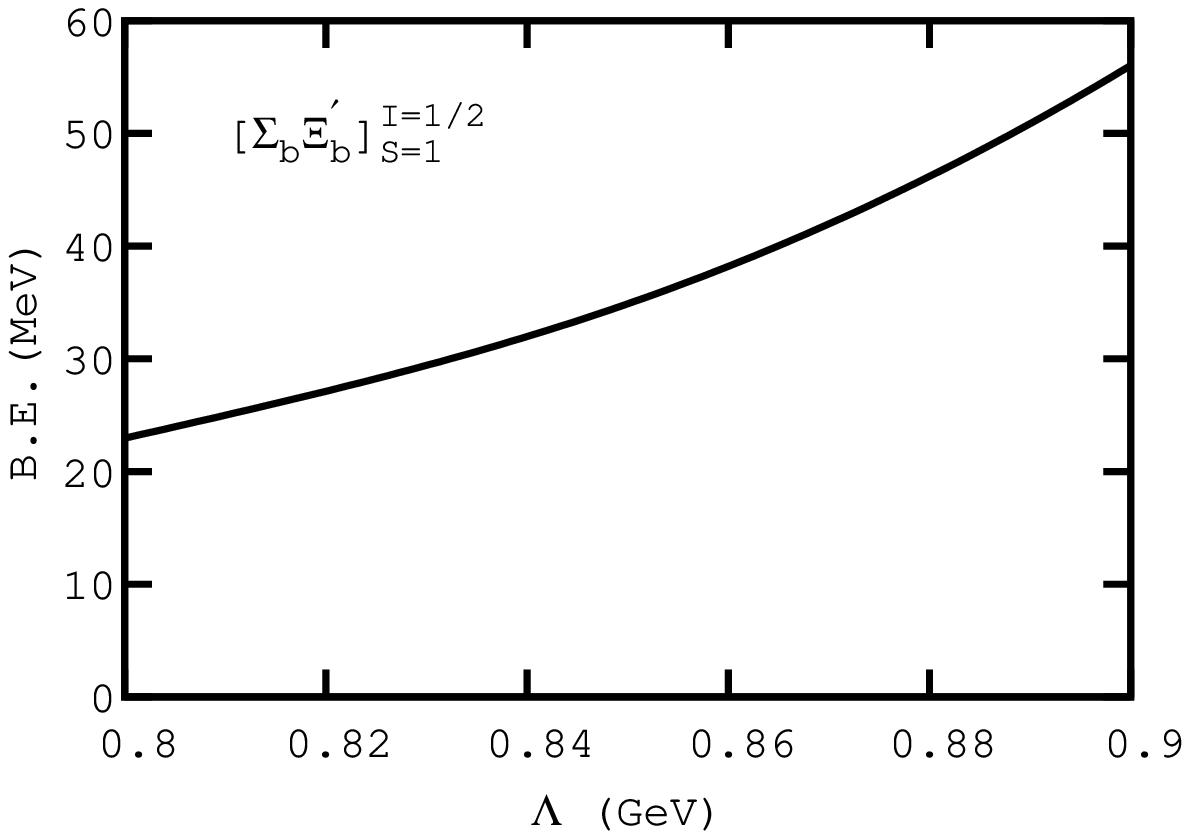}&
\includegraphics[width=0.245\textwidth]{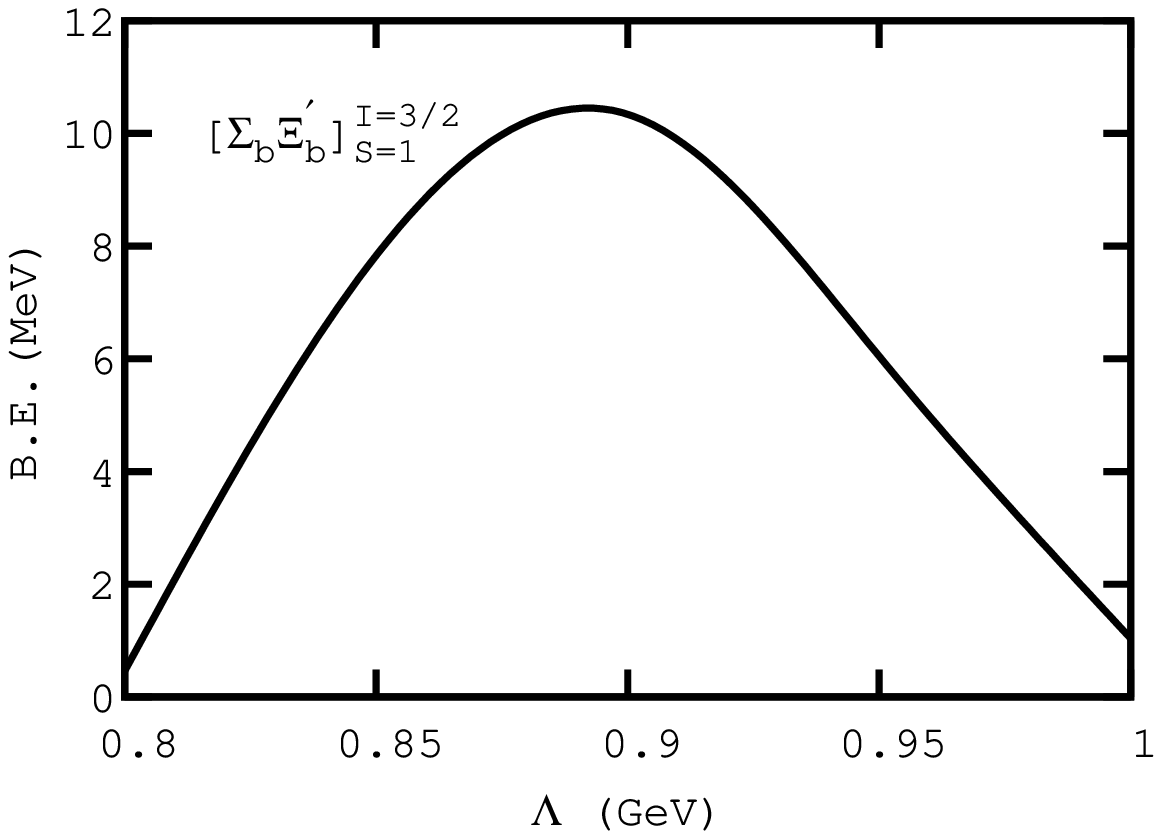}&
\includegraphics[width=0.245\textwidth]{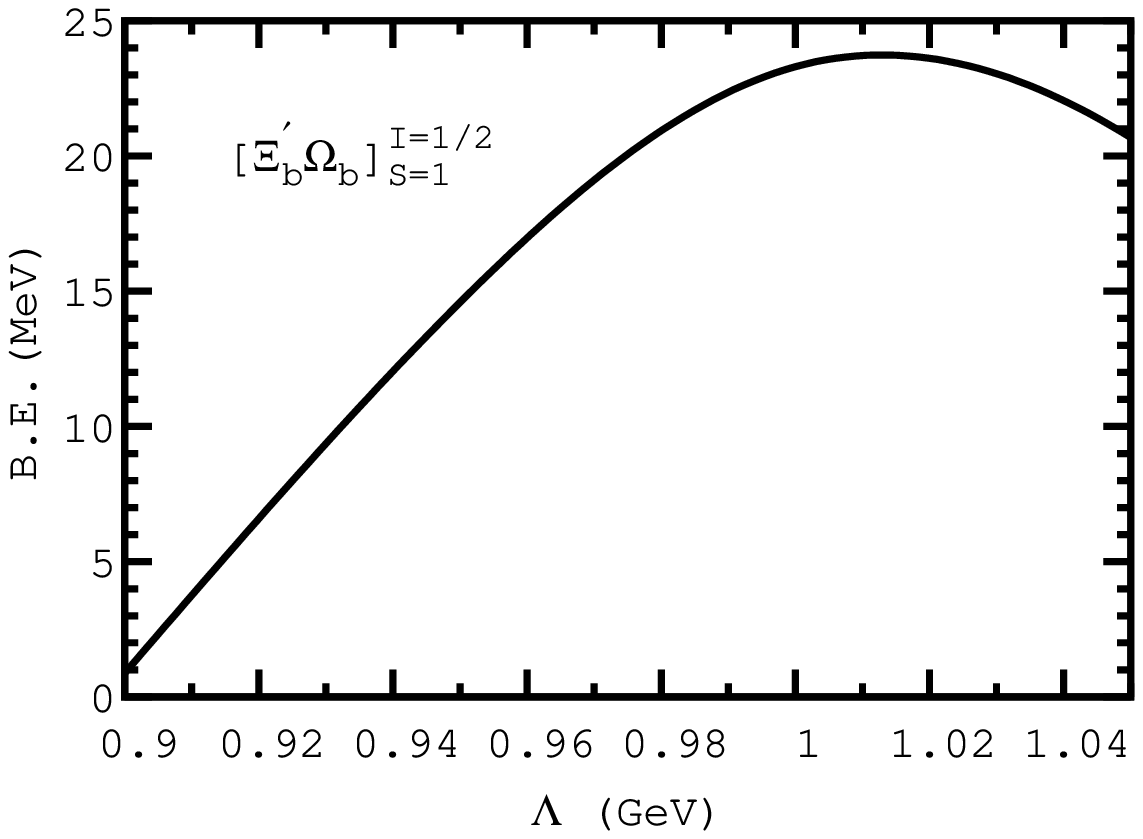}&
\includegraphics[width=0.245\textwidth]{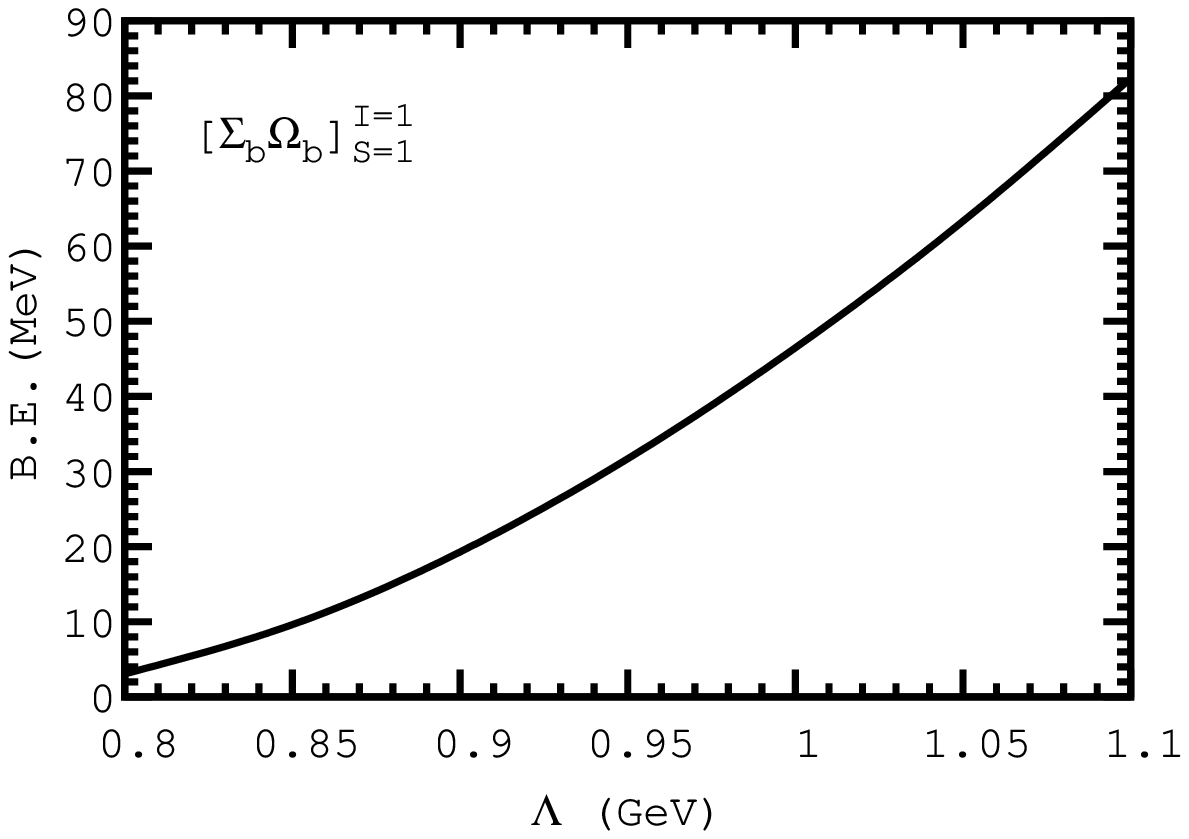}\\
    \includegraphics[width=0.245\textwidth]{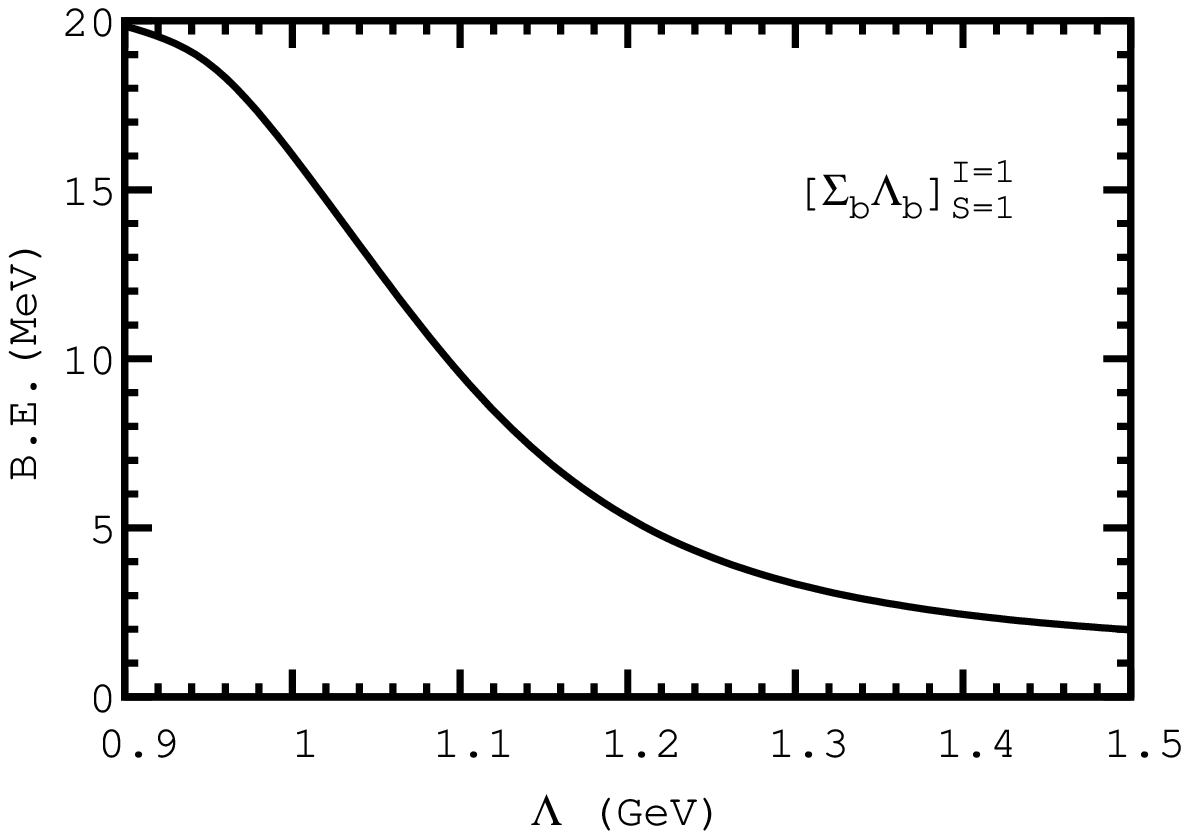}&
\includegraphics[width=0.245\textwidth]{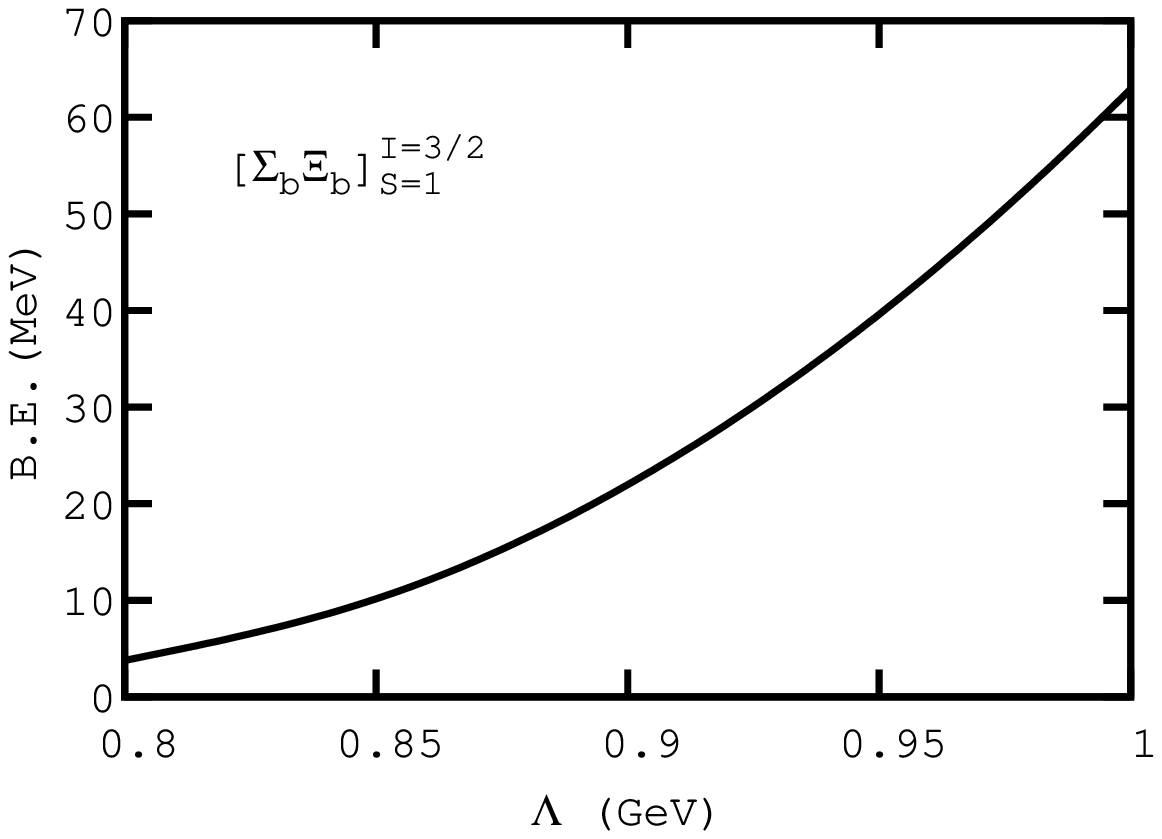}&
\includegraphics[width=0.245\textwidth]{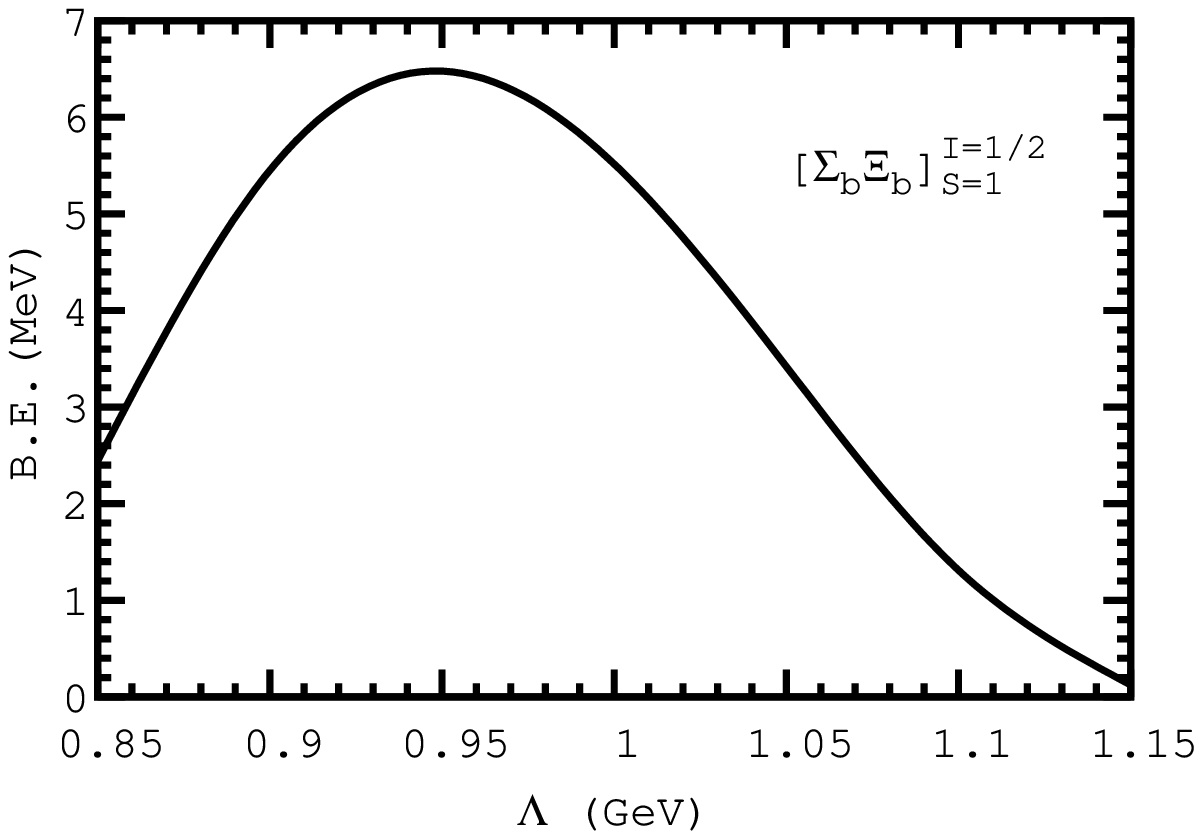}&
\includegraphics[width=0.245\textwidth]{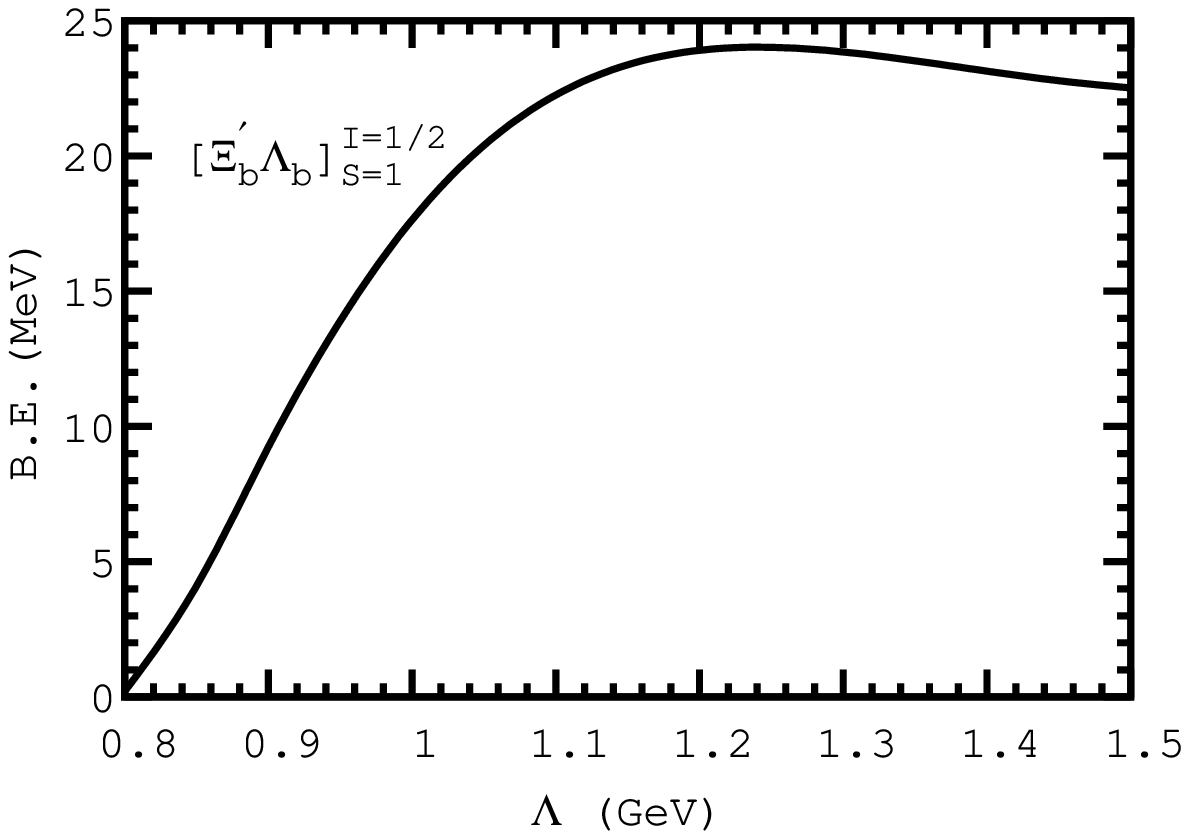}\\
    \includegraphics[width=0.245\textwidth]{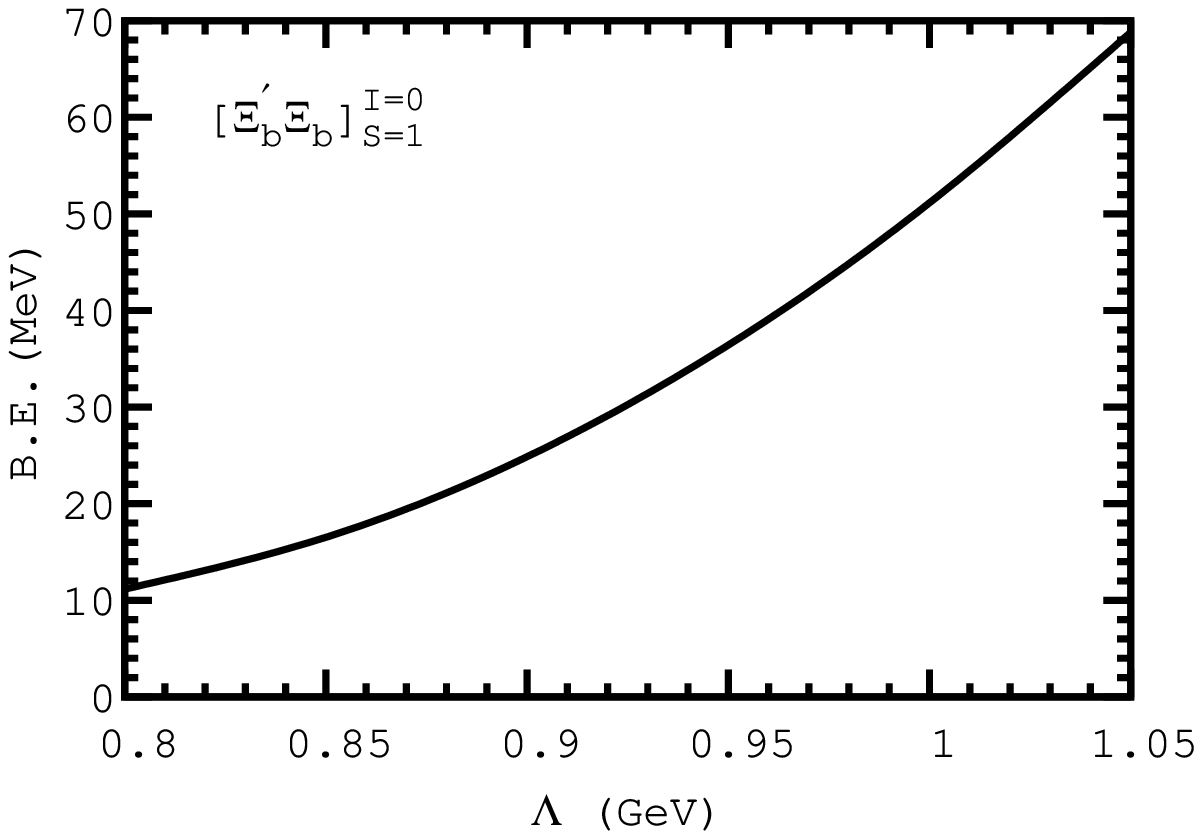}&
\includegraphics[width=0.245\textwidth]{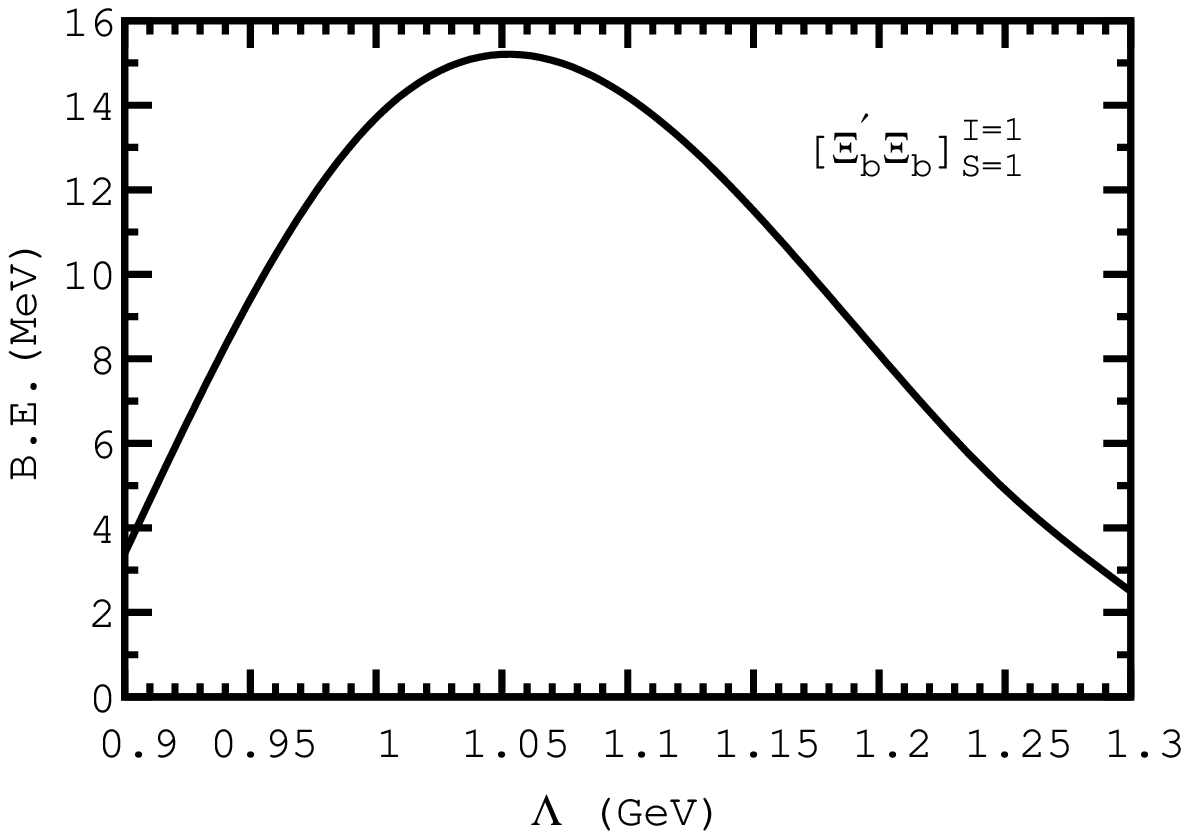}&
\includegraphics[width=0.245\textwidth]{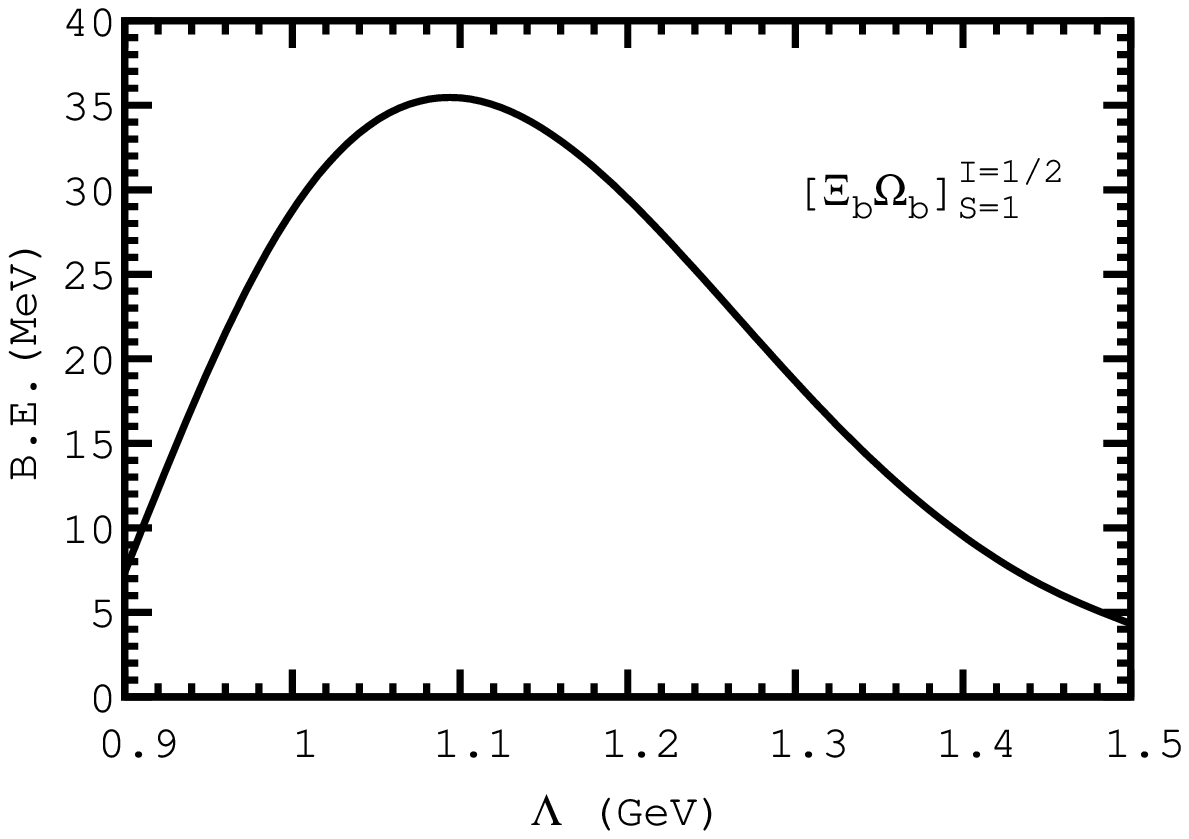}&
  \end{tabular}
    \caption{The dependence of the binding energy on the cutoff parameter
    for the spin-triplet ``$A_bB_b$" system with the OBE potential.}\label{plotBT}
\end{figure}

In the previous subsection, we give the numerical results with the
$\pi$ exchange potential which accounts for the long-range
interaction. Actually, only five systems out of the thirteen ones
allow the $\pi$ exchange. We find that the $\pi$ exchange is not
strong enough to form bound states for all the five spin-singlet
systems. In order to make the individual role of the exchanged
boson clear, we also give the numerical results within the
one-boson-exchange model (OBE) in Tables~\ref{SSOB}-\ref{STrOB},
and plot the dependence of binding energy on the cutoff parameter
in Figs.~(\ref{plotCS}-\ref{plotBT}).

In the spin-singlet case, we still find no binding solutions for
the state $[\S_Q\Xi_Q^{'}]^{I=1/2}_{S=0}$(Q=b,c) even if we add
the contributions of the heavier vector and scalar meson exchange.
Therefore, our results disfavor the existence of the molecules
$[\S_Q\Xi_Q^{'}]^{I=1/2}_{S=0}$ (Q=b,c). However, for the other
isospin multiplet of this state with $I=\f{3}{2}$, we find binding
solutions for both the charmed and bottomed cases. A bound state
of $[\S_c\Xi_c^{'}]^{I=3/2}_{S=1}$ appears with binding energy
about $3.54\sim67.46$ MeV when the cutoff parameter is around
$0.8\sim1.0$ GeV. The binding energy of the corresponding bottomed
state $[\S_b\Xi_b^{'}]^{I=3/2}_{S=0}$ is about 156.78 MeV when the
cutoff parameter is 1.0 GeV. Such a large binding energy seems too
deep for a loosely bound molecular state. For the
$[\S_Q\L_Q]^{I=1}_{S=0}$ system, the binding energy of the state
$[\S_c\L_c]^{I=1}_{S=0}$ is $0.28\sim47.34$ MeV with the cutoff
parameter $0.90\sim1.10$ MeV while the binding energy of
$[\S_b\L_b]^{I=1}_{S=0}$ is $0.34\sim62.13$ MeV with a cutoff
parameter $0.80\sim1.00$ GeV. We obtain bound states for both the
charmed and the bottomed cases for $[\Xi_Q\Xi_Q^{'}]$ with $I=0$
and $I=1$.

For the other systems without the $\pi$ exchange, we also find
binding solutions. The most interesting one may be
$[\Xi_Q\L_Q]^{I=1/2}_{S=0}$, which allows the $\sigma$ and
$\omega$ exchanges in the direct channel and $K^*/\bar{K}^*$
exchange in the cross channel. For the charmed case, a very
loosely bound states with binding energy $1.91\sim3.03$ MeV
appears when the cutoff parameter is $1.10\sim1.50$ GeV. For the
bottomed case, a bound state emerges with binding energy
$10.33\sim28.65$ MeV when the cutoff parameter is between 0.90 GeV
and 1.50 GeV. They are very good molecule candidates. For the
states $[\S_c\Xi_c]^{I=1/2}_{S=0}$ and $[\S_c\O_c]^{I=1}_{S=0}$,
we also obtain small binding energies and large root-mean-square
radii with reasonable cutoff parameter $1.0\sim1.50$ GeV as shown
in Table~\ref{SSOB}. Our results are in favor of the existences of
these molecular states. The binding energy of
$[\Xi_b^{'}\O_b]^{I=1/2}_{S=0}$ is $80.49\sim107.01$ MeV with
cutoff parameter $0.90\sim1.00$ GeV. Again, such a large binding
energy seems too deep for a loosely bound molecular state.

\begin{figure}
\centering
\begin{tabular}{ccc}
\includegraphics[width=0.33\textwidth]{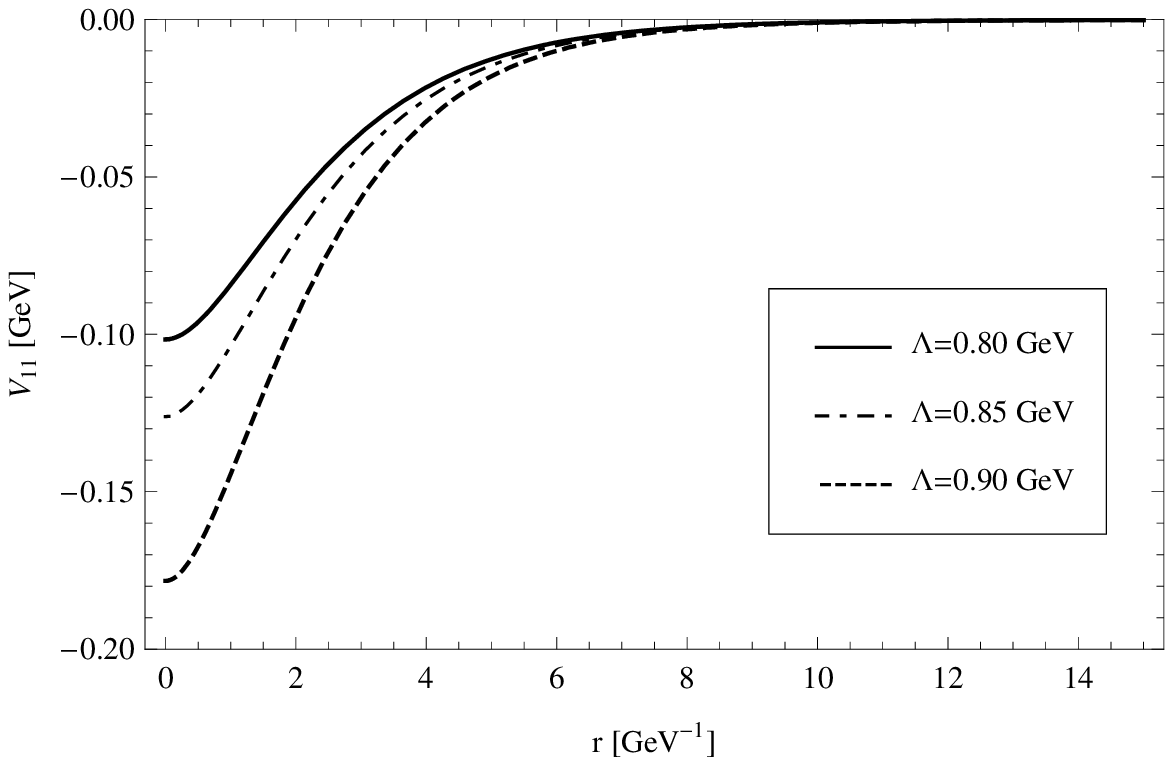}&
\includegraphics[width=0.33\textwidth]{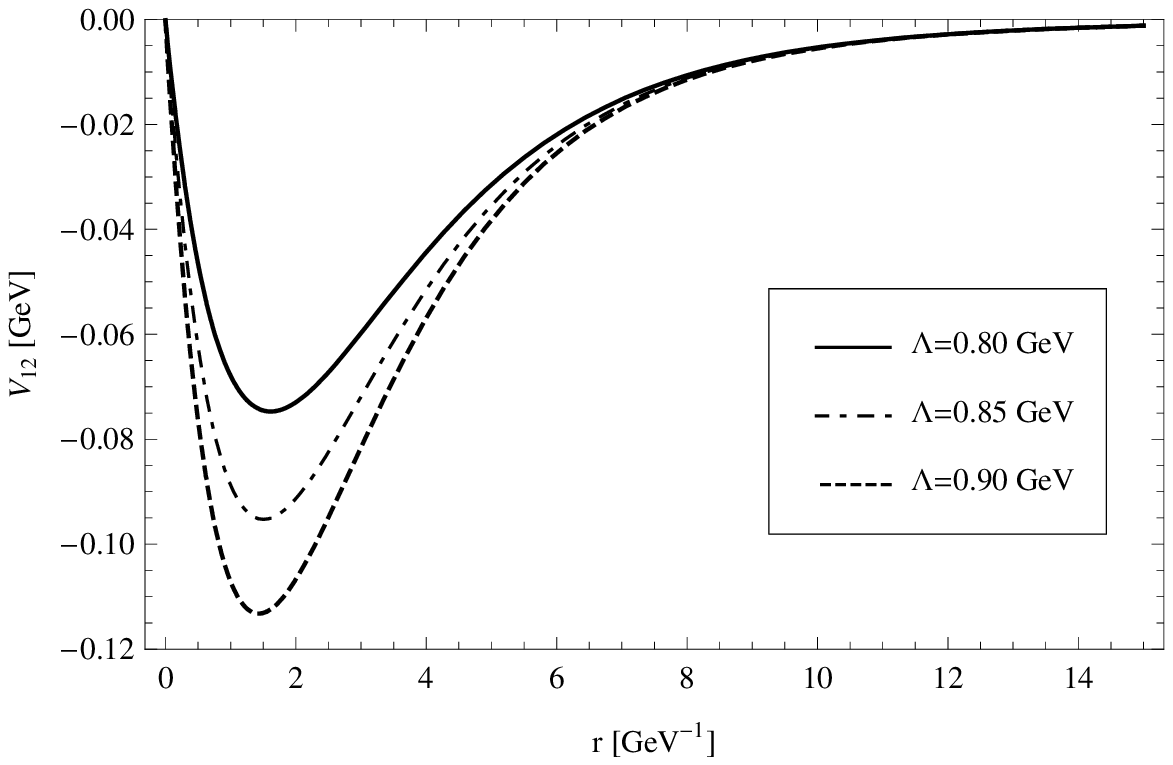}&
\includegraphics[width=0.33\textwidth]{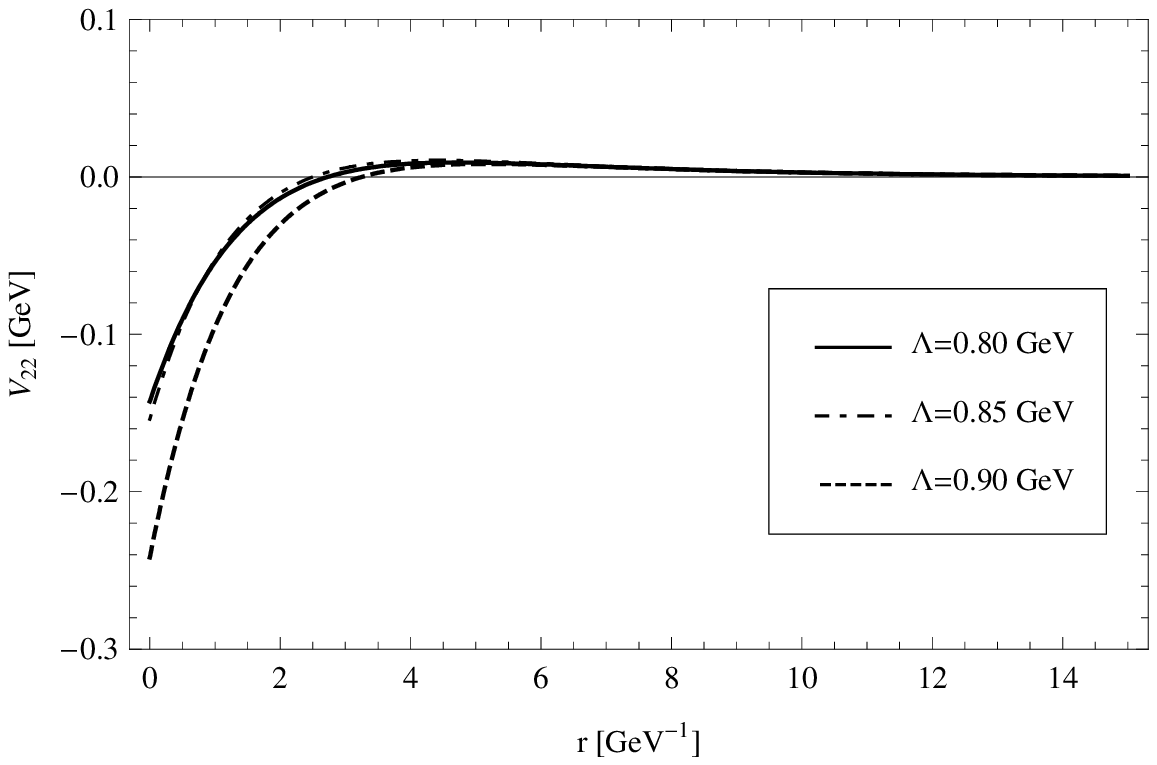}
\end{tabular}
\caption{The potential of system
$\S_c\Xi_c^{'}\left[I(J^P)=\f{1}{2}(1^+)\right]$ within the OBE model. $V_{11}$, $V_{12}$ and
$V_{22}$ are for the transitions $^3S_1$$\leftrightarrow$$^3S_1$,
$^3S_1$$\leftrightarrow$$^3D_1$ and $^3D_1$$\leftrightarrow$$^3D_1$,
respectively.}\label{one:potential}
\end{figure}

In the spin-triplet sector, it is interesting to compare with the deuteron
case. We plot the interaction potential of the
$\S_c\Xi^{'}_c\left[I(J^P)=\f{1}{2}(1^+)\right]$ system in
Fig.~\ref{one:potential}. From Fig.~\ref{deu:potential} and
Fig.~\ref{one:potential}, it is clear that the potentials of the two systems
are similar. Their binding solutions are also similar except that the
$\S_c\Xi_c^{'}$ system has even shallower binding energy and smaller D wave
probability, as can be seen from Tables~\ref{deu:numerical} and ~\ref{STrOB}.

The bound state of $[\S_c\Xi_c^{'}]^{I=3/2}_{S=1}$ disappears if
we take the heavier scalar and vector meson exchanges into
account. There is still no binding solution for the state
$[\Xi_c\Xi^{'}_c]^{I=1}_{S=1}$ when we consider all the
contributions of the exchanged mesons. The binding energy of the
state $[\S_Q\L_Q]^{I=1}_{S=1}$ becomes shallower in the OBE model.
For the other systems with the $\pi$ exchange, the numerical
results within the OBE model are similar to those within the OPE
model except that the binding energy becomes deeper as shown in
Table \ref{STrOB}.

From Table \ref{STrOB}, one can see that there is no S-D mixing
for the two states $[\Xi_Q\L_Q]^{I=1/2}_{S=1}$ and
$[\L_Q\O_Q]^{I=0}_{S=1}$. Actually, for these two systems the
results are the same for both the spin-singlet and spin-triplet
cases because the potential is the same. For the states
$[\Xi_c^{'}\O_c]^{I=1/2}_{S=1}$ and $[\S_c\Xi_c]^{I=3/2}_{S=1}$,
there are no binding solutions. However, a very loosely bound
state $[\Xi_c^{'}\L_c]^{I=1/2}_{S=1}$ exists with binding energy
$0.17\sim0.69$ MeV when the cutoff parameter is around
$1.00\sim1.40$ GeV. The binding energy of its bottomed counterpart
$[\Xi_b^{'}\L_b]^{I=1/2}_{S=1}$ is $17.64\sim23.91$ MeV with cutoff
parameter around $1.0\sim1.40$ GeV. We also obtain a loosely bound
state $[\Xi_c\O_c]^{I=1/2}_{S=1}$ with binding energy
$2.63\sim4.51$ MeV when the cutoff parameter is between 1.00 GeV
and 1.20 GeV. Once these three molecule states are produced, they
should be very stable because their constituents $\L_Q$,
$\Xi_Q^{'}$, $\Xi_Q$ and $\O_Q$ decay via weak interaction.
Comparing the results of the OBE model with those of the OPE
model, one can see that the contribution of the D wave decreases
if we take into account the scalar and vector meson exchange,
which implies that the S-D mixing mainly comes from the $\pi$
exchange.

\begin{table}[htp]
\renewcommand{\arraystretch}{1.3}
\centering \caption{The numerical results for the spin-singlet
(S=0) case with the OBE potential. $\times$ means no binding
solutions exist.}\label{SSOB}
\begin{tabular*}{16cm}{@{\extracolsep{\fill}}cccccccc}
\toprule[1.0pt]\addlinespace[5pt]
&\multicolumn{6}{c}{S=0}\\
             &\multicolumn{3}{c}{Q=c}   & \multicolumn{3}{c}{Q=b}\\
Systems      & $\L$(GeV)   &  B.E.(MeV)    & $r_{rms}$(fm)
&  $\L$(GeV) & B.E.(MeV)      & $r_{rms}$(fm)  \\
\midrule[0.7pt]\addlinespace[5pt]
\multirow{3}*{$[\S_Q\Xi_Q^{'}]^{I=1/2}_S$}
& \multicolumn{3}{c}{\multirow{3}*{$\times$}}
& \multicolumn{3}{c}{\multirow{3}*{$\times$}} \\
&       &     \\
&       &     \\
\multirow{3}*{$[\S_Q\Xi_Q^{'}]^{I=3/2}_S$}
&  0.80 & 3.54   & 1.97   &  0.80 & 25.63  & 0.74   \\
&  0.90 & 14.53  & 1.14   &  0.90 & 53.38  & 0.57   \\
&  1.00 & 67.46  & 0.67   &  1.00 & 156.78 & 0.40   \\ [7pt]
\multirow{3}*{$[\Xi_Q^{'}\O_Q]^{I=1/2}_{S}$}
& 0.90  &  28.50 &  0.88   &  0.90      & 80.49  &  0.49   \\
& 0.95  &  29.65 &  0.86   &  0.95      & 80.91  &  0.49   \\
& 1.00  &  44.09 &  0.75   &  1.00      & 107.01 &  0.45   \\ [7pt]
\multirow{3}*{$[\S_Q\O_Q]^{I=1}_S$}
& 1.00  & 0.56   & 3.93    &  0.90      & 5.12   &   1.19   \\
& 1.20  & 13.26  & 1.09    &  1.00      & 18.07  &   0.77   \\
& 1.40  & 36.77  & 0.75    &  1.10      & 36.67  &   0.60   \\ [7pt]
\multirow{3}*{$[\Xi_Q\L_Q]^{I=1/2}_{S}$}
& 1.10  & 1.91   & 2.47    &  0.90      & 10.33  &   0.96   \\
& 1.30  & 3.03   & 2.03    &  1.20      & 28.65  &   0.67   \\
& 1.50  & 2.87   & 2.08    &  1.50      & 27.78  &   0.68   \\ [7pt]
\multirow{3}*{$[\S_Q\L_Q]^{I=1}_S$}
& 0.90  & 0.28   & 5.48    &  0.80      & 0.34   &  5.19    \\
& 1.00  & 14.81  & 1.09    &  0.90      & 16.02  &  0.79    \\
& 1.10  & 47.34  & 0.73    &  1.00      & 62.13  &  0.52    \\ [7pt]
\multirow{3}*{$[\S_Q\Xi_Q]^{I=1/2}_S$}
& 1.00  & 4.13   & 1.77    &  0.90      & 11.85  &  0.90    \\
& 1.30  & 20.99  & 0.95    &  1.00      & 32.21  &  0.64    \\
& 1.50  & 26.92  & 0.86    &  1.10      & 49.64  &  0.55    \\ [7pt]
\multirow{3}*{$[\S_Q\Xi_Q]^{I=3/2}_S$}
& 0.90  &  1.29  & 2.91    &  0.90      & 19.88  &  0.78    \\
& 1.00  &  9.15  & 1.32    &  0.95      & 29.69  &  0.68    \\
& 1.10  &  33.32 & 0.82    &  1.00      & 46.46  &  0.58    \\ [7pt]
\multirow{3}*{$[\Xi_Q^{'}\L_Q]^{I=1/2}_S$}
& 0.90  &  0.58  & 3.99    &  0.90      & 16.62  &  0.83    \\
& 1.00  &  7.08  & 1.47    &  0.95      & 26.45  &  0.71    \\
& 1.10  &  24.23 & 0.93    &  1.00      & 40.88  &  0.61    \\ [7pt]
\multirow{3}*{$[\Xi_Q^{'}\Xi_Q]^{I=0}_S$}
& 0.95  &  6.67  & 1.48    &  0.90      & 5.67   &  1.19    \\
& 1.00  &  23.80 & 0.92    &  0.94      & 28.80  &  0.67    \\
& 1.05  &  44.48 & 0.74    &  1.00      & 73.68  &  0.49    \\ [7pt]
\multirow{3}*{$[\Xi_Q^{'}\Xi_Q]^{I=1}_S$}
& 0.90  &  8.18  & 1.38    &  0.90      & 40.88  &  0.61    \\
& 1.00  &  22.78 & 0.95    &  0.95      & 53.05  &  0.56    \\
& 1.10  &  56.04 & 0.69    &  1.00      & 73.19  &  0.50    \\ [7pt]
\multirow{3}*{$[\Xi_Q\O_Q]^{I=1/2}_S$}
& 0.90  &  2.19  & 2.31    &  0.80      & 1.44   &  2.06    \\
& 0.95  &  12.13 & 1.19    &  0.90      & 20.88  &  0.78    \\
& 1.00  &  30.45 & 0.86    &  0.94      & 40.58  &  0.62    \\ [7pt]
\multirow{3}*{$[\L_Q\O_Q]^{I=0}_S$}
& 1.00  &  5.40  & 1.60    &  0.90      & 14.10  &  0.86    \\
& 1.10  & 16.55  & 1.04    &  1.00      & 36.07  &  0.62    \\
& 1.20  & 32.02  & 0.82    &  1.10      & 64.89  &  0.51    \\ [7pt]
\bottomrule[1.0pt]
\end{tabular*}
\end{table}

\begin{table}
\renewcommand{\arraystretch}{1.3}
\centering \caption{The numerical results for the spin-triplet
(S=1) case with the OBE potential. $\times$ indicates no binding
solutions exist.}\label{STrOB}
\begin{tabular*}{18cm}{@{\extracolsep{\fill}}cccccccccccc}
\toprule[1.0pt]\addlinespace[5pt]
&\multicolumn{10}{c}{S=1}\\
             &\multicolumn{5}{c}{Q=c}   & \multicolumn{5}{c}{Q=b}\\
Systems      & $\L$(GeV)   &  B.E.(MeV)    & $r_{rms}$(fm)&$P_S(\%)$  & $P_D(\%)$
&  $\L$(GeV) & B.E.(MeV)   & $r_{rms}$(fm) & $P_S(\%)$    &$P_D(\%)$   \\
\midrule[0.7pt]\addlinespace[5pt]
\multirow{3}*{$[\S_Q\Xi_Q^{'}]^{I=1/2}_S$}
& 0.80  & 2.59   & 2.28    &  96.48     &  3.52
& 0.80  & 22.96  & 0.85    &  90.73     &  9.27  \\
& 0.90  & 15.15  & 1.17    &  95.02     &  4.98
& 0.85  & 34.88  & 0.74    &  90.16     &  9.84  \\
& 1.00  & 55.44  & 0.74    &  95.17     &  4.83
& 0.90  & 56.02  & 0.63    &  90.37     &  9.63  \\ [7pt]
\multirow{3}*{$[\S_Q\Xi_Q^{'}]^{I=3/2}_S$}
&\multicolumn{5}{c}{\multirow{3}*{$\times$}}
& 0.80  & 0.46   & 3.22    &  90.74     &  9.26  \\
&       &        &         &            &
& 0.90  & 10.34  & 1.08    &  82.64     &  17.36 \\
&       &        &         &            &
& 1.00  & 1.04   & 2.40    &  86.31     &  13.69 \\ [7pt]
\multirow{3}*{$[\Xi_Q^{'}\O_Q]^{I=1/2}_S$}
& \multicolumn{5}{c}{\multirow{3}*{$\times$}}
& 0.90  & 0.88   &  2.38   &  96.64     &  3.36 \\
&       &        &         &            &
& 1.00  & 23.30  &  0.76   &  89.72     &  10.28 \\
&       &        &         &            &
& 1.10  & 10.78  &  0.98   &  89.47     &  10.53 \\ [7pt]
\multirow{3}*{$[\S_Q\O_Q]^{I=1}_S$}
& 0.90  & 1.20   & 2.93    &  99.91     &  0.09
& 0.80  & 3.07   & 1.47    &  99.73     &  0.27 \\
& 1.00  & 10.04  & 1.25    &  99.67     &  0.33
& 0.90  & 19.30  & 0.78    &  99.24     &  0.76 \\
& 1.10  & 25.78  & 0.89    &  99.39     &  0.61
& 1.00  & 46.47  & 0.58    &  98.71     &  1.29 \\ [7pt]
\multirow{3}*{$[\Xi_Q\L_Q]^{I=1/2}_S$}
& 1.00  & 0.72   & 3.69    &  100.00    & 0.00
& 0.90  & 10.33  &  0.96   &  100.00    & 0.00  \\
& 1.30  &  3.03  & 2.03    &  100.00    & 0.00
& 1.30  & 29.55  & 0.66    &  100.00    & 0.00  \\
& 1.50  &  2.87  & 2.08    &  100.00    & 0.00
& 1.50  &  27.78 & 0.68    &  100.00    & 0.00  \\ [7pt]
\multirow{3}*{$[\S_Q\L_Q]^{I=1}_S$}
& 0.80  & 0.01   & 6.58    &  96.76     & 3.24
& 0.90  & 19.85  & 0.98    & 80.86      & 19.14 \\
& 0.90  & 0.46   & 4.49    &  95.26     & 4.74
& 1.10  & 9.56   & 1.36    &  77.43     & 22.57 \\
& 1.00  & 0.03   & 6.48    &  96.59     & 3.41
& 1.30  & 3.35   & 2.12    &  73.50     & 26.50 \\ [7pt]
\multirow{3}*{$[\S_Q\Xi_Q]^{I=1/2}_S$}
& 0.90  & 1.63   & 2.62    & 99.95      & 0.05
& 0.80  & 3.78   & 1.37    & 99.90      & 0.10 \\
& 1.00  & 17.30  & 1.04    & 99.88      & 0.12
& 0.90  & 21.96  & 0.75    & 99.60      & 0.40 \\
& 1.10  & 49.98  & 0.72    & 99.90      & 0.10
& 1.00  & 62.92  & 0.53    & 99.59      & 0.41 \\ [7pt]
\multirow{3}*{$[\S_Q\Xi_Q]^{I=3/2}_S$}
& \multicolumn{5}{c}{\multirow{3}*{$\times$}}
& 0.90  & 5.45   & 1.19    &  97.78     & 2.22 \\
&       &        &         &            &
& 1.00  & 5.52   & 1.19    &  96.60     & 3.10 \\
&       &        &         &            &
& 1.10  & 1.31   & 2.06    &  97.61     & 2.09  \\ [7pt]
\multirow{3}*{$[\Xi^{'}_Q\L_Q]^{I=1/2}_S$}
& 1.00  &  0.17  &  5.44   &  99.94     & 0.06
& 1.00  &  17.64 &  0.80   &  98.81     & 1.19 \\
& 1.20  & 0.69   &  3.71   &  99.93     & 0.07
& 1.20  & 23.91  &  0.71   &  99.41     & 0.59 \\
& 1.40  & 0.36   &  4.60   &  99.97     & 0.03
& 1.40  &  23.14 &  0.71   &  99.92     & 0.08 \\ [7pt]
\multirow{3}*{$[\Xi_Q{'}\Xi_Q]^{I=0}_S$}
& 0.90  &  3.10  & 2.08    &  97.44     & 2.56
& 0.80  &  11.16 & 1.06    &  88.45     & 11.55 \\
& 1.00  &  13.75 & 1.16    &  98.60     & 1.40
& 0.90  & 24.87  & 0.77    &  94.63     & 5.37 \\
& 1.10  &  32.31 & 0.84    &  99.38     & 0.62
& 1.00  &  51.15 & 0.58    &  98.27     & 1.73 \\ [7pt]
\multirow{3}*{$[\Xi_Q^{'}\Xi_Q]^{I=1}_S$}
&\multicolumn{5}{c}{\multirow{3}*{$\times$}}
& 0.90  & 3.40   & 1.47    & 96.48      & 3.52 \\
&       &        &         &            &
& 1.00  &  13.70 & 0.89    & 95.83      & 4.17 \\
&       &        &         &            &
& 1.20 &  8.11  & 1.08    & 96.54      & 3.46 \\ [7pt]
\multirow{3}*{$[\Xi_Q\O_Q]^{I=1/2}_S$}
& 1.00  &  2.63  & 2.09    & 99.37      & 0.63
& 1.00  & 28.78  & 0.69    & 95.78      & 4.22 \\
& 1.10  & 4.51   & 1.08    & 99.33      & 0.67
& 1.10  & 35.44  & 0.63    & 96.52      & 3.48 \\
& 1.20  & 2.73   & 2.05    & 99.57      & 0.43
& 1.20  & 29.48  & 0.67    & 97.72      & 2.28 \\ [7pt]
\multirow{3}*{$[\L_Q\O_Q]^{I=0}_S$}
& 1.00  & 5.40   & 1.60    & 100.00     & 0.00
& 0.80  & 1.62   & 1.88    & 100.00     & 0.00 \\
& 1.10  & 16.54  & 1.04    & 100.00     & 0.00
& 0.90  & 14.10  & 0.86    & 100.00     & 0.00 \\
& 1.20  & 32.02  & 0.82    & 100.00     & 0.00
& 1.00  & 36.07  & 0.62    & 100.00     & 0.00 \\ [3pt]
\bottomrule[1.0pt]
\end{tabular*}
\end{table}

%%%%%%%%%%%%%%%%%%%%%%%%%%%%%%%%%%%%%%%%%%%%%%%%%%%
\section{Numerical results for the $\L_Q\L_Q$ system}\label{lambdanum}
%%%%%%%%%%%%%%%%%%%%%%%%%%%%%%%%%%%%%%%%%%%%%%%%%%%

We investigated the $\L_Q\L_Q\left[I(J^P)=0(0^+)\right]$ system
with the $\sigma$ and $\omega$ exchange potential, but without the
coupled-channel effect in the flavor space in Ref.
\cite{Lee:2011rk}. We find no binding solutions. Later, the
authors in Ref. \cite{Meguro:2011nr} considered the
coupled-channel effect and studied this state using the
pion-exchange potential. They found a bound state solution. It is
intriguing to study the variation of the bound state solution with
the heavy quark mass, the S-D mixing effect, the long-range OPE
force and medium-/short-range interaction respectively. In the
first subsection, we shall present the numerical results for the
OPE model with the coupled-channel effect. In the second
subsection, we will add the scalar and vector meson exchange force
which also contributes to the transition in the flavor space.

%%%%%%%%%%%%%%%%%%%%%%%%%%%%%%%%%%%%%%%%%%%%%%%%%%%%%%
\subsection{Numerical Results for The $\L_Q\L_Q$ System
with The OPE Potential}

\begin{figure}[htp]
\centering
\begin{tabular}{ccc}
\includegraphics[width=0.245\textwidth]{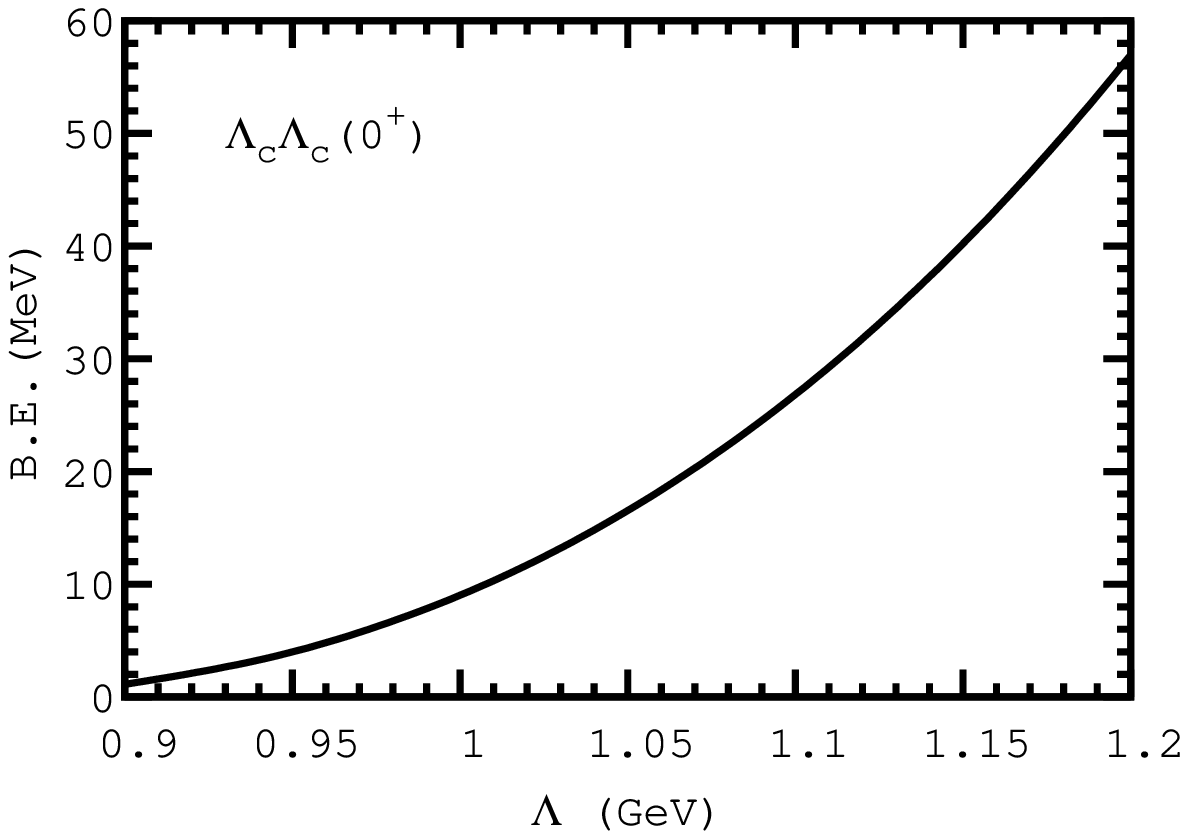}&
\includegraphics[width=0.245\textwidth]{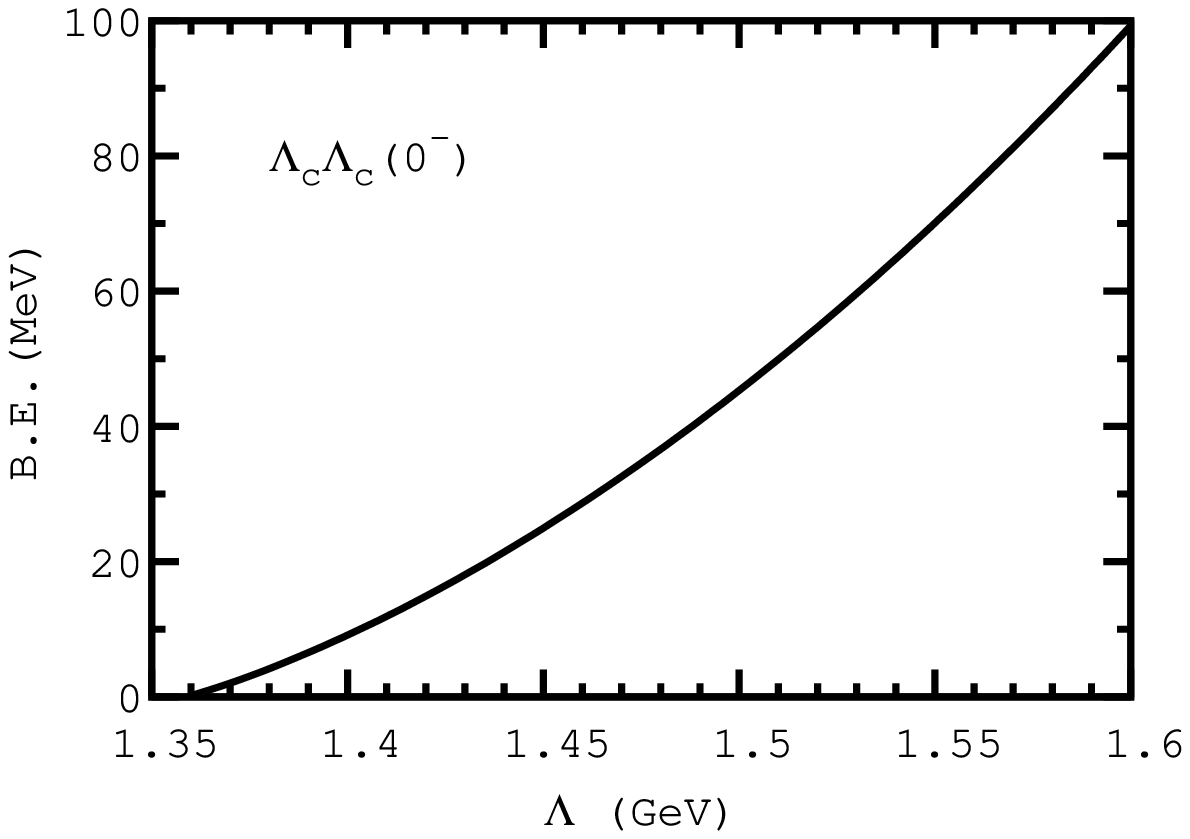}&
\includegraphics[width=0.245\textwidth]{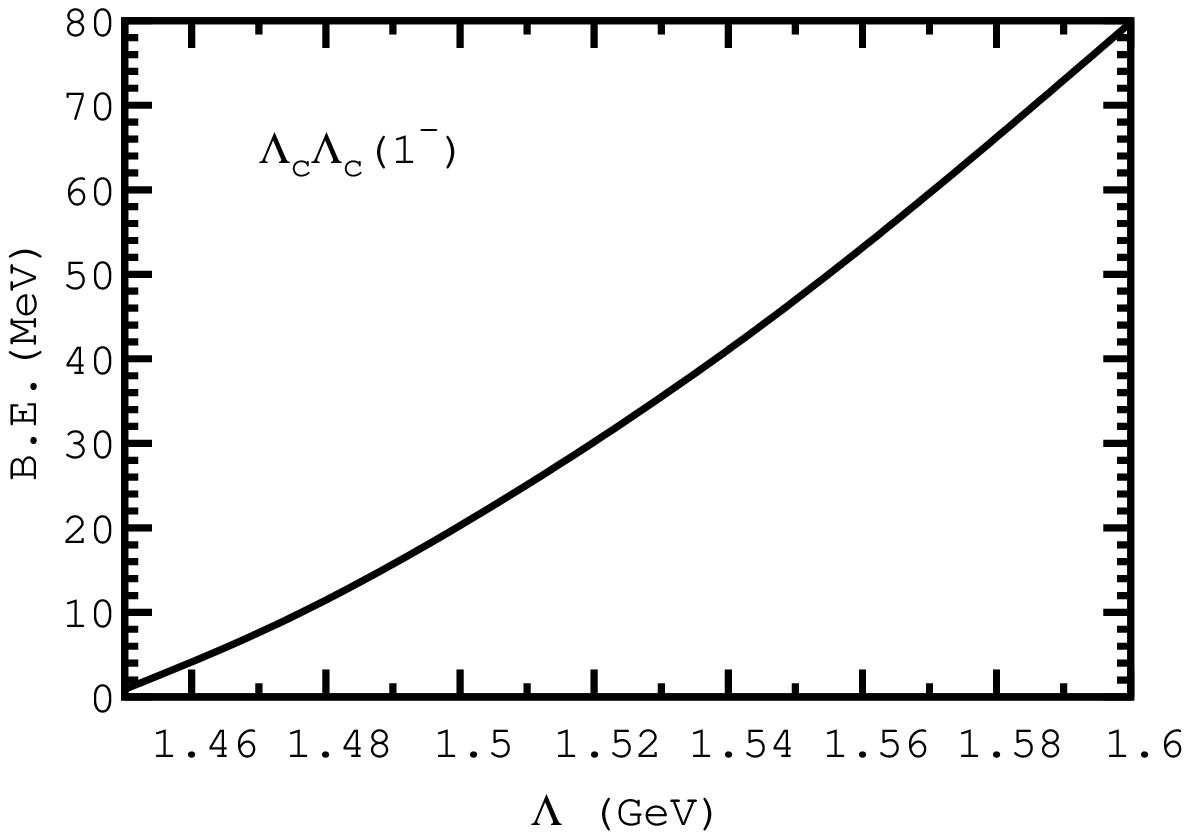}\\
\includegraphics[width=0.245\textwidth]{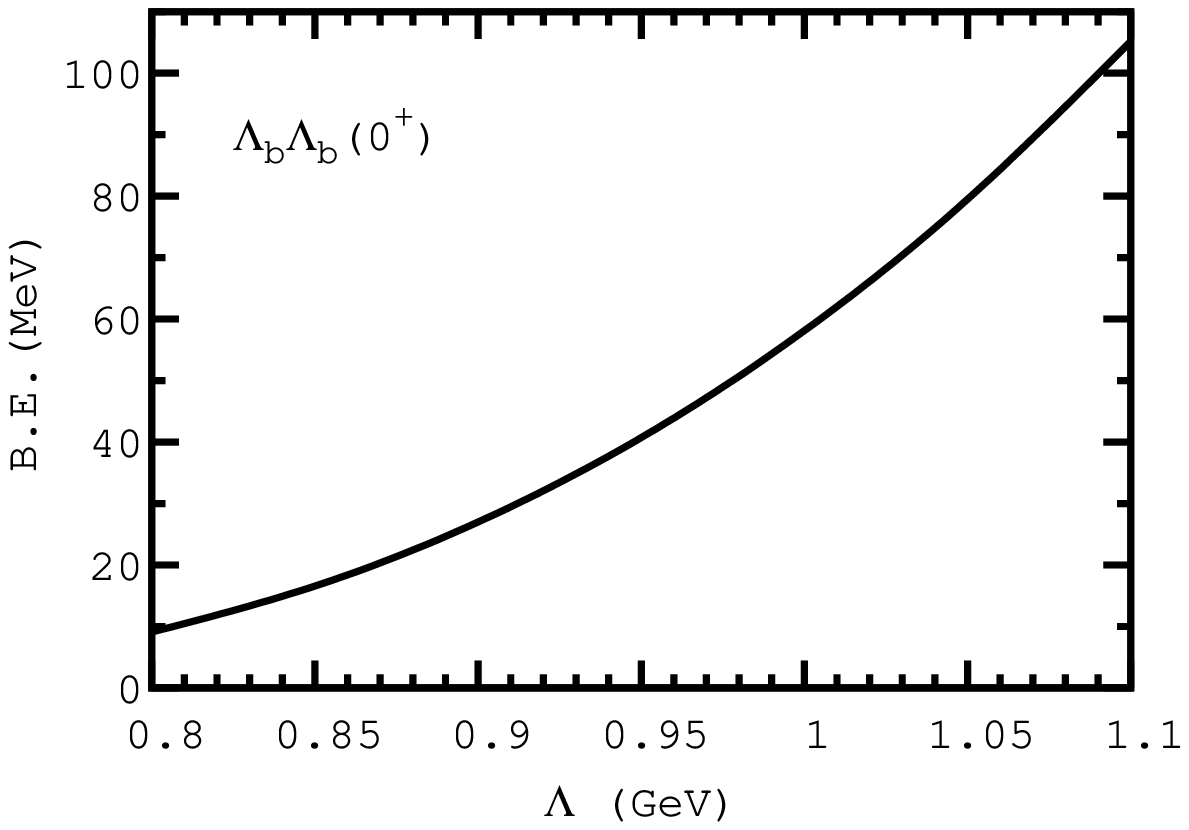}&
\includegraphics[width=0.245\textwidth]{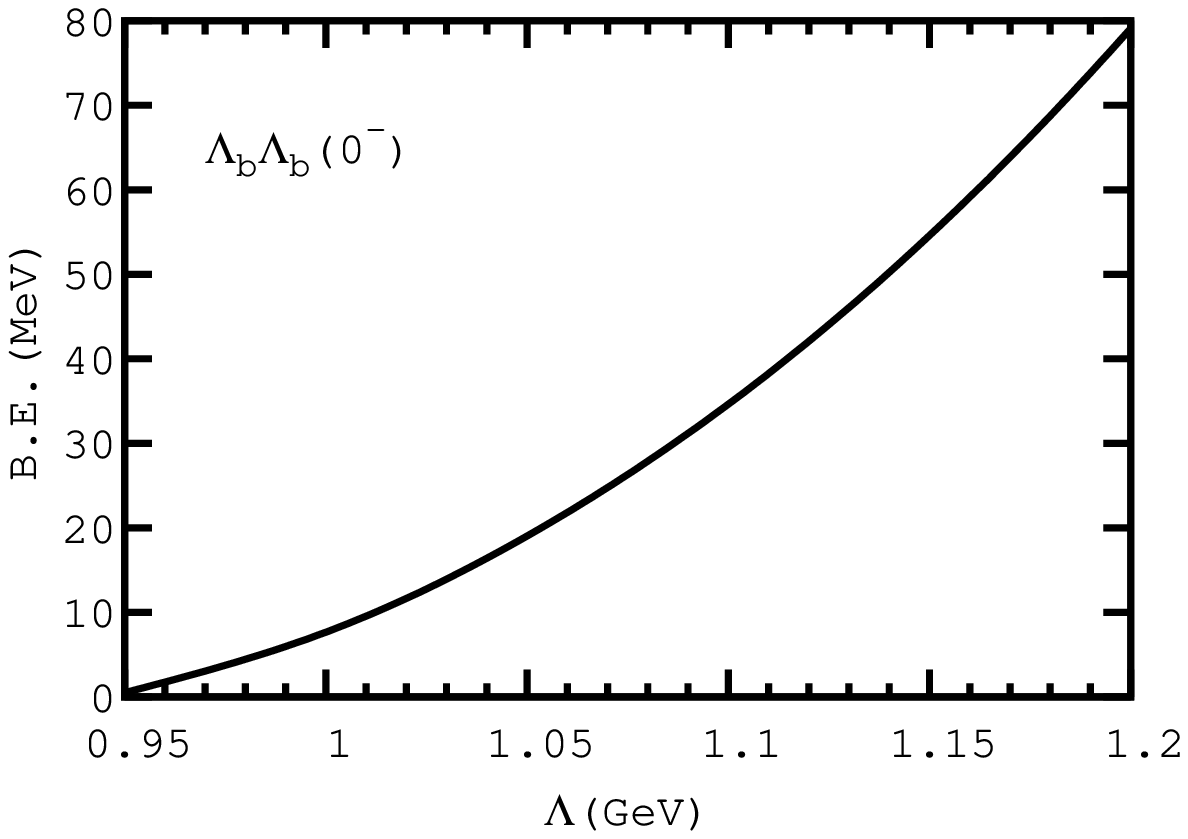}&
\includegraphics[width=0.245\textwidth]{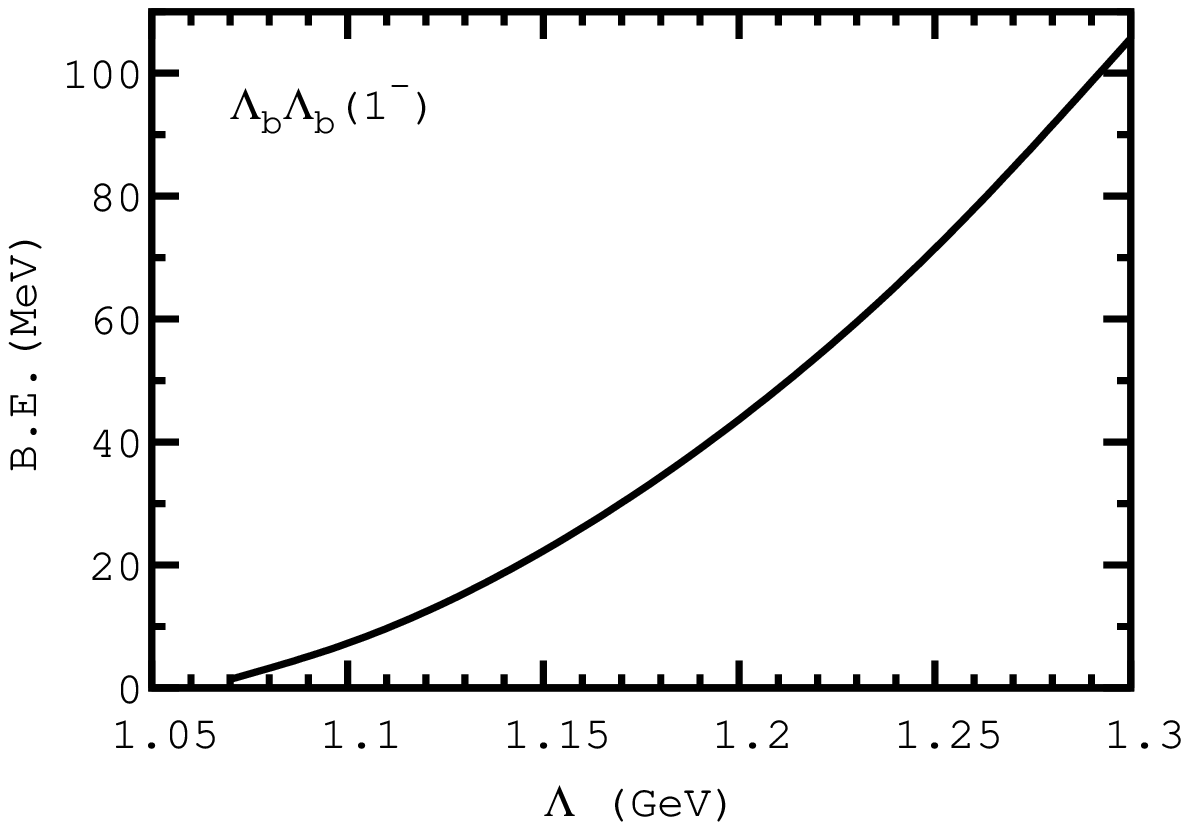}\\
\end{tabular}
\caption{The dependence of the binding energy for $\L_Q\L_Q$ on the cutoff
parameter with the OPE potential.}\label{plotLL}
\end{figure}

The state $\L_Q\L_Q$(Q=b,c) is the heavy analogue of the H
dibaryon. We use the $\pi$ exchange potential and include the
coupled-channel effect. Actually, the pion exchange is forbidden
for the $\L_Q\L_Q$ system due to the isospin conservation of the
strong interaction. The binding solution is mainly due to the
coupled-channel effect. We give the numerical results in Table
\ref{ReLLOPE} and plot the dependence of the binding energy on the
cutoff parameter in Fig.~\ref{plotLL}.

$\L_Q\L_Q\left[I(J^P)=0(0^+)\right]$. We reproduce the numerical
results of Ref. \cite{Meguro:2011nr} for the state $\L_c\L_c(^1S_0)$,
and list them in Table \ref{ReLLOPE}.
We also extend the
same formalism to the bottomed sector. A bound state
$\L_b\L_b(^1S_0)$ with binding energy $9.09\sim26.99$ MeV appears
when the cutoff parameter is chosen between 0.80 GeV and 0.90 GeV.
Correspondingly, the root-mean-square radius varies from 1.08 fm
to 0.77 fm. However, if the cutoff parameter increases to 1.10
GeV, the binding energy will reach as high as 105.17 MeV. There
are five channels shown in Table \ref{LL} for this state. The
$\L_b\L_b(^1S_0)$ component is dominant with a probability
about $90\%$. The contribution of the components $\S_b\S_b(^1S_0)$
and $\S_b^*\S_b^*(^1S_0)$ is very small, around 1\%.

$\L_Q\L_Q\left[I(J^P)=0(0^-)\right]$. We also extend the same
analysis to the case with the orbital excitation $L=1$ and obtain
a loosely bound state of $\L_c\L_c\left[0(0^-)\right]$ with
binding energy $0.19\sim24.97$ MeV when the cutoff parameter is
chosen between 1.36 GeV and 1.45GeV. The binding energy will
increase to 99.26 MeV when the cutoff parameter is 1.60 GeV. The
contribution of the dominant channel, $\L_c\L_c(^3P_0)$, is
$90.73\%\sim81.72\%$ when the cutoff parameter is around
$1.36\sim1.45$ GeV. The channel $\S_c^*\S_c^*(^3P_0)$ provides a
fairly small contribution, less than $1\%$. For the corresponding
bottomed state, its binding energy is $0.50\sim34.63$ MeV when the
cutoff parameter is $0.95\sim1.10$ GeV, which may also be a good
molecule candidate.

$\L_Q\L_Q\left[I(J^P)=0(1^-)\right]$. We consider seven channels
in this case, which are listed in Table \ref{LL}. We obtain a
shallow bound state with binding energy $0.91\sim20.27$ MeV in the
charmed sector when the cutoff parameter is $1.45\sim1.50$ GeV.
And the contribution of the dominant channel, $\L_c\L_c(^3P_1)$,
is $79.15\%\sim67.55\%$. If the cutoff parameter increases to 1.60
GeV, the binding energy will reach 79.78 MeV. The channel with the
second largest contribution is $\S_c\S_c(^3P_1)$, with a
probability of $14.55\%\sim22.90\%$ when the cutoff parameter is
around $1.45\sim1.50$ GeV. However, the probabilities of the other
three channels, $\S_c\S_c^*(^3P_1)$, $\S_c\S_c^*(^5P_1)$ and
$\S_c\S_c^*(^5F_1)$, are tiny as shown in Table \ref{ReLLOPE}. The
situation of the bottomed case is similar to that of the charmed
case except that the binding of the former is deeper.

\begin{table}[htp]
\renewcommand{\arraystretch}{1.3}
\centering
\caption{The binding solutions of $\L_Q\L_Q$ with OPE potential. ``$\L$" is the cutoff parameter.
``B.E." and ``$r_{rms}$" are the binding energy and the root-mean-square radius, respectively.
``$P_i$" is the probability of the individual channel which are given in Table \ref{LL}.}
\label{ReLLOPE}
\begin{tabular*}{18cm}{@{\extracolsep{\fill}}cc|ccccc|ccccc}
\toprule[1pt]\addlinespace[5pt]
$I(J^P)$   &          &\multicolumn{5}{c|}{Q=c}&
\multicolumn{5}{c}{Q=b}\\
\specialrule{0.8pt}{3pt}{3pt}
\multirow{8}*{$0(0^+)$}
 &$\L$(GeV)&  0.90    &   0.95   &  1.00    &  1.10   &  1.20   &
              0.80    &   0.90   &  0.95    &  1.00   &  1.10    \\
 &B.E.(MeV) &$1.11$   &$3.99$    &$9.02$    &$26.83$  &$56.97$  &
             $9.09$  &$26.99$   &$40.72$   &$58.11$  &$105.17$  \\
 &$r_{rms}$(fm)& 3.20  &  1.89    &  1.39    &  0.95   &  0.74   &
                 1.08  &  0.77    &  0.68    &  0.62   &  0.52    \\
 &$P_1(\%)$& 98.63    &  97.25   &  95.69   &  92.07  &  88.01  &
             96.27    &  92.77   &  90.75   &  88.59  &  84.07    \\
 &$P_2(\%)$&  0.08    &  0.19    &  0.35    &  0.81   &  1.48   &
              0.16    &  0.42    &  0.61    &  0.86   &  1.46     \\
 &$P_3$(\%)&  0.08    &  0.20    &  0.36    &  0.87   &  1.64   &
              0.25    &  0.67    &  0.98    &  1.37   &  2.37     \\
 &$P_4$(\%)&  0.86    &  1.68    &  2.57    &  4.44   &  6.29   &
              2.27    &  4.20    &  5.24    &  6.28   &  8.27     \\
 &$P_5$(\%)&  0.35    &  0.68    &  1.04    &  1.81   &  2.58   &
              1.05    &  1.94    &  2.42    &  2.91   &  3.84   \\
\specialrule{0.8pt}{3pt}{3pt}
\multirow{8}*{$0(0^-)$}
 &$\L$(GeV)& 1.36     &   1.40   &  1.45    &  1.50    &    1.60  &
             0.95     &   1.00   &  1.10    &  1.20    &    1.30   \\
 &B.E.(MeV)&$  0.19$  &$9.14$    &$24.97$   &$45.32$   &$ 99.26$ &
            $  0.50$  &$7.66$    &$34.63$   &$79.08$   &$ 142.32$  \\
 &$r_{rms}$(fm)&  2.39 &  1.06    &  0.84    & 0.73     &   0.60   &
                  1.94 &  1.03    &  0.72    & 0.59     &   0.50    \\
 &$P_1(\%)$   & 90.73    &  85.65   &  81.72   & 78.45    &  73.05   &
              95.70    &  91.97   &  85.76   & 80.01    &  74.89    \\
 &$P_2(\%)$   & 2.86     &  4.24    &  5.13    &  5.76    &   6.56   &
              1.26     &  2.27    &  3.70    &  4.75    &   5.46    \\
 &$P_3(\%)$   & 0.08     &  0.15    &  0.24    &  0.35    &  0.60    &
              0.01     &  0.02    &  0.10    &  0.25    &  0.48     \\
 &$P_4(\%)$   & 3.97     &  6.42    &  8.54    &  10.42   &  13.74    &
              1.35     &  2.74    &  5.55    &  8.59    &  11.56    \\
 &$P_5(\%)$   & 2.36     &  3.54    &  4.37    &  5.02    &  6.04    &
              1.69     &  3.00    &  4.89    &  6.40    &  7.62    \\
 \specialrule{0.8pt}{3pt}{3pt}
\multirow{10}*{$0(1^-)$}
 &$\L$(GeV)& 1.45     &   1.47   & 1.50     &  1.55    &  1.60 &
             1.07     &   1.10   & 1.15     &  1.20    &  1.30   \\
 &B.E.(MeV)& $0.91$   &$7.60$    &$20.27$   &$46.97$   & $79.78$ &
             $1.39$   &$7.26$   &$22.28$   &$43.68$   & $105.69$ \\
 &$r_{rms}$(fm)& 1.62  &  1.99    & 0.79     &  0.66    &  0.59  &
                 1.40  &  0.96    & 0.74     &  0.63    &  0.51     \\
 &$P_1(\%)$& 79.15    &  73.19   & 67.55    &  60.70   &  55.55 &
             91.48    &  87.06   & 80.52    &  74.33   &  63.52    \\
 &$P_2(\%)$& 14.55    &  18.82   & 22.90    &  27.85   &  31.51 &
              4.37    &  6.89    &  10.86    &  14.79   &  21.86    \\
 &$P_3(\%)$& 2.70     &   3.54   & 4.40     &  5.57    &  6.57  &
             1.14     &   1.81   & 2.92     &  4.05    &  6.25     \\
 &$P_4(\%)$& 3.53     &   4.36   &  5.04    &  5.77    &  6.26  &
             2.97     &   4.18   &  5.63    &  6.73    &  8.26     \\
 &$P_5(\%)$& 0.08     &   0.09   &  0.10    &  0.11    &  0.11  &
             0.04     &   0.06   &  0.08    &  0.09    &  0.10     \\
 &$P_6(\%)$& 0.00     &   0.00   &  0.00    &  0.01    &  0.01  &
             0.00     &   0.00   &  0.00    &  0.00    &  0.00     \\
 &$P_7(\%)$& 0.00     &   0.00   &  0.00    &  0.00    &  0.00  &
             0.00     &   0.00   &  0.00    &  0.00    &  0.00     \\
 \bottomrule[1pt]
\end{tabular*}
\end{table}

%%%%%%%%%%%%%%%%%%%%%%%%%%%%%%%%%%%%%%%%%%%%%%%%%%%%%%
\subsection{Numerical Results for The $\L_Q\L_Q$ System
with The OBE Potential}

In this subsection, we investigate the $\L_Q\L_Q$ system with the
OBE potential which not only includes the long-range $\pi$
exchange interaction but also the medium-/short-range $\eta$,
$\sigma$, $\rho$ and $\omega$ exchange interaction. The numerical
results are shown in Table \ref{ReLLOBE}.

We obtain a weakly bound state for $\L_c\L_c\left[I(J^P)=0(0^+)\right]$. The
binding energy is $2.53\sim55.11$ MeV when the cutoff parameter is around
$0.80\sim1.00$ GeV. Accordingly its root-mean-square radius is about
$2.31\sim0.73$ fm, which is comparable with the size of the deuteron. Similar
to the OPE potential case, the $\L_c\L_c(^1S_0)$ component is dominant with a
probability about $98.69\%\sim86.79\%$, and the total contributions of the
other channels are less than $15\%$. For the state
$\L_b\L_b\left[0(0^+)\right]$, the binding energy is much larger as expected.
Its binding energy is 27.30 MeV when the cutoff parameter is $0.80$ MeV. When
the cutoff parameter is 1.00 GeV, the binding energy reaches as high as 148.17
MeV.

For the state $\L_c\L_c\left[I(J^P)=0(0^-)\right]$, the binding energy is
$4.69$ MeV when the cutoff parameter is 1.15 GeV. When we increase the cutoff
parameter to 1.25 GeV, the binding energy is 61.36 MeV. The probability of the
dominant channel $\L_c\L_c(^3P_0)$ is about $85.12\%\sim71.08\%$. The
contribution of the second dominant channel is $\S_c\S_c^*(^3P_0)$, with a
probability about $11.48\%\sim23.57\%$. The results of the bottomed state
$\L_b\L_b\left[0(0^-)\right]$ are similar to those of the charmed case, but
with deeper binding energy. The binding energy is $2.80\sim100.54$ MeV with the
cutoff parameter around $0.85\sim1.05$ GeV. The probabilities of the channels
$\L_b\L_b(^3P_1)$ and $\S_b\S_b^*(^3P_1)$ are about $95.93\%\sim73.92\%$ and
$1.45\%\sim18.90\%$ respectively.

The state $\L_c\L_c\left[I(J^p)=0(1^-)\right]$ with binding energy around
$1.35\sim65.05$ MeV and cutoff parameter around $1.16\sim1.25$ GeV may also be
a loosely bound state. But the binding solutions depend sensitively on the
cutoff parameter. The binding energy of the state $\L_b\L_b\left[0(1^-)\right]$
is $0.24\sim74.65$ MeV when the cutoff parameter is $0.90\sim1.05$ GeV.

Besides the transition induced by the OPE force in the flavor space, we have
also considered the transitions caused by the eta meson and rho/omega meson
exchange, which greatly enhances the non-diagonal matrix element in the
Hamiltonian. With the same cutoff parameter, we can clearly see that the
binding energy in the OBE case is larger than that in the OPE case. For
example, the binding energy for the $\L_c\L_c\left[I(J^P)=0(0^+)\right]$ state
is 1.11 MeV in the OPE case if one fixes the cutoff at 0.90 GeV. However, the
binding energy increase to 16.61 MeV in the OBE case with the same cutoff. In
other words, the medium- and short-range attractive force plays a significant
role in the formation of the loosely bound $\L_c\L_c$ and $\L_b\L_b$ states.

\begin{table}[htp]
\renewcommand{\arraystretch}{1.3}
\centering
\caption{The binding solutions of $\L_Q\L_Q$ with the OBE potential. ``$\L$" is the cutoff parameter.
``B.E." and ``$r_{rms}$" are the binding energy and the root-mean-square radius, respectively.
``$P_i$" is the probability of the individual channel which are given in Table \ref{LL}.}
\label{ReLLOBE}
\begin{tabular*}{18cm}{@{\extracolsep{\fill}}cc|ccccc|ccccc}
\toprule[1pt]\addlinespace[5pt]
$I(J^P)$   &          &\multicolumn{5}{c|}{Q=c}&
\multicolumn{5}{c}{Q=b}\\
\specialrule{0.8pt}{3pt}{3pt}
\multirow{8}*{$0(0^+)$}
 &$\L$(GeV)&  0.80    &   0.85   &  0.90    &  0.95   &  1.00   &
              0.80    &   0.85   &  0.90    &  0.95   &  1.00    \\
 &B.E.(MeV) &$2.53$   &$8.04$    &$16.61$   &$30.18$  &$55.11$  &
             $27.30$  &$45.94$   &$69.61$   &$102.62$  &$148.17$  \\
 &$r_{rms}$(fm)& 2.31  &  1.48    &  1.13    &  0.91   &  0.73   &
                 0.78  &  0.66    &  0.58    &  0.50   &  0.44    \\
 &$P_1(\%)$& 98.69    &  97.38   &  95.43   &  92.26  &  86.79  &
             95.26    &  92.82   &  89.32   &  84.43  &  78.82    \\
 &$P_2(\%)$&  0.08    &  0.34    &  1.15    &  3.01   &  6.04   &
              0.27    &  0.81    &  2.24    &  4.81   &  8.21     \\
 &$P_3$(\%)&  0.06    &  0.18    &  0.47    &  1.03   &  2.68   &
              0.31    &  0.88    &  1.88    &  3.51   &  5.60     \\
 &$P_4$(\%)&  0.84    &  1.51    &  2.15    &  2.73   &  3.12   &
              2.89    &  3.73    &  4.44    &  4.85   &  4.84     \\
 &$P_5$(\%)&  0.33    &  0.58    &  0.81    &  0.97   &  1.37   &
              1.27    &  1.58    &  2.13    &  2.41   &  2.53   \\
\specialrule{0.8pt}{3pt}{3pt}
\multirow{8}*{$0(0^-)$}
 &$\L$(GeV)& 1.15     &   1.17   &  1.20    &  1.23    &    1.25  &
             0.85     &   0.90   &  0.95    &  1.00    &    1.05   \\
 &B.E.(MeV)&$  4.69$  &$13.34$    &$27.65$   &$46.59$   &$ 61.36$ &
            $  2.80$  &$17.97$   &$31.35$   &$59.76$   &$ 100.54$  \\
 &$r_{rms}$(fm)&  1.25 &  0.94    &  0.80    & 0.70     &   0.66   &
                  1.38 &  0.95    &  0.76    & 0.62     &   0.53    \\
 &$P_1(\%)$   & 85.12  &  81.38   &  76.85   & 73.16    &  71.08   &
              95.93    &  92.31   &  87.54   & 80.50    &  73.92    \\
 &$P_2(\%)$   & 0.64   &  0.57    &  0.37    &  0.19    &   0.18   &
              1.08     &  1.56    &  1.73    &  1.58    &   1.17    \\
 &$P_3(\%)$   & 0.75   &  1.03    &  1.44    &  1.84    &   2.11   &
              0.00     &  0.02    &  0.11    &  0.76    &  1.38     \\
 &$P_4(\%)$   & 11.48   &  14.67   &  18.64   &  21.87   &  23.57    &
              1.45     &  3.63    &  7.34    &  12.99   &  18.90    \\
 &$P_5(\%)$   & 2.01   &  2.36    &  2.70    &  2.93    &  3.06    &
              1.53     &  2.48    &  3.27    &  4.17    &  4.62    \\
 \specialrule{0.8pt}{3pt}{3pt}
\multirow{10}*{$0(1^-)$}
 &$\L$(GeV)& 1.16     &   1.18   & 1.20     &  1.23    &  1.25 &
             0.90     &   0.95   & 0.97     &  1.00    &  1.05   \\
 &B.E.(MeV)& $1.35$   &$11.32$   &$24.10$   &$47.32$   & $65.05$ &
             $0.24$   &$11.12$   &$18.82$   &$34.83$   & $74.65$ \\
 &$r_{rms}$(fm)& 1.57 &  0.94    & 0.78     &  0.67    &  0.62  &
                 2.28 &  0.92    & 0.80     &  0.68    &  0.55     \\
 &$P_1(\%)$& 82.41    &  75.18   & 70.21    &  64.57   &  61.64 &
             96.65    &  90.05   & 86.85    &  81.50   &  72.55    \\
 &$P_2(\%)$& 9.51     &  13.39   & 15.96    &  18.67   &  19.93 &
              1.51    &  4.71    &  6.26    &  8.76    &  12.56    \\
 &$P_3(\%)$& 1.96     &   2.70   & 3.17     &  3.68    &  3.93  &
             0.47     &   1.56   & 2.10     &  2.95    &  4.24     \\
 &$P_4(\%)$& 1.77     &   2.24   &  2.46    &  2.61    &  2.65  &
             1.29     &   2.89   &  3.40    &  4.03    &  4.64     \\
 &$P_5(\%)$& 4.26     &   6.36   &  8.02    &  10.24   &  11.60  &
             0.08     &   0.77   &  1.37    &  2.69    &  5.85     \\
 &$P_6(\%)$& 0.08     &   0.12   &  0.15    &  0.02    &  0.22  &
             0.00     &   0.01   &  0.03    &  0.05    &  0.13     \\
 &$P_7(\%)$& 0.01     &   0.02   &  0.03    &  0.03    &  0.04  &
             0.00     &   0.00   &  0.01    &  0.01    &  0.03     \\
 \bottomrule[1pt]
\end{tabular*}
\end{table}

%%%%%%%%%%%%%%%%%%%%%%%%%%%%%%%%%%%%%%%%%%%%%%
\section{Discussions and Conclusions}\label{SUMMARY}
%%%%%%%%%%%%%%%%%%%%%%%%%%%%%%%%%%%%%%%%%%%%%%

We have investigated the possible deuteron-like molecules composed
of two heavy flavor baryons with the form of ``$A_QB_Q$". We have
also performed an extensive analysis of the $\L_Q\L_Q$(Q=b,c)
system, which is the heavy analogue of the H dibaryon.

The weakly bound states are usually very sensitive to potential details
including the coupling constants and form factors etc. Sometimes small change
of the coupling constants may dismantle the bound state.

Throughout this work, we have adopted the root-mean-square radius and binding
energy of the system to judge whether the system is a loosely bound molecular
state. There exists another intuitive approach. The relative momentum of the
loosely bound system $p\sim \sqrt{2\mu E}$ probes distance around ${1\over p}$,
where $\mu$ is the reduced mass and $E$ is the binding energy. For a loosely
bound state, ${1\over p}$ should be much larger than the interaction range of
the potential, which is around ${1\over m_{\rho, \omega, \sigma}}\sim
(0.2-0.3)$ fm. In other words, the size of the system should be larger than
$(0.6\sim 1.0)$ fm. Accordingly, the binding energy should be much smaller than
${m_\rho^2\over M_B}$ where $M_B$ is the charmed or bottomed baryon mass.
Numerically, the binding energy should be much less than 240 MeV and 100 MeV
for the charmed and bottomed systems respectively. In other words, those states
in Tables~\ref{STrOP}-\ref{ReLLOBE} which do not satisfy the above criteria
should not be regarded as the loosely bound molecular states.

For the spin-singlet systems with the ``$A_QB_Q$" form, the pion
exchange force is not strong enough to form bound states for the
five systems, $[\S_Q\Xi_Q^{'}]^{(I=1/2,3/2)}_{S=0}$,
$[\S_Q\L_Q]^{I=1}_{S=0}$ and $[\Xi_Q\Xi_Q^{'}]^{(I=0,1)}_{S=0}$
(Q=b,c). When we add the contributions from the scalar and vector
meson exchanges, some bound states appear. The
following five states $[\S_c\O_c]^{I=1}_{S=0}$,
$[\Xi_c\L_c]^{I=1/2}_{S=0}$, $[\Xi_b\L_b]^{I=1/2}_{S=0}$,
$[\S_c\Xi_c]^{I=1/2}_{S=0}$ and $[\L_c\O_c]^{I=0}_{S=0}$ are all
very loosely bound with small binding energies and large
root-mean-square radii with cutoff parameter $1.0\sim1.50$ GeV.
They are good candidates of molecules.

In the spin-triplet case, the numerical results with the
one-pion-exchange potential alone indicate that
$[\S_Q\Xi_Q^{'}]^{I=1/2}_{S=1}$(Q=b,c),
$[\S_Q\L_Q]^{I=1}_{S=1}$(Q=b,c), $[\S_b\Xi_b^{'}]^{I=3/2}_{S=1}$
and $[\Xi_b\Xi_b^{'}]^{I=0}_{S=1}$ may be loosely bound states.
They have shallow binding solutions when the cutoff parameter is
around $0.90\sim1.50$ GeV. The three states
$[\S_c\Xi_c^{'}]^{I=3/2}_{S=1}$, $[\Xi_c\Xi_c^{'}]^{I=0}_{S=1}$
and $[\Xi_b\Xi_b^{'}]^{I=1}_{S=1}$ do not have binding solutions until the
cutoff parameter reaches 1.80 GeV. When taking the vector and
scalar boson exchanges into account, the numerical results do not
change significantly except that the bound state of
$[\S_c\Xi_c^{'}]^{I=3/2}_{S=1}$ disappear and the binding energy
in some channels becomes deeper. Therefore, we conclude that the
long-range one-pion-exchange interaction plays an dominant role in
forming these bound states. Comparing the results of OPE model
with those of the OBE model, one notices that the contribution of
the D wave is smaller for the latter, which implies that the S-D
mixing mainly comes from the pion exchange. Our results suggest
that the states $[\Xi_Q^{'}\L_Q]^{I=1/2}_{S=1}$(Q=b,c) and
$[\Xi_Q\O_Q]^{I=1/2}_{S=1}$(Q=b,c) with shallow binding solutions
and reasonable cutoff parameter may also be good candidates of
molecules.

For the heavy analogue of the H dibaryon, our results indicate
that $\L_Q\L_Q$(Q=b,c) with quantum numbers $I(J^P)=0(0^+)$,
$0(0^-)$ and $0(1^-)$ may all be molecules. The binding solutions
of $\L_Q\L_Q$ system with the OPE potential mainly come from the
coupled-channel effect. Besides the transition induced by the OPE
force in the flavor space, we have also considered the transitions
caused by the eta meson and rho/omega meson exchange. With the
same cutoff parameter, the binding energy in the OBE case is
larger than that in the OPE case. The medium- and short-range
attractive force plays a significant role in the formation of the
loosely bound $\L_c\L_c$ and $\L_b\L_b$ states.

The authors studied the $\L_Q\L_Q$ system at the quark level, and
obtained bound states with mass 4516 MeV for $\L_c\L_c$ and 9175
MeV for $\L_b\L_b$ \cite{Gerasyuta:2011zx}. Theoretical
investigations of these molecular states with other
phenomenological models are desirable.

If these states really exist as molecules, once produced, they
will be very stable because this system decays via weak
interaction. It is difficult to produce the states
with double charm or double bottom experimentally. However, there
is still hope to search for these interesting long-lived molecular
states with double heavy flavor at facilities such as the Large
Hadron Collider and RHIC.

All the molecule states (except those with $\S_Q$) are very stable
because their components have a long lifetime around
$10^{-13}\sim10^{-12}$s. On the other hand, the width of $\S_c$ is
about 2.2 MeV \cite{pdg2010}, this narrow width ensures relatively
long lifetime for the ``$\S_cX$"-type molecules. Such states can
decay into $X\L_c^+\pi$ followed  by $\L_c^+\rightarrow
pK^-\pi^+$. For the bottomed case, $\S_b X\rightarrow X\L_b^0\pi$
followed by $\L_b^0\rightarrow \L_c^+\pi^-$ and $\L_c^+\rightarrow
pK^-\pi^+$. These decay modes may be helpful to search for such
states in the future experiment.

%%%%%%%%%%%%%%%%%%%%%%%%%%%%%%%%
\section*{Acknowledgments}
%%%%%%%%%%%%%%%%%%%%%%%%%%%%%%%%

One of the authors (N. L.) is very grateful to Z. G. Luo and Dr. Y. R. Liu for
very helpful discussions. This project was supported by the National Natural
Science Foundation of China under Grants 11075004, 11021092 and Ministry of
Science and Technology of China (2009CB825200).

%\bibliographystyle{prd}
%\bibliography{coupling}

%\iffalse

%%%%%%%%%%%%%%%%%%%%%%%%%%%%%%%%%%%%%%%%%%%%%%
\section{APPENDIX}\label{APPENDIX}
%%%%%%%%%%%%%%%%%%%%%%%%%%%%%%%%%%%%%%%%%%%%%%

%%%%%%%%%%%%%%%%%%%%%%%%%%%%%%%%%%%%%%%%%%%%%%%%%
\subsection{Some Helpful Functions}\label{FUNCTION}

The functions $H_i$ etc are defined as,
\begin{eqnarray}
H_0(\L,m,r)&=&Y(u r)-\f{\l}{u}Y(\l r)
-\f{r\beta^2}{2u}Y(\l r), \qquad
H_1(\L,m,r)=Y(u r)-\f{\l}{u}Y(\l r)
-\f{r\l^2\beta^2}{2u^3}Y(\l r), \n\\
H_2(\L,m,r)&=&Z_1(u r)-\f{\l^3}{u^3}Z_1(\l r)
-\f{\l \beta^2}{2\mu^3}Y(\l r),\qquad
H_3(\L,m,r)=Z(u r)-\f{\l^3}{u^3}Z(\l r)
-\f{\l \beta^2}{2u^3}Z_2(\l r),\n\\
M_0(\L,m,r)&=&-\f{1}{\theta r}\left[\cos(\theta r)-e^{-\l r}\right]
+\f{\beta^2}{2\theta\l}e^{-\l r}, \qquad
M_1(\L,m,r)=-\f{1}{\theta r}\left[\cos(\theta r)-e^{-\l r}\right]
-\f{\l\beta^2}{2\theta^3}e^{-\l r},\n\\
M_3(\L,m,r)&=&-\left[\cos{(\theta r)}-\f{3\sin{(\theta r)}}{\theta r}
-\f{3\cos{(\theta r)}}{\theta^2r^2}\right]\f{1}{\theta r}
-\f{\l^3}{\theta^3}Z(\l r)-\f{\l\beta^2}{2\theta^3}Z_2(\l r),
\end{eqnarray}
where,
\begin{eqnarray*}
 \beta^2=\L^2-m^2,\quad u^2=m^2-Q_0^2,\quad \theta^2=-(m^2-Q_0^2),
\quad\l^2=\L^2-Q_0^2,
\end{eqnarray*}
and
\begin{eqnarray*}
 Y(x)=\f{e^{-x}}{x},\quad Z(x)=\left(1+\f{3}{x}+\f{3}{x^2}\right)Y(x),\quad
 Z_1(x)=\left(\f{1}{x}+\f{1}{x^2}\right)Y(x),\quad Z_2(x)=(1+x)Y(x).
\end{eqnarray*}
Fourier transformation formulae read:
\begin{eqnarray}
\f{1}{u^2+\bm{Q}^2}&\rightarrow&\f{u}{4\pi}H_0(\L,m,r),\quad
\f{\bm{Q}^2}{u^2+\bm{Q}^2}\rightarrow-\f{u^3}{4\pi}H_1(\L,m,r), \n\\
\f{\bm{Q}}{u^2+\bm{Q}^2}&\rightarrow&\f{iu^3}{4\pi}\bm{r}H_2(\L,m,r),\quad
\f{Q_iQ_j}{u^2+\bm{Q}^2}\rightarrow-\f{u^3}{12\pi}\left[H_3(\L,m,r)
k_{ij}+H_1(\L,m,r)\delta_{ij}\right], \label{FTformula}
\end{eqnarray}
where, $k_{ij}=3\f{r_ir_j}{r^2}-\delta_{ij}.$ If the form factor
is not introduced, there will be delta terms in the second and
fourth formule. In the above expressions, we employ another
function, $-\f{u^3}{4\pi}[H_1(\L,m,r)-H_0(\L,m,r)]$, to substitute
for the delta term. If one neglects the delta term, one should
adopt the following formulae,
\begin{eqnarray}
\f{\bm{Q}^2}{u^2+\bm{Q}^2}\rightarrow-\f{u^3}{4\pi}H_0(\L,m,r), \quad
\f{Q_iQ_j}{u^2+\bm{Q}^2}\rightarrow-\f{u^3}{12\pi}\left[H_3(\L,m,r)
k_{ij}+H_0(\L,m,r)\delta_{ij}\right].
\end{eqnarray}
If $u^2=m_{ex}^2-Q_0^2<0$, the last formula of the Eq.
(\ref{FTformula}) should be
\begin{eqnarray}
\f{Q_iQ_j}{u^2+\bm{Q}^2}\rightarrow-\f{\theta^3}{12\pi}\left[M_3(\L,m,r)
k_{ij}+M_1(\L,m,r)\delta_{ij}\right].
\end{eqnarray}
Accordingly, we make the replacement $M_1(\L,m,r)\rightarrow
M_0(\L,m,r)$ to neglect the delta term.

%%%%%%%%%%%%%%%%%%%%%%%%%%%%%%%%%%%%%%%%%%%%%%%%%%%%%%%%%%%%%%%%%
\subsection{The Numerical Results of The ``$A_QB_Q$" Systems
When The Contact Term Is Included} \label{numerical:contact}

For comparison, we collect the numerical results of the ``$A_QB_Q$" systems in
Tables~\ref{numerical:contact1} and~\ref{numerical:contact2} when the contact
term is included.
\begin{table}[htp]
\renewcommand{\arraystretch}{1.2}
\centering \caption{The binding solutions of the spin-singlet ``$A_QB_Q"$
systems when the interaction potential includes the contact
term.}\label{numerical:contact1}
\begin{tabular*}{16cm}{@{\extracolsep{\fill}}ccccccc}
\toprule[1pt]\addlinespace[3pt]
            &      \multicolumn{6}{c}{$S=0$}  \\ [3pt]
            &        \multicolumn{3}{c}{$Q=c$}
&\multicolumn{3}{c}{$Q=b$} \\ [3pt]
     States &  $\L$(GeV) &  B.E.(MeV)  &  $r_{rms}$(fm)
& $\L$(GeV) & B.E.(MeV)  &$r_{rms}$(fm)   \\
\specialrule{0.8pt}{3pt}{3pt}
$[\S_Q\Xi_Q^{'}]^{I=\f{1}{2}}_{S}$&  0.80      &$52.87$    &   0.64
&   0.80         &$162.52$  &   0.34   \\
                            &   0.85     &$82.54$    &   0.55
&   0.85         &$219.83$  &   0.30   \\
                            &   0.90     &$137.34$   &   0.45
&   0.90         &$321.82$  &   0.26   \\ [7pt]
$[\S_Q\Xi_Q^{'}]^{I=\f{3}{2}}_{S}$&   1.30     &$ 0.23$    &  5.44
&   1.00       &  $0.45$    &   3.49    \\
               &   1.40     &  $1.48$    &   3.03
&   1.10       & $ 0.45$    &  3.49     \\
               &   1.50     & $4.11$     &   2.10
&   1.20       &  $4.61$    &  1.66     \\  [7pt]
$[\Xi_Q^{'}\O_Q]^{I=\f{1}{2}}_{S}$&   1.10     & $0.94$    &  4.56
&  0.90        & $1.12$     &   2.39    \\
               &   1.20     &$1.03$      &  3.29
&  1.00        &  $5.36$    &  1.41     \\
               &   1.30     & $2.53$     &  2.35
&  1.10        & $7.59$     &  1.32     \\   [7pt]
$[\S_Q\O_Q]^{I=1}_{S}$ &   0.90     &  $2.55$    &  2.08
&  0.80        &  $4.89$    &  1.18     \\
               &   0.95     &$8.80$      &  1.26
&  0.85        & $15.30$    & 0.78      \\
               &   1.00     &$18.69$     &  0.94
&   0.90       &$31.25$     & 0.61      \\  [7pt]
$[\Xi_Q\L_Q]^{I=\f{1}{2}}_{S}$    &   1.00     & $0.73$    &   3.67
&  0.80        &   0.55     &   2.94   \\
               &   1.10     &$1.77$    &   2.55
&   0.85       &  4.93      &   1.25   \\
               &   1.20     &$2.31$    &   2.28
&   0.90       &   10.39    &  0.96    \\  [7pt]
$[\S_Q\L_Q]^{I=1}_{S=0}$ &   1.80     &$0.21$    &   5.96
& 0.80         &  $0.13$    &   6.53   \\
               &   1.90     & $0.83$   &  3.99
& 1.00         & $0.19$     &  6.06    \\
               & 2.00       &$2.00$    &  2.69
&  1.20        &  $0.32$    & 5.35     \\  [7pt]
$[\S_Q\Xi_Q]^{I=\f{1}{2}}_{S}$&   0.90   & $2.59$      &  2.12
&  0.80        & $6.13$     & 1.08    \\
               &   0.95     &$10.29$   &   1.22
&  0.85        & $14.88$    & 0.81    \\
               &   1.00     &$24.38$   &   0.88
&  0.90        & $30.65$    & 0.63    \\   [7pt]
$[\S_Q\Xi_Q]^{I=\f{3}{2}}_{S}$&  1.50       &$0.42$    &   4.55
&  0.85     &$ 0.85$     & 2.56       \\
            &  1.60      &$1.28$       &   3.06
&  1.00     &$ 2.25$     & 1.82       \\
            &  1.70      &$2.66$       &   2.30
&  1.20     &$ 3.69$     &  1.61      \\   [7pt]
$[\Xi_Q^{'}\L_Q]^{I=\f{1}{2}}_{S}$&  1.10    &$ 0.45$   &   4.39
&  0.85     & $1.55$     &   2.00     \\
            &  1.20      &$1.02$       &   3.32
&  1.00     &$7.16$     & 1.16       \\
            &  1.30      &$1.86$       &   2.70
&  1.20     &$11.56$     & 1.08       \\   [7pt]
$[\Xi_Q^{'}\Xi_Q]^{I=0}_{S}$&  1.10       & $1.85$   & 2.59
&   1.00    & $2.51$     &1.65        \\
            &  1.20      & $6.37$      &   1.60
&   1.10    & $11.11$    &  1.05      \\
            &  1.30      & $12.28$     &   1.27
&   1.20    & $21.83$    & 0.87       \\   [7pt]
$[\Xi_Q^{'}\Xi_Q]^{I=1}_{S}$ &  1.25       & $0.23$   &  5.31
&   0.95    &$0.80$      &  2.65      \\
            &  1.30      &  $0.47$     &  4.41
&  1.00     &  $1.79$    &  1.97      \\
            &  1.40      &  $1.25$     &   3.08
&  1.20     &$5.95$      &  1.38      \\   [7pt]
$[\Xi_Q\O_Q]^{I=\f{1}{2}}_{S}$ &  1.00       &  $0.52$  &  4.15
&  0.90     &  $1.34$    &  2.18      \\
            &  1.10      &  $1.99$     &   2.50
&  1.00     &  $7.82$    & 1.21       \\
            &  1.20      &  $3.41$     &   2.06
&  1.20     & $15.97$    &  1.06      \\   [7pt]
$[\L_Q\O_Q]^{I=0}_{S}$&  0.90    &  $0.20$     &   5.29
&0.80       &  1.64      &  1.87      \\
            &  0.95      &$2.11$     &  2.33
& 0.85      &  6.53      &  1.12      \\
            &  1.00      &$5.77$     &  1.55
& 0.90      & 14.20      & 0.86       \\   [7pt]
\bottomrule[1.pt]
\end{tabular*}
\end{table}

\begin{table}[htp]
\renewcommand{\arraystretch}{1.3}
\centering \caption{The binding solutions of spin-triplet ``$A_QB_Q$" systems
 when the interaction potential includes the contact
 term.}\label{numerical:contact2}
\begin{tabular*}{17cm}{@{\extracolsep{\fill}}ccccccccccc}
\toprule[1.pt]\addlinespace[3pt]
            &   \multicolumn{10}{c}{$S=1$} \\  [3pt]
            &     \multicolumn{5}{c}{$Q=c$}       & \multicolumn{5}{c}{$Q=b$}\\ [3pt]
     States &  $\L$(GeV) &  B.E.(MeV)  &  $r_{rms}$(fm) & $P_S(\%)$ & $P_D(\%)$
& $\L$(GeV) & B.E.(MeV)  &$r_{rms}$(fm)&  $P_S(\%)$     & $P_D(\%)$\\
\specialrule{0.8pt}{3pt}{3pt}
$[\S_Q\Xi_Q^{'}]^{I=\f{1}{2}}_S$&  0.90      &$1.39$ &   2.97    &   96.08 &$3.92$
&   0.80         &   5.24   &  1.51      & 88.01     &  11.99   \\
                            &   1.00     &  8.64     &  1.51     & 93.78   &  6.22
&   0.90         & 18.07    &   1.02     & 86.04     &  13.96   \\
                            &   1.10     & 20.31     &  1.14     & 92.74   & 7.26
&   0.95         & 27.54    &   0.90     & 85.53     &  14.47   \\ [8pt]
$[\S_Q\Xi_Q^{'}]^{I=\f{3}{2}}_S$&   0.85 & 3.04      &  1.98     &  97.19  & 2.81
&   0.80       &  16.92     &   0.82     &   92.42   &  7.58    \\
               &   0.90     &  8.66      &   1.30    &  96.77   &  3.23
&   0.85       &  36.05     &  0.65      &   91.49   &  8.51    \\
               &   0.95     &  16.21     &   1.00    &  97.31   &  2.69
&   0.90       & 56.13      &  0.54      &   92.82   &  7.18    \\ [8pt]
$[\Xi_Q^{'}\O_Q]^{I=\f{1}{2}}_S$&  0.90  &  2.37     &  2.12    & 99.38    & 0.62
&  0.85        & 11.23      &   0.80     &  99.56    &  0.44    \\
               &  0.95      &   10.33    &   1.19    &  98.51    &  1.49
&  0.90        &  31.96     &   0.61     &  97.38    &  2.62    \\
               &  1.00      &  20.87     &   0.91    &  98.08    &  1.92
&  0.95        & 56.35      &   0.52     &  95.42    &  4.58    \\ [8pt]
$[\S_Q\O_Q]^{I=1}_{S}$&  0.95&  $1.91$   &  2.42     &  99.84    & 0.16
&  0.80        &  1.35      &  2.03      &  99.27    & 0.23     \\
               &  1.00      & $5.15$     &1.62       &  99.71    & 0.29
&  0.85        &  5.53      & 1.20       &  99.48    & 0.52     \\
               &   1.10     &$15.31$     &1.08       &  99.40    & 0.60
&   0.90       &  12.11     &  0.92      &  99.17    &   0.83    \\  [8pt]
$[\Xi_Q\L_Q]^{I=\f{1}{2}}_{S}$&   1.00   & $0.73$    &   3.67   & 100.00  & 0.00
&   0.80       &   0.55     &   2.94     &  100.00   &   0.00    \\
               &   1.10     & 1.77       &   2.55    &  100.00   &  0.00
& 0.85         &   4.93     &  1.25      &  100.00   &  0.00    \\
               &   1.20     &  2.31      &   2.28    &  100.00   & 0.00
&  0.90        &   10.39    &  0.96      &  100.00   &  0.00     \\  [8pt]
$[\S_Q\L_Q]^{I=1}_S$ & 0.80 &  2.19    &   2.50      &  95.11      &  4.98
& 0.80         &  28.86     &  0.79    &  87.95      &  12.05    \\
               & 0.85       & $5.32$   &   1.75      &  95.23      &  4.77
& 0.85         & 41.93      &  0.67    &   89.83     &  10.17   \\
               &  0.90      &  $9.34$  &   1.39      &  95.67      &  4.33
&  0.90        &  56.53     &  0.58    &  92.07      &  7.93    \\   [8pt]
$[\S_Q\Xi_Q]^{I=\f{1}{2}}_S$&  0.90    &  $0.57$     &   3.99   &   99.96     &  0.04
&  0.80        & 1.10       &   2.22   &   99.12     &  0.08    \\
               &  1.00      & $9.52$   &  1.31       &   99.88     &  0.12
&  0.85        & 6.44       &  1.15    &   99.72     &  0.28    \\
               &  1.10      & $25.78$  &  0.92       &   99.88     &   0.12
&  0.90        &  15.96     &  0.84    &   99.56     &   0.44   \\   [8pt]
$[\S_Q\Xi_Q]^{I=\f{3}{2}}_S$&  0.95     &$0.49$      &   4.16      &   99.81  &  0.19
&  0.80     &   4.04     &  1.29       &   99.59  &  0.41  \\
            &  1.00      &$ 1.56$      &   2.60      &   99.68  &  0.32
&  0.85     &  0.90      &   0.98      &   98.73  &  1.27  \\
            &  1.10      &$ 6.02$      &   1.45      &   99.61  &  0.39
&  0.90     & 14.91      &  0.82       &   97.99  &  2.01  \\    [8pt]
$[\Xi_Q^{'}\L_Q]^{I=\f{1}{2}}_S$&  0.95     & $1.25$     &   2.91      &   99.88  &  0.12
&  0.80     & $3.15$     &  1.43       &   99.76  &  0.24  \\
            &  1.00      &$3.64$       &  1.85    &   99.83  & 0.17
&  0.85     & 8.40       &  1.01       &   99.40  & 0.60   \\
            &  1.10      &$12.23$      &   1.14   &   99.83  &  0.17
&  0.90     & 15.51      &  0.82       &   99.15  &  0.85  \\    [8pt]
$[\Xi_Q^{'}\Xi_Q]^{I=0}_S$&   0.90     & $4.89$   &1.67         &   98.42  & 1.58
&   0.80    & 5.71       & 1.39        &   86.06  & 13.94  \\
            &   0.95     & $14.60$     &  1.10    &   98.93  &  1.07
&   0.85    &   16.62    &  0.90       &   91.74  &  8.26   \\
            &   1.00     & $28.43$     &0.86      &   99.36  &  0.64
&   0.90    &  35.81     & 0.66        &   96.05  &  3.95   \\   [8pt]
$[\Xi_Q^{'}\Xi_Q]^{I=1}_S$&   0.90     &$0.81$    &  3.41   & 99.63 &  0.37
&   0.80    & 0.77       &  2.50       &   98.86  &  1.14   \\
            &  0.95     &  $4.39$     &  1.69      &   99.39  &  0.61
&  0.85     &  10.07    &  0.93       &   98.57    &  1.43  \\
            &  1.00     &$9.88$       &  1.22      &   99.31  &  0.69
&  0.90     &  23.63     &  0.70       &   98.19  &  1.81   \\   [8pt]
$[\Xi_Q\O_Q]^{I=\f{1}{2}}_{S}$ &  0.90 &  $0.45$  &  4.22   &99.75  &  0.25
&  0.85     &  3.01      &   1.45      &   97.79  &  2.21   \\
            &  0.95      &  $5.94$     &1.49      &   99.34  &  0.66
&  0.90     &  19.44     &  0.76       &   97.00  &  3.00   \\
            &  1.00      & $15.35$     &  1.04    &   99.23  &  0.77
&  0.95     &  41.92     &  0.58       &  97.13   &   2.87  \\  [8pt]
$[\L_Q\O_Q]^{I=0}_S$&  0.90    &  $0.20$     &   5.29  & 100.00   &  0.00
&  0.80     &  1.64      &  1.87       &   100.00 &  0.00   \\
            &  0.95      &  2.11       &  2.33         &  100.00  &  0.00
&  0.85     &  6.53      &  1.12       &   100.00 &  0.00   \\
            &  1.00      &  5.77       &   1.55   &  100.00       & 0.00
&  0.90     &  14.20     &  0.86       &   100.00 &  0.00   \\  [3pt]
\bottomrule[1pt]
\end{tabular*}\label{numerical:charmed2}
\end{table}

%%%%%%%%%%%%%%%%%%%%%%%%%%%%%%%%%%%%%%%%%%%%%%%%%%%%%%%
\subsection{The Numerical Results of The Baryon-antibaryon
Systems With the OPE potential}
\label{anti}

As a byproduct, we present the binding solutions of the heavy baryon-antibaryon
systems with the pion-exchange potential. It is straightforward to obtain the
potential via changing the sign of the potential for the baryon-baryon systems
since the G-parity of the pion is negative. Our results indicate that the
pion-exchange alone is strong enough to form some bound states. The numerical
results are collected in Tables~\ref{ReBBS}-\ref{ReBBT}.
\begin{table}[htp]
\renewcommand{\arraystretch}{1.1}
\centering \caption{The numerical results of the spin-singlet
heavy baryon-antibaryon systems with the OPE potential. ``$\L$" is
the cutoff parameter. ``B.E." is the binding energy, and
``$r_{rms}$"is the root-mean-square radius which reflects the size
of the bound state. ``$\times$" denotes no binding
solutions}\label{ReBBS}
\begin{tabular*}{16cm}{@{\extracolsep{\fill}}cccccccc}
\toprule[1.0pt]\addlinespace[5pt]
&\multicolumn{6}{c}{S=0}\\
             &\multicolumn{3}{c}{Q=c}   & \multicolumn{3}{c}{Q=b}\\
Systems      & $\L$(GeV)   &  B.E.(MeV)    & $r_{rms}$(fm)
&  $\L$(GeV) & B.E.(MeV)      & $r_{rms}$(fm)  \\
\specialrule{0.8pt}{3pt}{3pt}
\multirow{3}*{$[\S_Q\bar{\Xi}_Q^{'}]^{I=1/2}_S$}
& \multicolumn{3}{c}{\multirow{3}*{$\times$}}
& 1.80  &   0.01  &  6.75   \\
&       &         &
& 2.00  &   0.02  &  6.66   \\
&       &         &
&  2.50 &    0.03 &  6.50   \\   [5pt]
\multirow{3}*{$[\S_Q\bar{\Xi}_Q^{'}]^{I=3/2}_S$}
&\multicolumn{3}{c}{\multirow{3}*{$\times$}}
&\multicolumn{3}{c}{\multirow{3}*{$\times$}}  \\
&       &        &        &       &        &        \\
&       &        &        &       &        &        \\ [5pt]
\multirow{3}*{$[\S_Q\bar{\L}_Q]^{I=1}_{S}$}
&\multicolumn{3}{c}{\multirow{3}*{$\times$}}
&  0.80 & 0.06   &  9.34   \\
&       &        &         &  1.00      & 0.13   &  4.94   \\
&       &        &         &  1.30      & 0.29   &  2.44   \\ [5pt]
\multirow{3}*{$[\Xi_Q\bar{\Xi}_Q^{'}]^{I=0}_S$}
&\multicolumn{3}{c}{\multirow{3}*{$\times$}}
&\multicolumn{3}{c}{\multirow{3}*{$\times$}}  \\
&       &        &         &            &        &          \\
&       &        &         &            &        &          \\ [5pt]
\multirow{3}*{$[\S_Q\bar{\Xi}_Q^{'}]^{I=3/2}_{S}$}
&\multicolumn{3}{c}{\multirow{3}*{$\times$}}
&\multicolumn{3}{c}{\multirow{3}*{$\times$}}  \\
&       &        &         &            &        &         \\
&       &        &         &            &        &         \\ [5pt]
\bottomrule[1.0pt]
\end{tabular*}
\end{table}

\begin{table}
\renewcommand{\arraystretch}{1.3}
\centering \caption{The numerical results of the spin-triplet
heavy baryon-antibaryon systems with the OPE potential. ``$\L$" is
the cutoff parameter. ``B.E." is the binding energy while
``$r_{rms}$" is the root-mean-square radius. ``$P_S$" and ``$P_D"$
are the probabilities of the S wave and the D wave, respectively.
``$\times$" denotes no binding solutions.}\label{ReBBT}
\begin{tabular*}{18cm}{@{\extracolsep{\fill}}cccccccccccc}
\toprule[1.0pt]\addlinespace[5pt]
&\multicolumn{10}{c}{S=1}\\
             &\multicolumn{5}{c}{Q=c}   & \multicolumn{5}{c}{Q=b}\\
Systems      & $\L$(GeV)   &  B.E.(MeV)   & $r_{rms}$(fm)&$P_S$  & $P_D$
&  $\L$(GeV) & B.E.(MeV)      & $r_{rms}$(fm)& $P_S(\%)$     & $P_D(\%)$   \\
\specialrule{0.8pt}{3pt}{3pt}
\multirow{3}*{$[\S_Q\bar{\Xi}_Q^{'}]^{I=1/2}_S$}
& 1.20  & 1.37   & 2.80    &  89.79     &  10.21
& 0.80  & 4.05   & 1.60    &  74.87     &  25.13  \\
& 1.30  & 4.95   & 1.69    &  84.74     &  15.26
& 0.90  & 9.86   & 1.21    &  70.37     &  29.63  \\
& 1.40  & 11.24  & 1.25    &  81.80     &  18.92
& 1.00  & 19.09  & 0.98    &  67.15     &  32.85  \\ [7pt]
\multirow{3}*{$[\S_Q\bar{\Xi}_Q^{'}]^{I=3/2}_{S}$}
& \multicolumn{5}{c}{\multirow{3}*{$\times$}}
& 1.20  & 0.72   & 2.68    &  94.08     &  5.92  \\
&       &        &         &            &
& 1.30  & 1.75   & 1.88    &  92.19     &  7.81  \\
&       &        &         &            &
& 1.50  & 5.80   & 1.18    &  89.44     &  10.56  \\ [7pt]
\multirow{3}*{$[\S_Q\bar{\L}_Q]^{I=1}_S$}
& 1.60  & 0.03   &  6.65   &  97.47     &  2.53
& 0.90  & 1.41   &  2.75   &  87.92     &  12.08 \\
& 1.80  & 1.38   &  2.99   &  94.34     &  5.66
& 1.10  & 4.42   &  1.59   &  86.20     &  13.80 \\
& 2.00  & 5.87   &  1.56   &  92.02     &  7.98
& 1.30  & 11.71  &  1.04   &  84.88     &  15.12 \\ [7pt]
\multirow{3}*{$[\Xi_Q\bar{\Xi}_Q^{'}]^{I=0}_{S}$}
& 2.10  & 0.46   & 4.13    &  97.15     & 2.85
& 1.00  & 0.24   & 4.52    &  93.44     & 6.56  \\
& 2.30  & 3.16   & 2.89    &  94.51     & 5.49
& 1.20  & 1.89   & 1.91    &  90.07     & 9.93 \\
& 2.50  & 9.15   & 1.20    &  92.55     & 7.45
& 1.40  & 6.22   & 1.20    &  87.85     & 12.15 \\ [7pt]
\multirow{3}*{$[\Xi_Q\bar{\Xi}_Q^{'}]^{I=1}_{S}$}
& \multicolumn{5}{c}{\multirow{3}*{$\times$}}
& \multicolumn{5}{c}{\multirow{3}*{$\times$}}\\
&       &        &         &            &
&       &        &         &            &     \\
&       &        &         &            &
&       &        &         &            &     \\   [3pt]
\bottomrule[1.0pt]
\end{tabular*}
\end{table}

\subsection{The Dependence of The Binding Energy on The Cutoff Parameter}\label{PLOT}

\end{document}